\documentclass[traditabstract]{aa} 
\usepackage{graphicx}
\usepackage{multirow}
\usepackage{txfonts}
\usepackage{arydshln}

\usepackage{appendix}
\usepackage{bigdelim}
\usepackage{soul,ulem}
\usepackage[comma,authoryear]{natbib}
\def\cm#1{\ifmmode {\,{\rm cm^{-#1}}}                  
        \else \hbox{$\,${\rm cm$^{\rm -#1}$}}\fi}
\def\raw {\ifmmode\rightarrow\else$\rightarrow$\fi}
\def\ex#1{\ifmmode {\times 10^{#1}}         
        \else \hbox{{$\times 10^{\rm #1}$}}\fi}

\newcommand{\ioe}{intermediate/outer envelope}
\newcommand{\water}{\mbox{H$_2$O}}
\newcommand{\nmols}{HNCO, HNCS, HC$_3$N, and NO}
\newcommand{\iram}{IRAM-30\,m}
\newcommand{\hso}{{\it Herschel}}

\newcommand{\kms}{\mbox{km~s$^{-1}$}}
\newcommand{\s}{\mbox{$''$}}
\newcommand{\mloss}{\mbox{$\dot{M}$}}
\newcommand{\my}{\mbox{$M_{\odot}$~yr$^{-1}$}}
\newcommand{\ls}{\mbox{$L_{\odot}$}}
\newcommand{\msun}{\mbox{$M_{\odot}$}}

\newcommand{\rs}{\mbox{$R_{\star}$}}
\newcommand{\rd}{\mbox{$R_{\rm c}$}}

\newcommand{\vexp}{\mbox{$V_{\mathrm{exp}}$}}

\newcommand{\vlsr}{\mbox{$V_{\mathrm{LSR}}$}}

\newcommand{\eu}{\mbox{$E_{\mathrm{u}}$}}
\newcommand{\tex}{\mbox{$T_{\mathrm{ex}}$}}

\newcommand{\trot}{\mbox{$T_{\mathrm{rot}}$}}
\newcommand{\ntot}{\mbox{$N_{\mathrm{tot}}$}}
\newcommand{\tkin}{\mbox{$T_{\mathrm{kin}}$}}
\newcommand{\dens}{\mbox{$n_{\mathrm{H_2}}$}}
\newcommand{\nc}{\mbox{$n_{\mathrm{crit}}$}}

\newcommand{\trece}{$^{13}$CO\,($J$=1$-$0)}

\newcommand{\docem}{$^{12}$CO}
\newcommand{\trecem}{$^{13}$CO}

\newcommand{\intensity}{\mbox{erg\,s$^{-1}$\,cm$^{-2}$}}

\newcommand{\oh}{\mbox{OH\,231.8$+$4.2}}

\newcommand{\invsec}{\mbox{s$^{-1}$}}
\newcommand{\microns}{\mbox{$\mu$m}}

\begin{document}

   \title{New N-bearing species towards \oh:}

   \subtitle{HNCO, HNCS, HC$_3$N, and NO\thanks{Based on observations carried out with the \iram\, Telescope. IRAM is supported by INSU/CNRS (France), MPG (Germany), and IGN (Spain).}}

   \author{L.~Velilla Prieto\inst{1,2}
          \and
          C.~S\'anchez Contreras\inst{2}
          \and
          J.~Cernicharo\inst{1,3}
          \and
          M.~Ag\'undez\inst{1,3,4}
          \and
          G.~Quintana-Lacaci\inst{1,3}
          \and
          J.~Alcolea\inst{5}
          \and
          V.~Bujarrabal\inst{6}
          \and
          F.~Herpin\inst{4}
          \and
          K.~M.~Menten\inst{7}
          \and
          F.~Wyrowski\inst{7}
          }

   \institute{Grupo de Astrof\'isica Molecular. Instituto de Ciencia de Materiales de Madrid, CSIC, c/ Sor Juana In\'es de la Cruz 3, 28049 Cantoblanco, Madrid, Spain\\
              \email{lvelilla@icmm.csic.es}
         \and 
         Centro de Astrobiolog\'ia, INTA-CSIC, E-28691 Villanueva de la Ca\~nada, Madrid, Spain
         \and
         Centro de Astrobiolog\'ia, INTA-CSIC, Ctra. de Torrej\'on a Ajalvir km 4, 28850 Torrej\'on de Ardoz, Madrid, Spain
         \and
         Universit\'e de Bordeaux, LAB, UMR 5804, F-33270, Floirac, France
         \and
         Observatorio Astron\'omico Nacional (IGN), Alfonso XII No 3, 28014 Madrid, Spain
         \and
         Observatorio Astron\'omico Nacional (IGN), Ap 112, 28803 Alcal\'a de Henares, Madrid, Spain
         \and
         Max-Planck-Institut f\"{u}r Radioastronomie, Auf dem H\"{u}gel 69, 53121 Bonn, Germany
             }

   \date{Received 7 August 2014 / Accepted 4 December 2014}

  \abstract{
Circumstellar envelopes (CSEs) around asymptotic giant
branch (AGB) are the main sites of
molecular formation. \oh\ is a well studied oxygen-rich CSE around an
intermediate-mass evolved star that, in dramatic contrast to most AGB
CSEs, displays bipolar molecular outflows accelerated up to
$\sim$400\,\kms. \oh\ also presents an exceptional molecular richness
probably due to shock-induced chemical processes. We report the first
detection in this source of four nitrogen-bearing species,
HNCO, HNCS, HC$_3$N, and NO, which have been observed with the
\iram\ radiotelescope in a sensitive mm-wavelength survey towards this
target. HNCO and HNCS are also first detections in CSEs.  The observed
line profiles show that the emission arises in the massive
($\sim$0.6\,\msun) central component of the envelope, expanding with
low velocities of \vexp$\sim$15-30\,\kms, and at the base of the fast
lobes. The NO profiles (with FWHM$\sim$40-50\,\kms) are broader than
those of HNCO, HNCS, and HC$_3$N and, most importantly, broader than
the line profiles of \trecem, which is a good mass tracer. This
indicates that the NO abundance is enhanced in the fast lobes relative
to the slow, central parts. From LTE and non-LTE excitation analysis,
we estimate beam-average rotational temperatures
of \trot$\sim$15-30\,K (and, maybe, up to $\sim$55\,K for HC$_3$N) and
fractional abundances of $X$(HNCO)$\sim$[0.8-1]$\times$10$^{-7}$,
$X$(HNCS)$\sim$[0.9-1]$\times$10$^{-8}$,
$X$(HC$_3$N)$\sim$[5-7]$\times$10$^{-9}$, and
$X$(NO)$\sim$[1-2]$\times$10$^{-6}$.
NO is, therefore, amongst the most abundant N-bearing species in \oh.
We performed thermodynamical chemical equilibrium and chemical
kinetics models to investigate the formation of these N-bearing
species in \oh.  The model underestimates the observed abundances for
HNCO, HNCS, and HC$_3$N by several orders of magnitude, which
indicates that these molecules can hardly be products of standard
UV-photon and/or cosmic-ray induced chemistry in \oh and that other
processes (e.g.\,shocks) play a major role in their formation. For NO,
the model abundance, $\approx$10$^{-6}$, is compatible with the
observed average value; however, the model fails to reproduce the NO
abundance enhancement in the high-velocity lobes (relative to the slow
core) inferred from the broad NO profiles.  The new detections
presented in this work corroborate the particularly rich chemistry
of \oh, which is likely to be profoundly influenced by shock-induced
processes, as proposed in earlier works.

}


   \maketitle

\section{Introduction}\label{sec:intro}
For about 40 years, circumstellar chemistry has been a fertile field as a
source of new molecular discoveries and the development of physical and
chemical models.  
Circumstellar envelopes (CSEs) around asymptotic giant
branch (AGB) stars are formed as the result of the intense mass
loss process undergone by these objects. AGB CSEs are composed
of molecular gas and dust, standing among the most complex chemical
environments in space \citep[][and references therein]{cer00,ziu06}.

Circumstellar envelopes are classified according to their elemental
[C]/[O] ratio, which are carbon-rich or oxygen-rich if the ratio is
$\textgreater$ 1 or $\textless$ 1, respectively (objects with
[C]/[O]$\sim$1 are designed as S-type stars). The chemistry of CSEs
is very dependent on the relative abundances of oxygen and carbon. In
the case of oxygen-rich CSEs, carbon plays the role of ``limiting
reactant'' and is supposed to be almost fully locked up in CO, which
is a very abundant and stable species, while the remaining oxygen is
free to react with other atoms, thereby forming additional oxygen-bearing
molecules. This is why O-rich envelopes are relatively poor in
C-bearing molecules other than CO, while C-rich ones show low
abundances of O-bearing species \citep[e.g.][]{buj94}. 

To date, most
of the observational efforts to detect new circumstellar molecules
have focused on C-rich sources, which are believed to have a more
complex and rich chemistry than their oxygen counterparts. The most
studied object of this kind is the carbon-rich evolved star
IRC$+$10216 in whose envelope $\sim$80 molecules have been discovered
\citep[e.g.][]{sol71,morr75,cer87,cer00,cab13}.  Recent works suggest,
however, that O-rich shells may be more chemically diverse than
originally thought. For example, some unexpected chemical compounds
(e.g. HNC, HCO$^{+}$, CS, CN)  have been identified in a number of
O-rich late-type stars, including the object \oh\ studied in this work
\citep{san97,ziu09}.
The chemical processes that lead to the formation of these and other 
species in O-rich CSEs remain poorly known. 

In this paper, we present our recent results for the study of \oh: an
O-rich CSE around an intermediate-mass evolved star that, to date,
displays the richest chemistry amongst the objects in its class. We
report the detection of HNCO, HNCS, HC$_3$N, and NO as part of a
sensitive molecular line survey of this object in the mm-wavelength
range with the \iram\, telescope \citep[][full survey data to be
  published soon by Velilla et al., in prep.]{vel13}.  We have
detected hundreds of molecular transitions, discovering $>$30 new
species (including different isotopologues) and enlarging the sequence
of rotational transitions detected for many others, in this source. This has led to
very detailed information on the physico-chemical global structure of
this envelope.
 
\begin{figure}[hbtp!] 
\centering
\includegraphics[width=0.90\hsize]{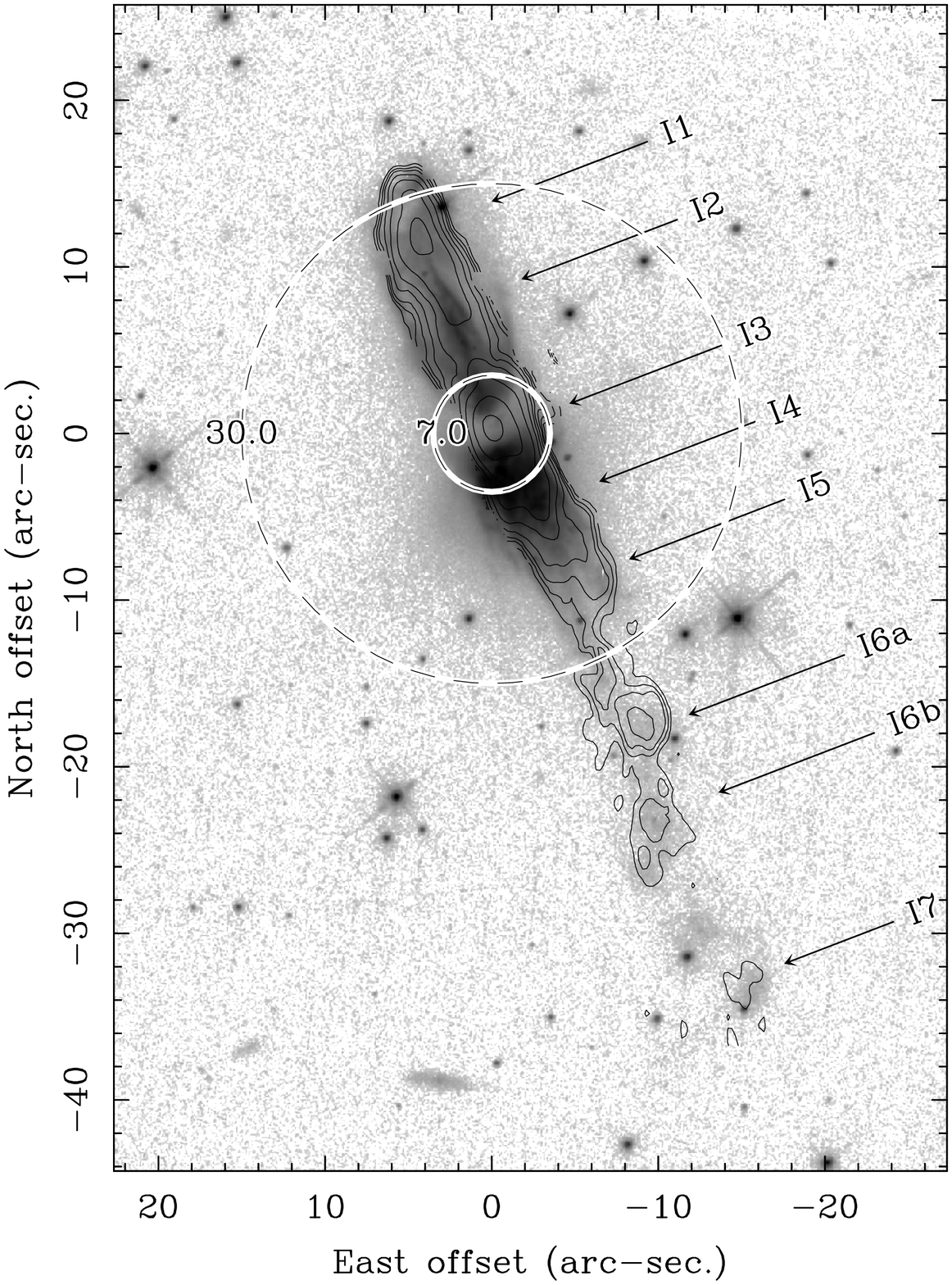}
\caption{
Composite image of \oh\ displaying: (grey scale) the dust
distribution as observed with $HST$/WFPC2 and the broad-band F791W
filter \citep{buj02}; (contours) the molecular outflow as traced by
the \docem\,(2--1) emission (velocity integrated) mapped with
1\farcs5$\times$0\farcs7-resolution \citep[see Fig.\,4
in][]{alc01}. The dashed circles show the area covered by the largest
and smallest telescope beams (HPBW) of the \iram\ observations
presented in this work.  The different emitting clumps, I1-I7, are
labelled as in
\cite{alc01}. 
The \vlsr\ range (in \kms) of each clump is I1) [$-$80:$-$30], I2)
[$-$30:$+$10], I3) [$+$10:$+$55], I4) [$+$55:$+$80], I5)
[$+$80:$+$150], I6a) [$+$150:$+$205], I6b) [$+$205:$+$230], and I7)
[$+$230:$+$285] \citep[see Table\,2 in][for more details on the
physical properies of the clumps]{alc01}.}
\label{fig:oh231}
\end{figure}

\oh\ (Fig.\,\ref{fig:oh231}), discovered by \cite{tur71}, is a well-studied
bipolar nebula around an OH/IR source\footnote{OH/IR objects are
infrared-bright evolved stellar objects with a dense envelope showing
prominent OH maser emission.}. Although its evolutionary stage is not
clear owing to its many unusual properties, it is believed to be a
planetary nebula (PN) precursor probably caught in a short-lived
transitional phase.  The obscured central star, named QX\,Pup, is
classified as M9-10\,III and has a Mira-like variability consistent
with an evolved AGB star
\citep{coh81,fea83,kas92,san04}. The late evolution of this object may 
have been complex since it has a binary companion star (of type A0\,V)
that has been indirectly identified from analysis of the spectrum
of the hidden central source reflected by the nebular
dust \citep{coh85,san04}. The system, located at $\sim$ 1500
pc \citep{choi12}, has a total luminosity of $\sim$10$^4$\ls,\,
and its systemic velocity relative to the Local Standard of Rest
is \vlsr$\sim$34\,\kms. \oh\ is very likely a member of the open cluster 
M\,46 with a progenitor  mass of $\sim$3\,\msun\ \citep{jur85}. 

Most of the nebular material of \oh\ is in the form of dust and
molecular gas, which are best traced by scattered starlight and by the
emission from rotational transitions of CO, respectively \citep[see
e.g.][]{san97,alc01,buj02}. The molecular gas is cool ($\sim$10--40\,K
over the bulk of the nebula) and massive ($\sim$1\,\msun). With a spatial distribution similar to that of dust, this gas is located in a
very elongated and clumpy structure with two major components
(Fig.\,\ref{fig:oh231}): ($i$) a central core (clump I3) with an
angular diameter of $\sim$6-8\arcsec, a total mass of
$\sim$0.64\,\msun, and low expansion velocity ($\sim$6-35\,\kms), and
($ii$) a highly collimated $\la$6\s$\times$57\s\ bipolar outflow, with
a total mass of $\sim$0.3\msun\ and expansion velocities that increase
linearly with the distance from the centre, reaching values of up to
$\sim$200 and 430\,\kms\ at the tip of the northern and southern lobes,
respectively. The temperature in the lobes is notably low,
$\sim$10-20\,K \citep{san97,alc01}.  A shock-excited atomic/ionized
gas nebulosity, hotter ($\sim$10,000\,K) but far less massive
($\sim$2$\times$10$^{-3}$\,\msun), surrounds the front edges of the
molecular outflow delineating two inflated bubble-like, asymmetric
lobes \citep[not shown in Fig.\,\ref{fig:oh231};
see][]{rei87,san00opt,buj02,san04}.

The molecular envelope of \oh\ is remarkably different from the slow,
roughly round winds of most AGB stars; however, its pronounced axial symmetry,
high expansion velocities, and the presence of shocks are
common in objects that have left the AGB phase and are evolving to the
PN stage, so-called pre-PNs \citep{neri98,buj01,cc10,san12}. It is
believed that the nebula of \oh\ was created as the result of a huge
mass loss that occurred during the late-AGB evolution of the primary at a
rate of \mloss$\approx$10$^{-4}$\,\my.
With a total linear momentum of $\sim$27\,\msun\kms,
the bipolar flow is interpreted as the result of a sudden axial acceleration of the
envelope. It is probable that such an acceleration resulted from the
violent collision between underlying jets (probably emanating from the
stellar companion) on the slowly expanding AGB
envelope \citep{san00,alc01,buj02,san04}; this is one plausible
scenario that has been proposed to explain the shaping and
acceleration of bipolar pre-PNs and PNs \citep[e.g.][]{sah98,bal02}.
Recently, \cite{sab14} have found indications of a well-organized
magnetic field parallel to the major axis of the CO-outflow that could
point to a magnetic-outflow launching mechanism. As mentioned by these
authors, the magnetic field could have, alternatively, been dragged by
the fast outflow, which may have been launched by a different
mechanism. The linear distance-velocity relation observed in the
CO-outflow (with a projected velocity gradient of $\nabla
V$$\sim$6.5\,\kms\,arcsec$^{-1}$) suggests that the acceleration of
the lobes took place $\sim$\,800\,yr ago in less than
$\sim$150\,yr. The low-velocity central core is thought to be the
fossil remnant of the AGB CSE.

\oh\ is the chemically richest CSE around an O-rich low/intermediate mass 
evolved star. In addition to the typical oxygen-rich content, with
molecules such as H$_2$O, OH, or
SiO \citep{bow84,morr87,zij01,san02,des07}, it displays strong lines
of many different molecular species, including many containing
carbon. The full inventory of the molecules reported in \oh\ prior to
our survey are $^{12}$CO, $^{13}$CO, SO, SO$_2$, \water, OH, SiO,
H$_2$S, HCN, H$^{13}$CN, HNC, CS, HCO$^+$, H$^{13}$CO$^+$,OCS,
H$_2$CO, NH$_3$, and NS \citep[][and references
therein]{uki83,gui86,morr87,omo93,san97,lin92,san00}. High-angular
resolution mapping of the HCO$^+$\,($J$=1--0) emission indicates that
this ion is present in abundance in the fast
lobes \citep{san00}. Based on single-dish maps of the SiO\,($J$=5--4)
emission, the abundance of this molecule could also be enhanced in the
lobes \citep{san97}. The spectrum of \oh\ is unusually rich, even for
O-rich CSEs' standards, in lines from S- and N-bearing molecules (for
example, it was the first O-rich CSE in which H$_2$S, NS, CS, and OCS
were detected.) Some of these S- and N-compounds are present in the
envelope at relatively high levels, for example, SO$_2$ and HNC (see
references above). It is believed that extra Si and S are released
into the gas phase from dust grains by shocks. Shocks might also
initiate (endothermic) reactions that trigger the N and S chemistry
and could also be additional suppliers of free atoms and
ions \citep[][]{morr87}.

\section{Observations}\label{sec:obs}

The observations presented in this paper are part of a sensitive
mm-wavelength ($\sim$79-356\,GHz) survey carried out with the \iram\,
telescope (Pico Veleta, Granada, Spain) towards the CSEs of two
O-rich evolved stars: \oh\, and IK\,Tau. Preliminary results from
this survey are reported in \cite{san11} and \cite{vel13}. 

We used the new-generation heterodyne Eight MIxer Receiver
(EMIR)\footnote{\tt
  \tiny{http://www.iram.es/IRAMES/mainWiki/EmirforAstronomers}}, which
works at four different mm-wavelength bands, E090=3\,mm,
E150=2\,mm, E230=1\,mm, and E330=0.9\,mm \citep[][]{car12}.  EMIR was
operated in single-sideband (SSB) mode for band E150 and in dual
sideband (2SB) mode for bands E090, E230, and E330.  E090 and E150 were
observed simultaneously providing 8~GHz and 4~GHz instantaneous bandwidths,
respectively.  E330 was also observed simultaneously with E150,
providing 16 GHz of instantaneous bandwidth, and E230 was observed
alone, providing 16 GHz of instantaneous bandwidth.

Each receiver band was connected to different spectrometers; here we
report data observed with the WILMA autocorrelator, which provides a
spectral resolution of 2\,MHz (i.e.\,7.5-1.7\,\kms\ in the observed
frequency range, 79-356\,GHz), and the fast Fourier transform
spectrometer (FTS) in its 195 kHz spectral resolution mode
(i.e.\,0.7-0.2\,\kms). Both spectrometers provide full coverage of the
instantaneously available frequencies.

 For each band, the two orthogonal polarizations were observed
 simultaneously for a series of tuning steps until we covered the
 total frequency range accessible to each band. The central
 frequencies of the different tuning steps were chosen to provide a
 small frequency overlap between adjacent tunings. The average
 rejection of the image band signal was measured to be $\sim$14\,dB,
 in agreement with the typical values for EMIR; this implies that the
 peak intensity of a line entering through the image band is only
 $\sim$4\,\%\ of its real value.

Observations were performed towards the centre of the nebula
(R.A.2000=$07^h 42^m 16^s.93$, Dec.2000=$-14^{\circ} 42^{'}
50^{''}.20$).  We used the wobbler switching mode, with a wobbler
throw of 120\arcsec\ in azimuth.  The beamwidth of the antenna is in
the range $\sim$30\arcsec-7\arcsec at the observed frequencies
(Table\,\ref{tab:iram_param}). 
These observations thus provide spectra that are spatially integrated over the
slow central core of \oh\ (clump I3, from which the bulk of the
molecular emission arises), and more or less depending on
the observed frequency, from the fast bipolar outflows, always
  leaving out the emission from the most distant and, thus, fastest
  and most tenuous clumps (I6-I7) in the southern lobe (see
  Fig.\,\ref{fig:oh231}).

\begin{table}[hbt!] 
\caption{Main parameters of the \iram\ EMIR receiver at 
representative frequencies.}
\label{tab:iram_param}
\begin{tabular}{c c c c c}
\hline\hline
Frequency & Beam eff. & Forward eff. & HPBW & S/T$_a^*$ \\
(GHz)     &    (\%)         &   (\%)             & (")  & (Jy/K) \\
\hline
86      & 81 & 95 & 29  & 5.9  \\ 
145     & 74 & 93 & 17  & 6.4 \\ 
210     & 63 & 94 & 12  & 7.5 \\
340     & 35 & 81 & 7   & 10.9 \\ 
\hline 
\end{tabular}
\tablefoot{
(Col. 1) Representative frequency;
(Col. 2) beam efficiency; 
(Col. 3) forward efficiency; 
(Col. 4) half power beam width; 
(Col. 5) Flux to antenna temperature conversion factor in Jansky per Kelvin. 
}
\end{table}

Pointing and focus were checked regularly (every $\sim$1.5 and
$\sim$4\,h, respectively) on strong nearby sources.  On-source
integration times per tuning step were $\sim$1\,h.  Additional
information on the observations is provided in
Table\,\ref{tab:survey}.

\begin{table}[hbt!] 
\caption{Relevant observational information.}
\label{tab:survey}
\begin{tabular}{c c c c c c}
\hline\hline
 Band & Mode & IBW & $\nu_{obs}$  &  rms  & Opacity   \\
      & & (GHz) & (GHz)       & (mK)  &           \\
\hline
E090 & 2SB & 8  & 79.3 - 115.7   & 1 - 3   & 0.07 - 0.38\\ 
E150 & SSB & 4  & 128.4 - 174.8  & 2 - 8   & 0.03 - 0.39\\ 
E230 & 2SB & 16 & 202.1 - 270.7  & 5 - 10  & 0.12 - 0.30\\ 
E330 & 2SB & 16 & 258.4 - 356.2  & 6 - 24  & 0.07 - 0.76\\ 
\hline 
\end{tabular}
\tablefoot{ 
(Col. 1) EMIR receiver band;  
(Col. 2) observing mode single sideband (SSB) or dual sideband (2SB);
(Col. 3) instantaneous bandwidth (IBW);
(Col. 4) observed frequency windows in GHz,
(Col. 5) root mean square (rms) noise in units of T$_a^*$ for a spectral resolution of 2\,MHz;
(Col. 6) zenith atmospheric opacities at the observed frequency.}
\end{table}

Calibration scans on the standard two load + sky system were taken
every $\sim$18\,min; the atmospheric transmission is modelled at
\iram\ using ATM \citep{cernicharo1,pardo1}.  All spectra have been
calibrated on the antenna temperature (T$^{*}_{\rm A}$) scale, which
is related to the mean brightness temperature of the source ($T_{\rm
  B}$) via the equation
\begin{equation}\label{eq:tmb}
T_{\rm B} = T^{*}_{\rm A} \; \frac{F_{\rm eff}}{B_{\rm eff}} \; \delta^{-1}
= T_{\rm mb} \; \delta^{-1} 
\end{equation}
where T$_{\rm mb}$ is the main-beam temperature, $F_{eff}$ and
$B_{eff}$ are the forward efficiency and
the main-beam efficiency of the telescope, respectively, and $\delta$
is the beam-filling factor. 
The ratio between $F_{eff}$ and $B_{eff}$ is described 
by the equation
\begin{equation}\label{eq:etaeff}
\frac{F_{\rm eff}}{B_{\rm eff}} = 1.1 e^{\,(\nu(GHz)/398.5)^{2}} 
.\end{equation}

The molecular outflow of \oh\ has been assumed to be a uniform
elliptical source with major and minor axes $\theta_{\rm a}$ and
$\theta_{\rm b}$. In this case, the beam-filling factor is given by
\citep[see e.g.][]{kra97}:

\begin{equation}\label{eq:dilu}
\delta = 1 - e^{-\ln2\frac{\theta_{\rm a}\times\theta_{\rm b}}{HPBW^2}} 
\end{equation} 

\noindent where $HPBW$ is the half power beam width of an elliptical Gaussian
beam. Based on previous maps of CO and other molecules, we adopt an
angular source size of $\theta_{\rm a}$$\times$$\theta_{\rm
  b}$=4\arcsec$\times$12\arcsec.

We have checked the relative calibration between adjacent frequency
tunings by comparing the intensities of the lines in the overlap
regions and in frequency tunings that were observed in
different epochs. An extra check of the calibration has been
made by comparing the intensities of the \docem\ and \trecem\ lines
from this survey with those measured in previous observations
\citep{morr87,san97}.
Errors in the absolute flux calibration are expected to be $\la$25\%.

Data were reduced using CLASS\footnote{CLASS is a world-wide
software to process, reduce, and analyse heterodyne line observations
maintained by the Institut de Radioastronomie Millim\'etrique (IRAM)
and distributed with the GILDAS software, see \tt
\tiny{http://www.iram.fr/IRAMFR/GILDAS}} to obtain the final spectra.
We followed the standard procedure, which includes flagging of bad channels,
flagging of low-quality scans, baseline substracting, averaging individual scans,
and channel smoothing to a typical spectral 
resolution of ($\sim$2\,MHz).

\begin{table*}[hbtp!] 
\caption{Parameters of the lines detected in \oh\ reported in this work.
}
\label{tab:measures}
\centering    
\begin{tabular}{c c c c c c c c c c}
\hline\hline
Molecule & Transition & $\nu _{\mathrm{rest}}$ & $E_{\mathrm{u}}$ & $A_{\mathrm{ul}}$ & $\int T_{\mathrm{mb}}^{*}\,d\mathrm{v}$  & FWHM & T$_{\mathrm{mb,peak}}$ \\    
         & quantum numbers & (MHz)  & (K) &  s$^{-1}$ & (K km s$^{-1}$) & (km s$^{-1}$) & (mK) \\
\hline \\[-1.0em]
$^{13}$CO                         & 1 $\rightarrow$ 0     &  110201.35 & 5.3     &   6.336$\times$10$^{-8}$ & 12.81(0.05)  & 36.1(0.4)  & 324(2) \\ 
$J$                               & 2 $\rightarrow$ 1     &  220398.68 & 15.9    &   6.082$\times$10$^{-7}$ & 69.63(0.13)  & 32.4(0.4)  & 1818(9)  \\ 
                                  & 3 $\rightarrow$ 2     &  330587.96 & 31.7    &   2.199$\times$10$^{-6}$ & 87.6(0.6)    & 29.5(0.4)  & 2420(50)  \\ 
\hline \\[-1.0em]
HNCO &  4$_{0,4}$ $\rightarrow$ 3$_{0,3}$                                        &  87925.24 &  10.5 & 9.025$\times$10$^{-6}$  & 1.31(0.05) & 32.0(1.3) & 32.8(1.6) \\ 
$J_{K_{\mathrm{a}},K_{\mathrm{c}}}$     &  5$_{0,5}$ $\rightarrow$ 4$_{0,4}$     & 109905.75 &  15.8 & 1.802$\times$10$^{-5}$  & 2.54(0.06) & 33.2(0.7) & 61.3(1.8) \\ 
     &  6$_{0,6}$ $\rightarrow$ 5$_{0,5}$                                        & 131885.73 &  22.2 & 3.163$\times$10$^{-5}$  & 3.64(0.05) & 29.4(0.5) & 93.7(2.1) \\ 
     &  7$_{0,7}$ $\rightarrow$ 6$_{0,6}$                                        & 153865.09 &  29.5 & 5.078$\times$10$^{-5}$  & 4.83(0.06) & 28.6(0.4) & 140(3) \\ 
     &  10$_{0,10}$ $\rightarrow$ 9$_{0,9}$                                      & 219798.27 &  58.0 & 1.510$\times$10$^{-4}$  & 6.67(0.14) & 24.5(0.9) & 200(9) \\  
     &  11$_{0,11}$ $\rightarrow$ 10$_{0,10}$                                    & 241774.03 &  69.6 & 2.019$\times$10$^{-4}$  & 7.32(0.14) & 21.5(0.5) & 260(8) \\ 
     &  12$_{0,12}$ $\rightarrow$ 11$_{0,11}$                                    & 263748.62 &  82.3 & 2.630$\times$10$^{-4}$  & 4.73(0.22) & 21.3(1.0) & 179(14)  \\ 
     &  13$_{0,13}$ $\rightarrow$ 12$_{0,12}$                                    & 285721.95 &  96.0 & 3.355$\times$10$^{-4}$  & 6.3(0.4)\tablefootmark{\dag} & \tablefootmark{*} & \tablefootmark{*} \\ 
     &  14$_{0,14}$ $\rightarrow$ 13$_{0,13}$                                    & 307693.90 & 110.8 & 4.200$\times$10$^{-4}$  & 2.48(0.20) & 19.9(1.6) & 114(16)  \\ 
     &  5$_{1,5}$ $\rightarrow$ 4$_{1,4}$     & 109495.99 &  59.0 & 1.692$\times$10$^{-5}$ & 0.08(0.04) & 13(6) & 4.8(2.1) \\ 
     &  6$_{1,6}$ $\rightarrow$ 5$_{1,5}$     & 131394.23 &  65.3 & 3.006$\times$10$^{-5}$ & 0.22(0.02) & 14(4) & 7.6(1.4) \\ 
     &  6$_{1,5}$ $\rightarrow$ 5$_{1,4}$     & 132356.70 &  65.5 & 3.072$\times$10$^{-5}$ & 0.16(0.03) & 14(9) & 5.5(1.8) \\ 
     &  7$_{1,7}$ $\rightarrow$ 6$_{1,6}$     & 153291.94 &  72.7 & 4.863$\times$10$^{-5}$ & 0.25(0.04) &  9(4) & 15.8(2.2) \\
     &  7$_{1,6}$ $\rightarrow$ 6$_{1,5}$     & 154414.76 &  72.9 & 4.971$\times$10$^{-5}$ & 0.28(0.03) & 18(7) & 6.5(2.1) \\ 
     &  11$_{1,11}$ $\rightarrow$ 10$_{1,10}$ & 240875.73 & 112.6 & 1.957$\times$10$^{-4}$ & 0.35(0.10) & 12(2) & 34(8) \\ 
     &  12$_{1,12}$ $\rightarrow$ 11$_{1,11}$ & 262769.48 & 125.3 & 2.554$\times$10$^{-4}$ & 0.31(0.09) & 12(3) & 22(11) \\ 
     &  13$_{1,13}$ $\rightarrow$ 12$_{1,12}$ & 284662.17 & 138.9 & 3.261$\times$10$^{-4}$ & 0.28(0.07) & 14(3) & 20(10) \\         
\hline \\[-1.0em]                                   
HNCS &  8$_{0,8}$ $\rightarrow$ 7$_{0,7}$                                            &  93830.07 & 20.3 & 1.217$\times$10$^{-5}$ & 0.19(0.04) &  27(9)& 4.4(2.0)  \\ 
$J_{K_{\mathrm{a}},K_{\mathrm{c}}}$     &  9$_{0,9}$ $\rightarrow$ 8$_{0,8}$         & 105558.08 & 25.3 & 1.744$\times$10$^{-5}$ & 0.23(0.04) & \tablefootmark{*} & 4.0(1.6)  \\ 
     &  11$_{0,11}$ $\rightarrow$ 10$_{0,10}$                                        & 129013.26 & 37.2 & 3.215$\times$10$^{-5}$ & 0.32(0.04) &  33(5) & 8.5(2.2)  \\ 
     &  12$_{0,12}$ $\rightarrow$ 11$_{0,11}$                                        & 140740.38 & 43.9 & 4.189$\times$10$^{-5}$ & 0.26(0.04) &  34(9) & 4.7(2.3)  \\ 
     &  13$_{0,13}$ $\rightarrow$ 12$_{0,12}$                                        & 152467.14 & 51.2 & 5.342$\times$10$^{-5}$ & 0.30(0.04) &  18(2) & 14.8(2.1) \\ 
     &  14$_{0,14}$ $\rightarrow$ 13$_{0,13}$                                        & 164193.52 & 59.1 & 6.690$\times$10$^{-5}$ & 0.34(0.02) &  21(1) & 15.1(1.4) \\                                               
\hline \\[-1.0em]                                   
HC$_3$N &   9 $\rightarrow$ 8  &  81881.46 & 19.6 & 4.215$\times$10$^{-5}$  & 0.30(0.06) &  27(6) & 7.5(2.3) \\ 
$J$     &  10 $\rightarrow$ 9  &  90978.99 & 24.0 & 5.812$\times$10$^{-5}$  & 0.38(0.03) &  26(5) & 8.5(1.5) \\ 
        &  11 $\rightarrow$ 10 & 100076.38 & 28.8 & 7.770$\times$10$^{-5}$  & 0.44(0.04) &  28(3) & 12.4(1.7) \\ 
        &  12 $\rightarrow$ 11 & 109173.64 & 34.1 & 1.012$\times$10$^{-4}$  & 0.15(0.03) &  \tablefootmark{*} & \tablefootmark{*}  \\ 
        &  15 $\rightarrow$ 14 & 136464.40 & 52.4 & 1.993$\times$10$^{-4}$  & 0.35(0.03) &  16(2) & 17.8(1.9) \\ 
        &  16 $\rightarrow$ 15 & 145560.95 & 59.4 & 2.424$\times$10$^{-4}$  & 0.29(0.04) &  17(3) & 12(3) \\ 
        &  17 $\rightarrow$ 16 & 154657.29 & 66.8 & 2.912$\times$10$^{-4}$  & 0.20(0.04) &  20(4) & 9.6(2.3) \\ 
        &  18 $\rightarrow$ 17 & 163753.40 & 74.7 & 3.463$\times$10$^{-4}$  & 0.25(0.03) &  22(5) & 10.7(2.1) \\                              
\hline \\[-1.0em]                                                                                    
NO      & $\Pi^{+}$(3/2,5/2) $\rightarrow$ (1/2,3/2)                            & 150176.48 & 7.2  & 3.310$\times$10$^{-7}$ \ldelim\}{5}{0.1pt}[] & \multirow{5}{*}{1.15(0.08)\tablefootmark{\ddag}} & \multirow{5}{*}{55(10)\tablefootmark{\ddag}} & \multirow{5}{*}{9(3)\tablefootmark{\ddag}} \\ 
$\Pi^{\mathrm{band}}(J,F)$        & $\Pi^{+}$(3/2,3/2) $\rightarrow$ (1/2,1/2)  & 150198.76 & 7.2  & 1.839$\times$10$^{-7}$ & & & \\ 
        & $\Pi^{+}$(3/2,3/2) $\rightarrow$ (1/2,3/2)                            & 150218.73 & 7.2  & 1.471$\times$10$^{-7}$ & & & \\ 
        & $\Pi^{+}$(3/2,1/2) $\rightarrow$ (1/2,1/2)                            & 150225.66 & 7.2  & 2.943$\times$10$^{-7}$ & & & \\ 
        & $\Pi^{+}$(3/2,1/2) $\rightarrow$ (1/2,3/2)                            & 150245.64 & 7.2  & 3.679$\times$10$^{-8}$ & & & \\
        & $\Pi^{-}$(3/2,3/2) $\rightarrow$ (1/2,1/2)                            & 150644.34 & 7.2  & 1.853$\times$10$^{-7}$ & 0.20(0.03) & 39(9) & 4.9(2.1) \\ 
        & $\Pi^{+}$(5/2,7/2) $\rightarrow$ (3/2,5/2)                            & 250436.85 & 19.2 & 1.841$\times$10$^{-6}$ \ldelim\}{5}{0.1pt}[] & \multirow{5}{*}{5.08(0.16)\tablefootmark{\ddag}} & \multirow{5}{*}{54(2)\tablefootmark{\ddag}} & \multirow{5}{*}{82(9)\tablefootmark{\ddag}} \\  
        & $\Pi^{+}$(5/2,5/2) $\rightarrow$ (3/2,3/2)                            & 250440.66 & 19.2 & 1.547$\times$10$^{-6}$ & & & \\  
        & $\Pi^{+}$(5/2,3/2) $\rightarrow$ (3/2,1/2)                            & 250448.53 & 19.2 & 1.381$\times$10$^{-6}$ & & & \\  
        & $\Pi^{+}$(5/2,3/2) $\rightarrow$ (3/2,3/2)                            & 250475.41 & 19.2 & 4.420$\times$10$^{-7}$ & & & \\  
        & $\Pi^{+}$(5/2,5/2) $\rightarrow$ (3/2,5/2)                            & 250482.94 & 19.2 & 2.947$\times$10$^{-7}$ & & & \\
        & $\Pi^{-}$(5/2,7/2) $\rightarrow$ (3/2,5/2)                            & 250796.44 & 19.3 & 1.849$\times$10$^{-6}$ \ldelim\}{3}{0.1pt}[] & \multirow{3}{*}{4.77(0.15)\tablefootmark{\ddag}} & \multirow{3}{*}{64(2)\tablefootmark{\ddag}} & \multirow{3}{*}{68(9)\tablefootmark{\ddag}} \\ 
        & $\Pi^{-}$(5/2,5/2) $\rightarrow$ (3/2,3/2)                            & 250815.59 & 19.3 & 1.554$\times$10$^{-6}$ & & & \\ 
        & $\Pi^{-}$(5/2,3/2) $\rightarrow$ (3/2,1/2)                            & 250816.95 & 19.3 & 1.387$\times$10$^{-6}$ & & & \\
        & $\Pi^{+}$(7/2,9/2) $\rightarrow$ (5/2,7/2)                            & 350689.49 & 36.1 & 5.418$\times$10$^{-6}$ \ldelim\}{5}{0.1pt}[] & \multirow{5}{*}{7.2(0.5)\tablefootmark{\ddag}} & \multirow{5}{*}{40(4)\tablefootmark{\ddag}} & \multirow{5}{*}{149(40)\tablefootmark{\ddag}} \\
        & $\Pi^{+}$(7/2,7/2) $\rightarrow$ (5/2,5/2)                            & 350690.77 & 36.1 & 4.976$\times$10$^{-6}$ & & & \\ 
        & $\Pi^{+}$(7/2,5/2) $\rightarrow$ (5/2,3/2)                            & 350694.77 & 36.1 & 4.815$\times$10$^{-6}$ & & & \\ 
        & $\Pi^{+}$(7/2,5/2) $\rightarrow$ (5/2,5/2)                            & 350729.58 & 36.1 & 5.897$\times$10$^{-7}$ & & & \\
        & $\Pi^{+}$(7/2,7/2) $\rightarrow$ (5/2,7/2)                            & 350736.78 & 36.1 & 4.423$\times$10$^{-7}$ & & & \\
        & $\Pi^{-}$(7/2,9/2) $\rightarrow$ (5/2,7/2)                            & 351043.52 & 36.1 & 5.433$\times$10$^{-6}$ \ldelim\}{3}{0.1pt}[] & \multirow{3}{*}{5.2(0.5)\tablefootmark{\ddag}} & \multirow{3}{*}{41(5)\tablefootmark{\ddag}} & \multirow{3}{*}{122(40)\tablefootmark{\ddag}} \\
        & $\Pi^{-}$(7/2,7/2) $\rightarrow$ (5/2,5/2)                            & 351051.47 & 36.1 & 4.990$\times$10$^{-6}$ & & & \\
        & $\Pi^{-}$(7/2,5/2) $\rightarrow$ (5/2,3/2)                            & 351051.70 & 36.1 & 4.830$\times$10$^{-6}$ & & & \\
\hline 
\end{tabular}
\tablefoot{ 
For each transition:
(Col. 1) name of the molecule and set of quantum numbers used;
(Col. 2) specific quantum numbers;
(Col. 3) rest frequency;
(Col. 4) upper level energy;
(Col. 5) Einstein spontaneus emission coefficient;
(Col. 6) line flux obtained integrating the area below the profile in main-beam temperature scale;
(Col. 7) line full width at half maximum (FWHM); 
(Col. 8) main-beam peak temperature and rms for a spectral resolution of $\sim$2\,MHz (in parentheses).
(Col. 7) and (Col. 8), were obtained by fitting a Gaussian function to the line core profile, masking the wings. 
The formal errors given (within parentheses) do not include absolute flux calibration and baseline subtraction uncertainties.\\
\tablefoottext{\dag}{Line blend with SO$_2$ $17_{3,15}-17_{2,16}$; the value of the flux provided here accounts only for the HNCO line flux, which represents a 75\%\ of the total flux of the blend.}
\tablefoottext{\ddag}{The individual hyperfine components of NO are spectrally unresolved, therefore, one single value for the line blend is provided.}
\tablefoottext{*}{Unreliable and, thus, not measured value due to line blending, low signal-to-noise ratio, and/or poor baseline subtraction.}  
}
\end{table*}

\section{Observational results}\label{sec:obsres}
Line identification over the full frequency range covered in this
survey was done using the public line catalogues from the Cologne
Database for Molecular Spectroscopy \citep[CDMS,][]{mul05} and the Jet
Propulsion Laboratory \citep[JPL,][]{pick98}, together with 
a private spectroscopic catalogue that assembles information for
almost five thousand spectral entries (molecules and atoms), 
including isotopologues and vibrationally excited states, compiled
from extensive laboratory and theoretical works by independent
teams \citep{madex}.

We have identified hundreds of transitions from
more than 50 different molecular species including their main
isotopologues ($^{13}$C-, $^{18}$O-, $^{17}$O-, $^{33}$S-, $^{34}$S-,
$^{30}$Si, and $^{29}$Si) in \oh, confirming the chemical richness of
this source, which is unprecedented amongst O-rich AGB and post-AGB stars.

\begin{figure}[hbtp!] 
\centering
\includegraphics[width=0.75\hsize]{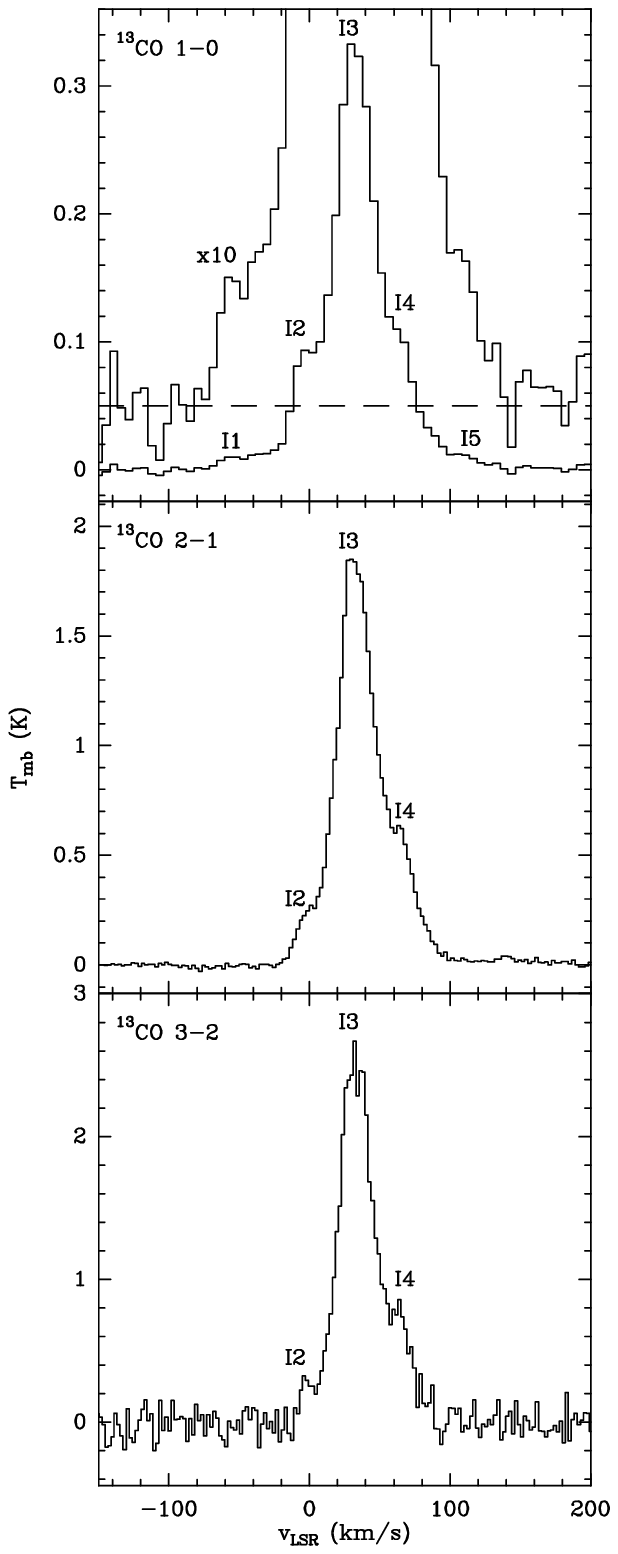}
\caption{$^{13}$CO mm-wavelength transitions in \oh.  We labelled
  the different spectral features in the \trecem\ profiles as I1-I5 to
  identify them with the corresponding regions/clumps of CO-outflow
  where the emission features mainly arise (see
  Fig.\,\ref{fig:oh231}). In the top panel, the $^{13}$CO\,$J$=1--0
  spectrum is also plotted using a larger T$_{\mathrm{mb}}$ scale for
  an improved view of the weak broad wings.}
\label{fig:13co}
\end{figure}

First detections from this survey include the N-bearing species HNCO,
HNCS, HC$_3$N, and NO, which are the focus of this paper. Together with
these, we present the spectra of the \trecem\,$J$=1--0, $J$=2--1, and
$J$=3--2 transitions, which are excellent tracers of the mass
distribution and dynamics in \oh: \trecem\ lines are optically thin
(or, at most, moderately opaque towards the nebula centre) and are
expected to be thermalised 
over the bulk of the outflow and, certainly,
in the regions that lie within the telescope beam in these
observations, characterized by average densities always above
10$^{4}$\,cm$^{-3}$ \citep{san97,alc01}. Spectra of different
rotational transitions from the ground vibrational state ($v=0$) of
\trecem, HNCO, HNCS, HC$_3$N, and NO are shown in
Figs.\,\ref{fig:13co}-\ref{fig:no}, and main line parameters are reported
in Table\,\ref{tab:measures}. 

\subsection{\trecem\ spectra}\label{sec:13co}

The \trecem\ lines (Fig.\,\ref{fig:13co}) show broad, structured
profiles with two main components: (1) the intense, relatively narrow
(FWHM$\sim$30-35\,\kms) core centred at \vlsr=33.4$\pm$0.9\kms, which
arises in the slow, dense central parts of the nebula (clump I3); and
(2) weak broad wings, with full widths of up to $\sim$220\,\kms\ in
the $J$=1--0 line, which originate in the fast bipolar lobes clumps
I1-I2 and I4-I5). The most intense spectral component in the
\trecem\ wings arises at clump I4, i.e.\ the base of the southern lobe, and
the I4/I3 feature peak-intensity ratio is I4/I3$\sim$0.3. The
single-dish profiles of the $J$=1--0 and 2--1 lines are already known
from previous observations \citep{morr87,san97} and, within the
expected calibration errors, are consistent with those observed in
this survey.

The full width of the wings is largest for the \trece\ transition,
which is observed over a velocity range of \vlsr=[$-$80:$+$140]\,\kms,
and decreases for higher-$J$ transitions down to
\vlsr=[$-$10:$+$90]\,\kms\, for the $J$=3--2 line.  The different
width of the wings is partially explained by the increase in the
expansion velocity with the distance to the centre along the
CO outflow and the smaller beam for higher frequencies. 
Also, as can be observed (Fig.\,\ref{fig:13co} and
Table\,\ref{tab:measures}), the FWHM of the \trecem\, transitions
decrease as the upper energy level increases. This suggests that the
envelope layers with higher excitation conditions (i.e.\,warmer and,
thus, presumably closer to the central star) are characterized by
lower expansion velocities. This trend is confirmed by higher
frequency transitions of \trecem\ (and of most molecules) observed
with \hso\ with a larger telescope beam \cite[e.g.][]{buj12,san14}.

\subsection{New N-bearing molecules}\label{sec:nbear}
\begin{figure*}[hbtp!] 
\centering
\includegraphics[width=0.80\hsize]{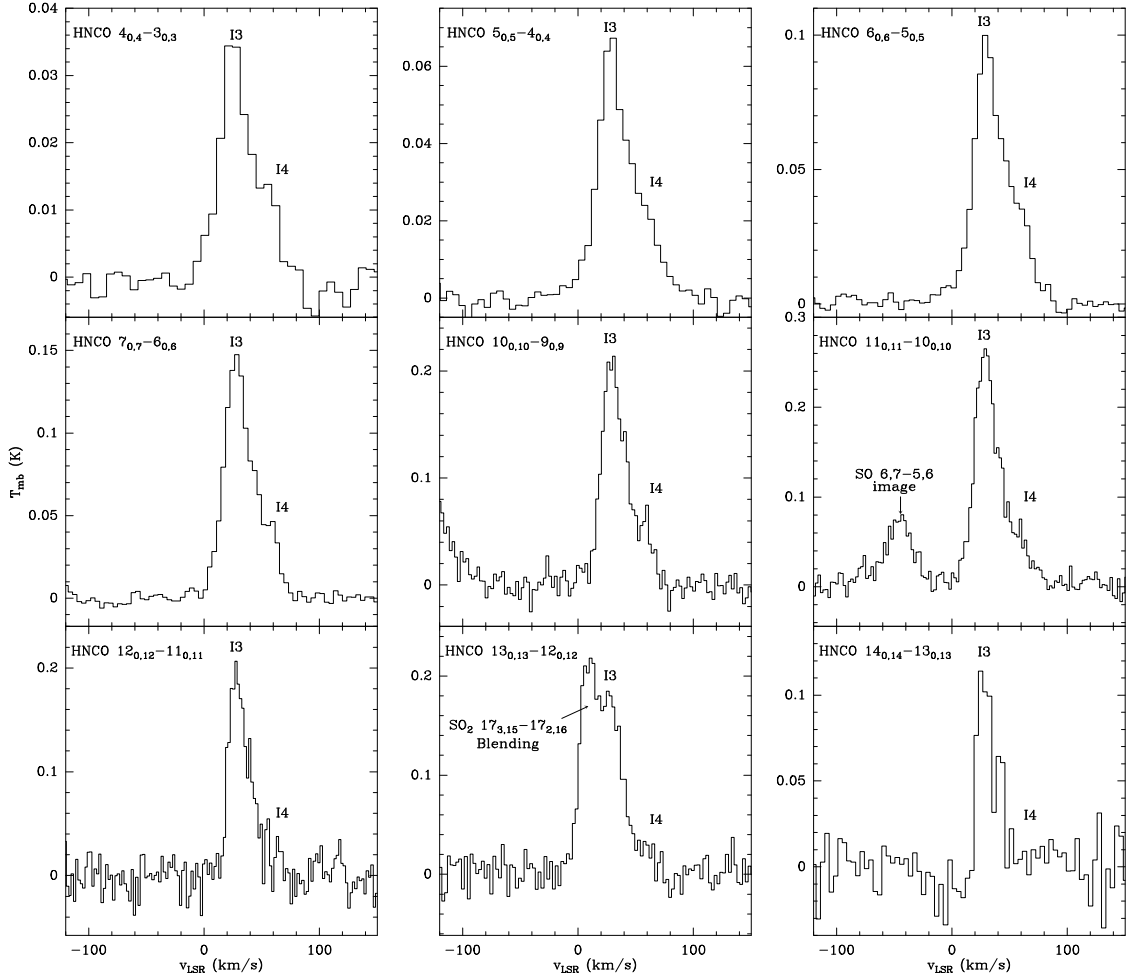}
\caption{HNCO transitions of the $K_{\mathrm{a}}$=0 ladder detected in
  \oh. Spectral features have been labelled as in Fig.\,\ref{fig:13co}. The
  HNCO $13_{0,13}-12_{0,12}$ transition is blended with the SO$_2$
  $17_{3,15}-17_{2,16}$ line (central, bottom panel).
}
\label{hncoK0}
\end{figure*}

\begin{figure*}[hbtp!] 
\centering
\includegraphics[width=0.90\hsize]{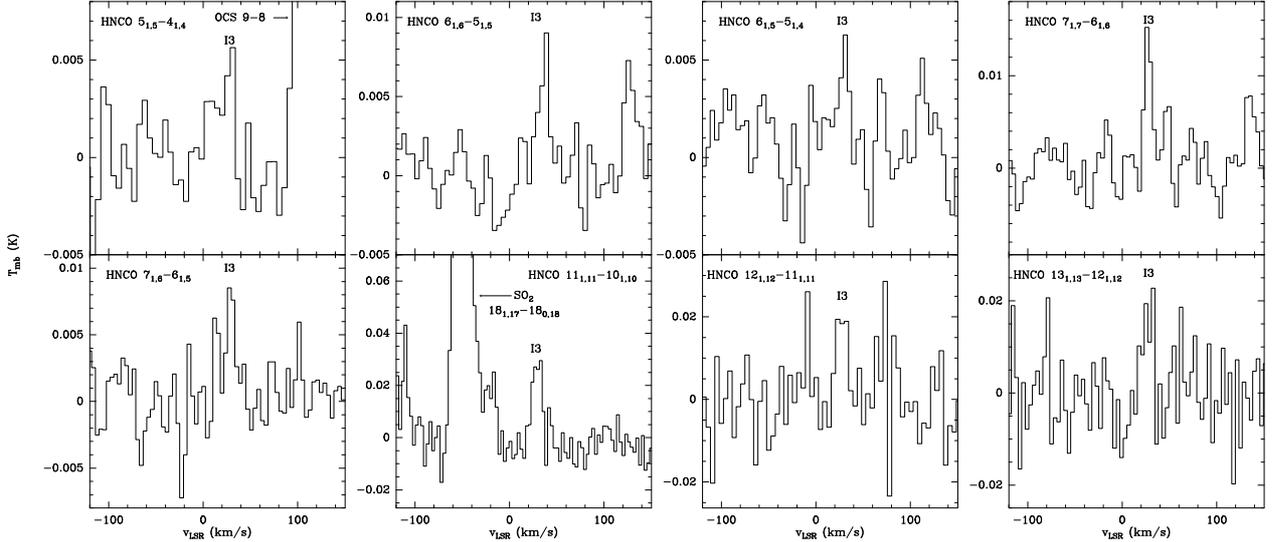}
\caption{Same as in Fig.~\ref{hncoK0} but for the $K_{\mathrm{a}}=1$
  ladder. The S/N of some of these transitions is low, so they
  may be considered as tentative detections; however, in
  spite of the large errobars, the intensity, centroid \vlsr, and FWHM
  of the tentative lines are consistent with the trend and values
  deduced for transitions detected with higher S/N.
}
\label{hncoK1}
\end{figure*}

Isocyanic acid (HNCO) is a quasi-linear asymmetric rotor whose
structure was first determined by \cite{jon50a}.  It contains a
nitrogen atom that has a nuclear spin ($I$=1) leading to a splitting
of each rotational level. This hyperfine (hpf) structure is not
resolved since the maximum separation in velocity of the hpf
components from the most intense one is $\la$2\,\kms, that is, much
less than the expansion velocity of the envelope (and, in some
cases, even smaller than the spectral resolution of our
observations). The rotational levels of HNCO are expressed in terms of
three quantum numbers: $J$ the rotational quantum number,
$K_{\mathrm{a}}$ and $K_{\mathrm{c}}$, which are the projections of $J$
onto the A and C molecular axes, respectively. We have detected several
a-type transitions (i.e.\ with $\Delta K_{\mathrm{a}}$=0 and $\Delta
K_{\mathrm{c}}$=$\pm$1) in the $K_{\mathrm{a}}$=0 and
$K_{\mathrm{a}}$=1 ladders (Figs.\,\ref{hncoK0},\ref{hncoK1}); the
difference between the $K_{\mathrm{a}}$=0 zero level and the 
$1_{1,1}$ level is $\Delta
E$=44.33\,K. 
The profiles of the HNCO transitions detected of the
$K_{\mathrm{a}}$=0 and $K_{\mathrm{a}}$=1 ladders show notable
differences: the $K_{\mathrm{a}}$=0 transitions, which are stronger
than the $K_{\mathrm{a}}$=1 ones, show an intense central core
component centred at \vlsr=29.5$\pm$1.5\,\kms\ and with a line width
of FWHM$\sim$20-33\,\kms; as for \trecem, the linewidth
decreases as the transition upper level energy increases. In addition
to the line core (arising in the central parts of the nebula, clump
I3), the HNCO $K_{\mathrm{a}}$=0 profiles show red-wing emission from
the base of the southern lobe, clump I4; the I4/I3 feature peak-intensity ratio
is $\sim$0.3. The HNCO $K_{\mathrm{a}}$=1 transitions, centred on
\vlsr=29.5$\pm$1.1\kms, are not only weaker
but narrower (FWHM$\sim$13\,\kms) than the $K_{\mathrm{a}}$=0 lines.

\begin{figure}[hbtp!] 
\centering \includegraphics[width=0.95\hsize]{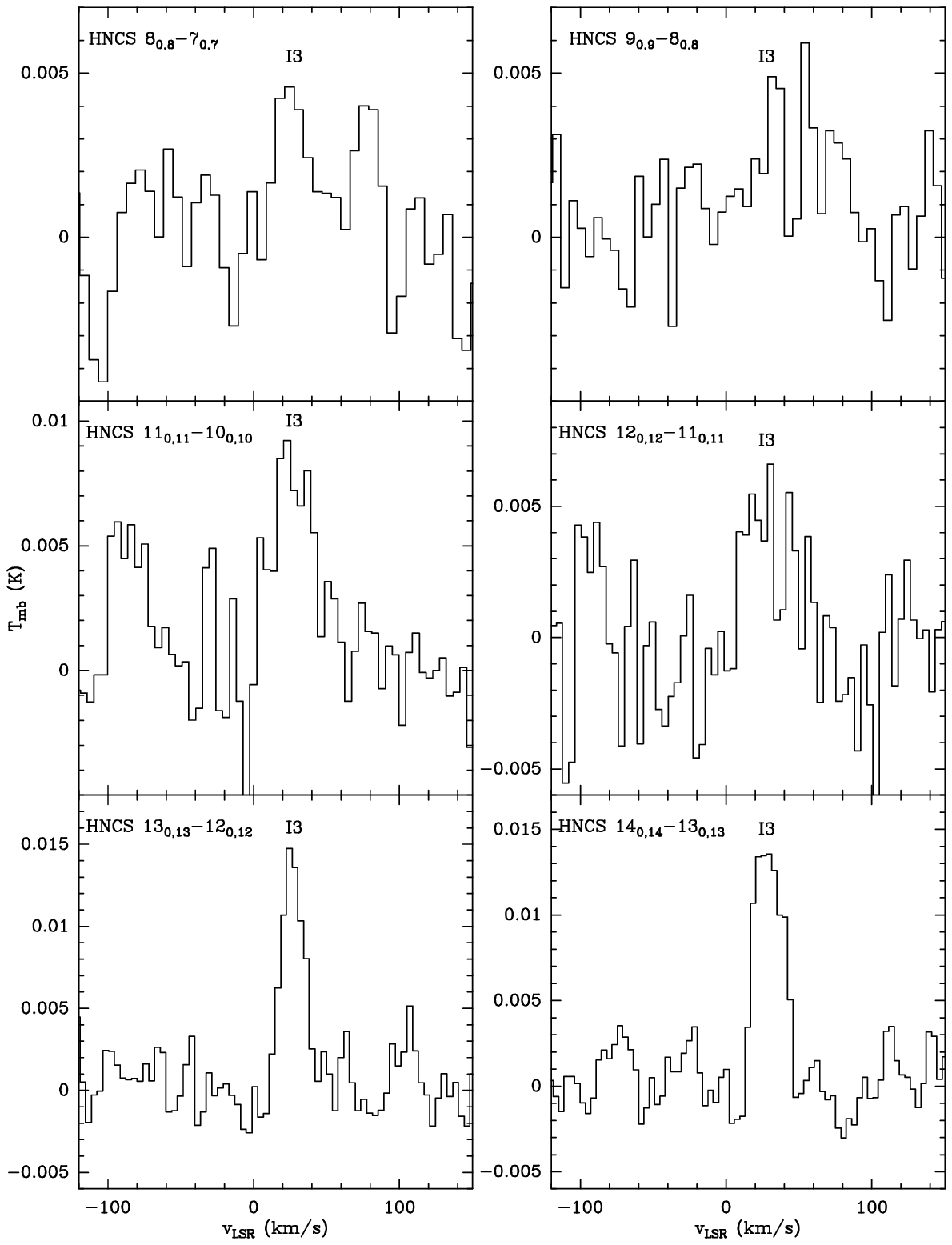}
\caption{Same as in Fig.~\ref{hncoK0} but for HNCS.
}
\label{hncs}
\end{figure}

Isothiocyanic acid (HNCS) presents a structure similar to HNCO; i.e.,\,
it is a slightly asymmetric rotor \citep[][]{jon50b}.
Its hpf structure due to the nitrogen nuclear spin is not spectrally
resolved in our data. We have detected several a-type (i.e.\,with
$\Delta K_{\mathrm{a}}$=0 and $\Delta K_{\mathrm{c}}$=$\pm$1)
transitions of the $K_{\mathrm{a}}$=0 ladder (Fig.\,\ref{hncs}). These
lines are weak and narrow, with a median FWHM$\sim$25\,\kms, and are
centred at \vlsr=28.2$\pm$0.4\,\kms, indicating that the emission
observed arises mainly in the slow central parts of the nebula. HNCS
wing emission from the fast flow (if present) is below the noise
level. We have not detected any of the $K_{\mathrm{a}}$=1 transitions
of HNCS in the frequency range covered by us. These transitions have
upper-level state energies \eu$\ga$78\,K and expected intensities well
below our detection limit.

\begin{figure*}[hbtp!] 
\centering
\includegraphics[width=0.95\hsize]{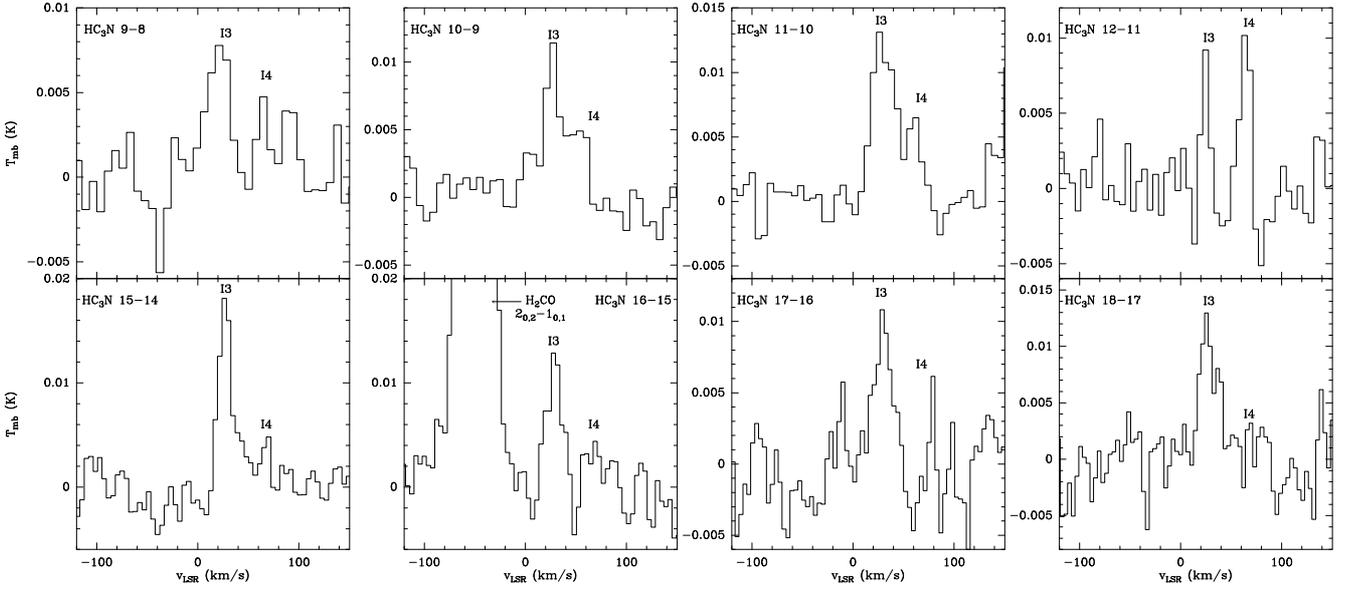}
\caption{Same as in Fig.~\ref{hncoK0} but for HC$_3$N. The poor
  baseline in the spectrum of the $J$=12--11 line makes the
  detection of this transition tentative.
}
\label{hc3n}
\end{figure*}

Cyanoacetylene (HC$_3$N) is a linear molecule that belongs to the
nitriles family. We do not resolve its hpf structure
spectrally (which is due to the nitrogen nuclear spin), so its rotational levels
are described only by the rotational number $J$ \citep[][]{wes50}.
The spectra of the HC$_3$N transitions detected in \oh\ are shown in
Fig.\,\ref{hc3n}. The line profiles are centred at
\vlsr=28.0$\pm$0.9\,\kms\ and are relatively narrow, with typical line
widths of FWHM$\sim$22\,\kms. Tentative emission from clump I4 (at
\vlsr$\sim$60-65\,\kms) is observed in most profiles, with a I4/I3 
feature peak-intensity ratio of $\sim$0.2-0.4. 

\begin{figure*}[hbtp!] 
\centering
\includegraphics[width=0.75\hsize]{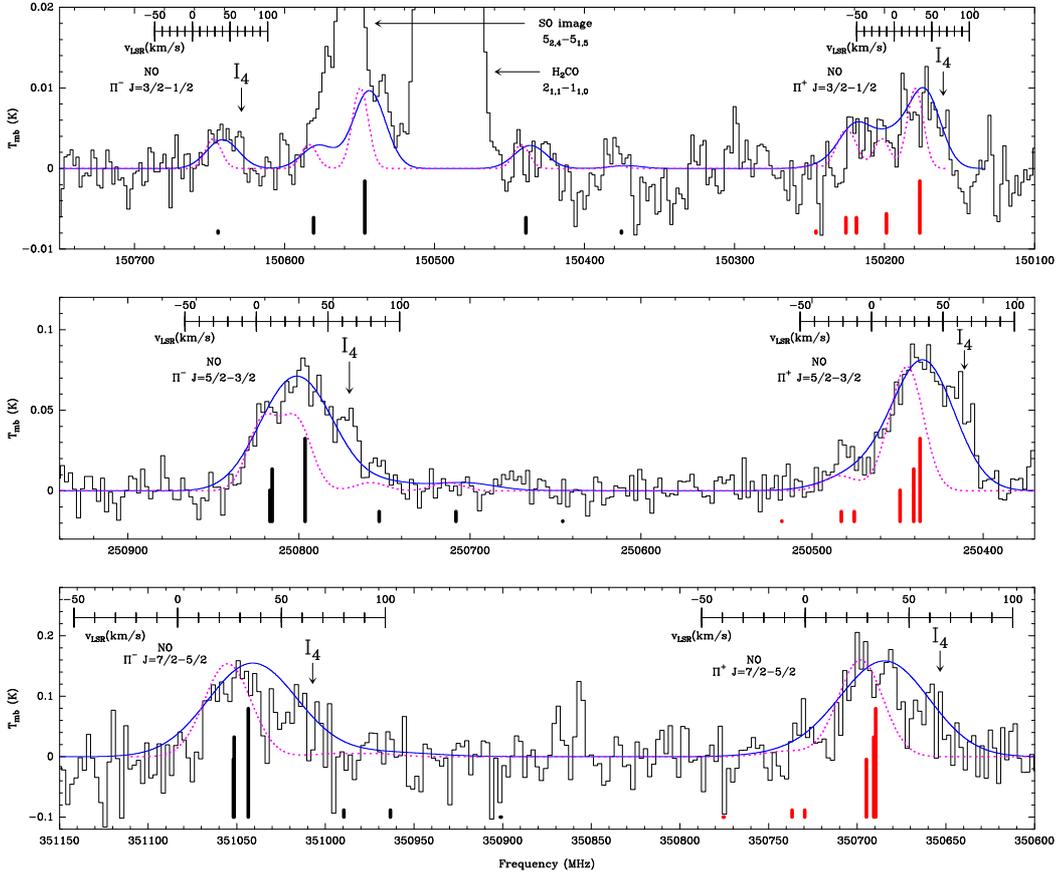}
\caption{Spectra of the NO transitions detected in \oh. The quantum
  numbers indicated in this plot for NO are $\Pi$ and $J$
  (Table\,\ref{tab:measures}). Vertical thick segments indicate the
  rest frequency and relative strength of the individual
  hyperfine (hpf) components within each $\Pi^+$ and $\Pi^-$ blend
  (red and black, respectively).
  The LSR velocity scale is shown on top of the NO transitions that
  have been detected (the velocity scale refers to the strongest
  hpf component in each case). The LSR velocity of the emission
  feature associated to clump I4, at the base of the southern lobe, is
  indicated. The dotted line is the model spectrum
  (\S\,\ref{sec:nbear} and Table\,\ref{tab:abun}) assuming that the
  individual hpf components are centred at \vlsr=28\,\kms,\ and have
  an intrinsic width of FWHM=25\,\kms, similar to the dominant
  narrow component of the HNCO, HNCS, and HC$_3$N profiles
  (Figs.\,\ref{hncoK0}-\ref{hc3n}). These line parameters cannot
  explain the broad profiles and the apparent overall redshift of the
  NO blends, which are reproduced better by adopting larger FWHM and
  \vlsr\ (solid blue line: representative model with
  FWHM=50\,\kms\ and \vlsr=40\,\kms).
}
\label{fig:no}
\end{figure*}

Nitric oxide (NO) is a radical with a $^{2}\Pi$ ground state that
splits into two different ladders, $\Omega$=1/2 ($^{2}\Pi_{1/2}$) and
$\Omega$=3/2 ($^{2}\Pi_{3/2}$), owing to the unpaired electron by
spin-orbit coupling.  Additionally,  $\Lambda$-doubling means that each
transition is split into two different bands of opposite parity
$\Pi^{+}$ and $\Pi^{-}$. NO also presents hpf structure owing
to the nitrogen nuclear spin that interacts with the total angular
momentum, thereby splitting each single rotational level into $2(I,J)_{\rm
min}$+1 levels described by the quantum number $F$ \citep[for a
complete description of the structure and spectroscopy of the NO
rotational transitions see ][and references therein]{liq09}. We have
detected several transitions of the $^{2}\Pi_{1/2}$ ladder around 150,
250, and 350\,GHz (see Fig.\,\ref{fig:no}).  Transitions of the
$^{2}\Pi_{3/2}$ ladder, with upper level state energies above 180\,K,
are not detected. As shown in Fig.\,\ref{fig:no}, each of the
$\Lambda$-doublets ($\Pi^{+}$ and $\Pi^{-}$) of NO is composed of
several hpf components that are blended in our spectra, except
for the $\Pi^{-}$\,$(3/2,3/2)-(1/2,1/2)$ line at 150.644\,GHz, which
is isolated.

The NO blends appear on average redshifted by a few \kms\ from the
source systemic velocity and are broader than the profiles of the
other N-bearing species discussed in this work. The broadening is
partially (but not only) due to the hpf structure of the NO
transitions. To constrain the intrinsic linewidth and centroid of the
individual hpf components contributing to the observed profile, we
have calculated and added together the emergent spectrum of the hpf
components. The synthetic spectrum was calculated using our code
MADEX \citep[][see also Appendix\,\ref{sec:app}]{madex} and also the
task MODSOURCE of CLASS, both giving similar results at LTE (non-LTE
calculations are available in MADEX but not in MODSOURCE). We adopted a
Gaussian profile for the hpf lines. 
We find that, first, the shape of
the NO blends cannot be reproduced by adopting sharply centrally peaked
profiles such as those of HNCO, HNCS, and HC$_3$N, with line centroids at
\vlsr$\sim$28-29\,\kms\ and widths of FWHM$\sim$15-30\,\kms\ (dotted
line in Fig.\,\ref{fig:no}). In order to match the profiles of the NO
blends, the individual hpf components must have a larger width of
FWHM$\sim$40-50\,\kms\ and must be centered at
\vlsr$\sim$35-40\,\kms\ (Fig.\,\ref{fig:no}). In support
of this conclusion, a Gaussian fit to the
$\Pi^{-}$\,$(3/2,3/2)-(1/2,1/2)$ line at 150.644\,GHz, which is
unblended, also indicates a broad profile, with a FWHM=40$\pm$8\,\kms,
centred at \vlsr=41$\pm$4\,\kms.  Although the intrinsic linewidth 
is uncertain, the broad profiles of NO indicate that this
molecule is present in abundance in the high-velocity lobes and that
the wing-to-core emission contribution is greater for NO than for HNCO,
HNCS, and HC$_3$N. In particular, an important part of the NO emission
profile arises at clump I4.  The
emission from feature I4 is indeed comparable to that of the narrow
core (I3), with an estimated I4/I3 feature peak-intensity ratio of $\sim$1 at
150\,GHz, $\sim$0.6-0.8 at 250\,GHz, and $\sim$0.5-0.6 at 350\,GHz.
This significant NO emission contribution from clump I4 to the total
profile is one explanation for the apparent overall redshift of the NO
lines to an intermediate velocity between that of I3 and I4.


\section{Data analysis: molecular abundances}\label{sec:analysis}
In the following sections, beam-averaged column densities (\ntot)
and rotational temperatures (\trot) are obtained from the population
diagram method (\S\,\ref{sec:popdiag}) and from non-LTE excitation
calculations (\S\,\ref{sec:nolte}).  We derive fractional
abundances ($X$) relative to molecular hydrogen (H$_2$) for the
different molecular species detected in \oh. As a reference, we have
used the fractional abundance of \trecem, for which we adopt
$X$(\trecem)=5$\times$10$^{-5}$ as calculated by \cite{morr87}.
The \trecem\ abundance adopted is in the high end of
    the typical range of values for O-rich stars; in the case of \oh,
    it reflects the particularly low $^{12}$C/$^{13}$C isotopic ratio,
    $\sim$5-10, measured in this object \citep[and other O-rich
      CSEs;][and references therein]{san00,tey06,mil09,ram14}.

The fractional abundances of the N-bearing species reported in this
work have been calculated as

\begin{equation} 
\label{eq:abun}
\frac{X(m)}{X(^{13}CO)}\,=\,\frac{\ntot(m)}{\ntot(^{13}CO)},
\end{equation}
where $m$ represents the name of the analysed molecule. 

\subsection{Population diagram analysis}\label{sec:popdiag}
Population (or rotation) diagrams are used to obtain first-order,
beam-averaged column densities (\ntot) and rotational temperatures
(\trot) from the integrated intensities of multiple rotational
transitions of a given molecule in the same vibrational state.  This
method, which is described in detail and extensively discussed by \cite{gol99}, among others, relies on two main assumptions: $i$) lines are
optically thin, and $ii$) all levels involved in the transitions used
are under local thermodynamical equilibrium (LTE) conditions. 
This
assumption implies that the level populations are described by the
Boltzmann distribution with a single rotational temperature, \trot,
which is equal to the kinetic temperature of the gas (\tkin=\trot).
Under these assumptions, the line integrated intensity or line flux
($W$) is related to \ntot\ and \trot\ by the following expression:

\begin{equation} 
\label{eq:rotdiag}
\ln\,\left(\frac{N_{u}}{g_{u}}\right)=\ln\,\left(\frac{3k_{B}W}{8\pi^{3}\, \nu \, S_{ul}\,\mu^{2}}\right)=\ln\,\left(\frac{N}{Z}\right)-\frac{E_{u}}{k_B T_{rot}}
\end{equation}
where $N_{\mathrm{u}}$ is the column density of the upper level,
$g_{\mathrm{u}}$  the degeneracy of the upper level, $W$  the
velocity-integrated intensity of the transition, $k_{\mathrm{B}}$ 
the Boltzmann constant, $\nu$  the rest frequency of the line,
$S_{\mathrm{ul}}$  the line strength of the transition, $\mu$  the
dipole moment of the corresponding transition, $Z$  the partition
function, and \eu\ is the upper level energy of the transition.

The partition function, Z$_{\rm rot}$ has been computed for each molecule
by explicit summation of 

\begin{equation}\label{eq:zrot} 
Z_{\rm rot} = \displaystyle\sum_{i=0}^\infty g_i e^{\frac{-E_i}{kT}},
\end{equation} 
for enough levels to obtain accurate values,
using the code MADEX (Appendix\,\ref{sec:app}). 
At low temperatures ($\lesssim$50\,K), this ensures
moderate uncertainties in the column density ($<$5\%) as derived from
the low-\eu\ transitions detected, since the contribution of
high-energy levels to the partition function is negligible.

The line flux ($W$) has been obtained by integrating the area below
the emission profile, typically within the range
\vlsr$\sim$[0-100]\,\kms, and is given in a source brightness
temperature scale ($W$=$\int$T$_{\rm b}$dv), obtained from T$^*_a$ via
Eq.\,\ref{eq:tmb}.
In the case of \trecem, $W$ does not include the weak emission from
the wings beyond \vlsr$\pm$70\,\kms\ 
since this high-velocity component is not
detected in the other molecules.
The values for $W$ used to build the population
diagrams are given in Table\,\ref{tab:measures}.

The beam-filling factor ($\delta$; see Eq.\,\ref{eq:dilu})
has been computed by adopting a characteristic size for the emitting
region of $\Omega_{\mathrm{s}}$=4\arcsec$\times$12\arcsec\ for all
molecules. This size is comparable to but slightly less than the
angular size (at half intensity) of the CO-outflow measured by
\cite{alc01} -- see Fig.\,\ref{fig:oh231}. 
In any case, we have checked that the parameters derived from the
population diagram do not vary significantly for a range of reasonable
values of $\theta_{\rm a}$$\times$$\theta_{\rm
  b}$$\sim$(3-6)\arcsec$\times$(10-18)\arcsec.

\begin{figure}[hbtp!] 
\centering
\includegraphics[width=0.85\hsize]{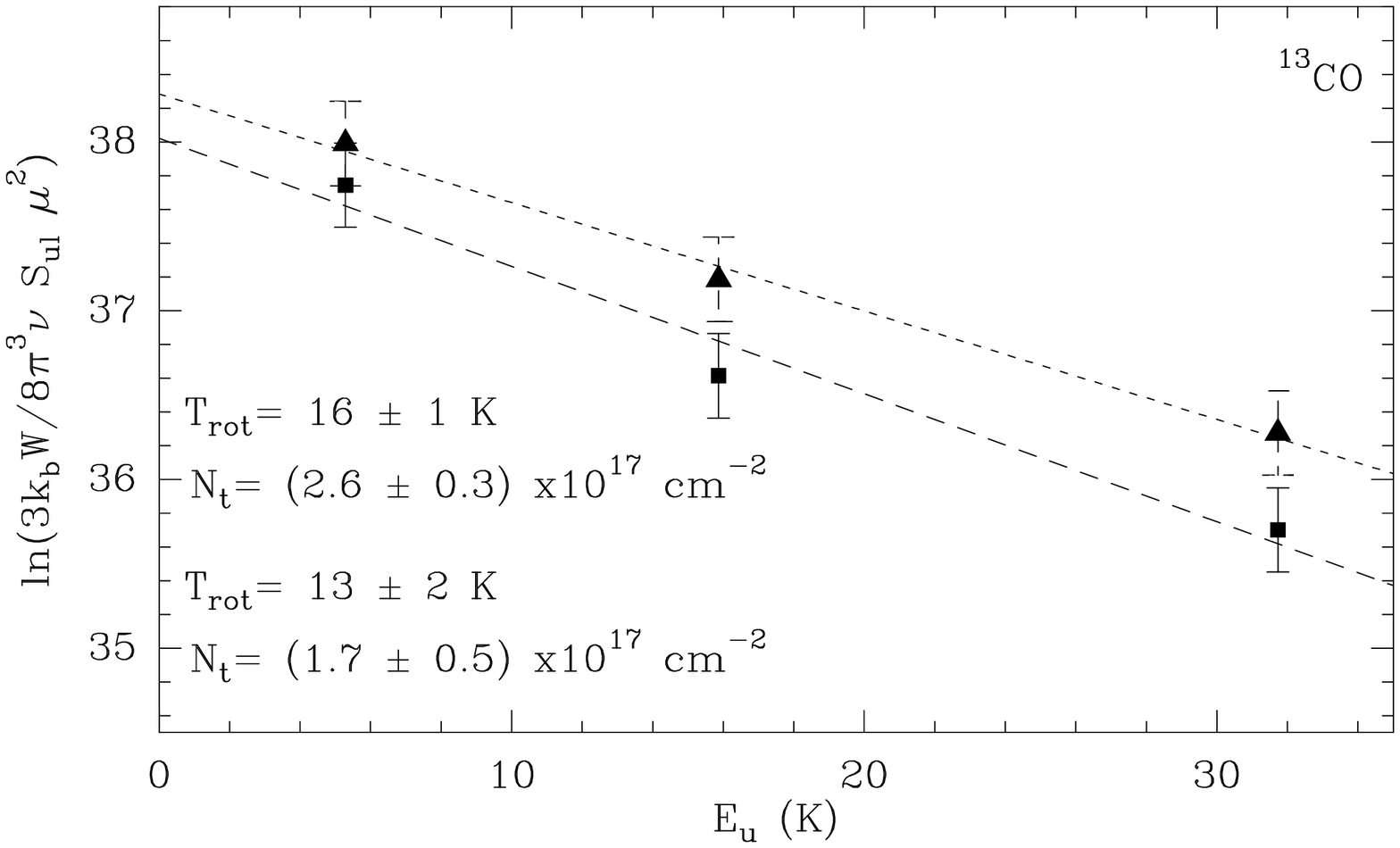}
\caption{Population diagram for $^{13}$CO and best linear fit (dashed
  line) to the data (filled symbols). Triangles and squares represent
  data points with and without the opacity correction $C_\tau$
  applied, respectively (\S\,\ref{sec:popdiag}). The values of \trot\ and
  \ntot\ derived from the fits are indicated in the bottom left corner
  of the box for the optically thin (lower) and optical depth
  corrected (upper) approximations. The error bars of the data points
  include flux uncertainties due to the rms of the spectra and
  absolute flux calibration, up to $\sim$25\%\,(\S\,\ref{sec:obs}).}
\label{fig:13co_rot}
\end{figure}

\begin{figure}[hbtp!]
\centering
\includegraphics[width=0.85\hsize]{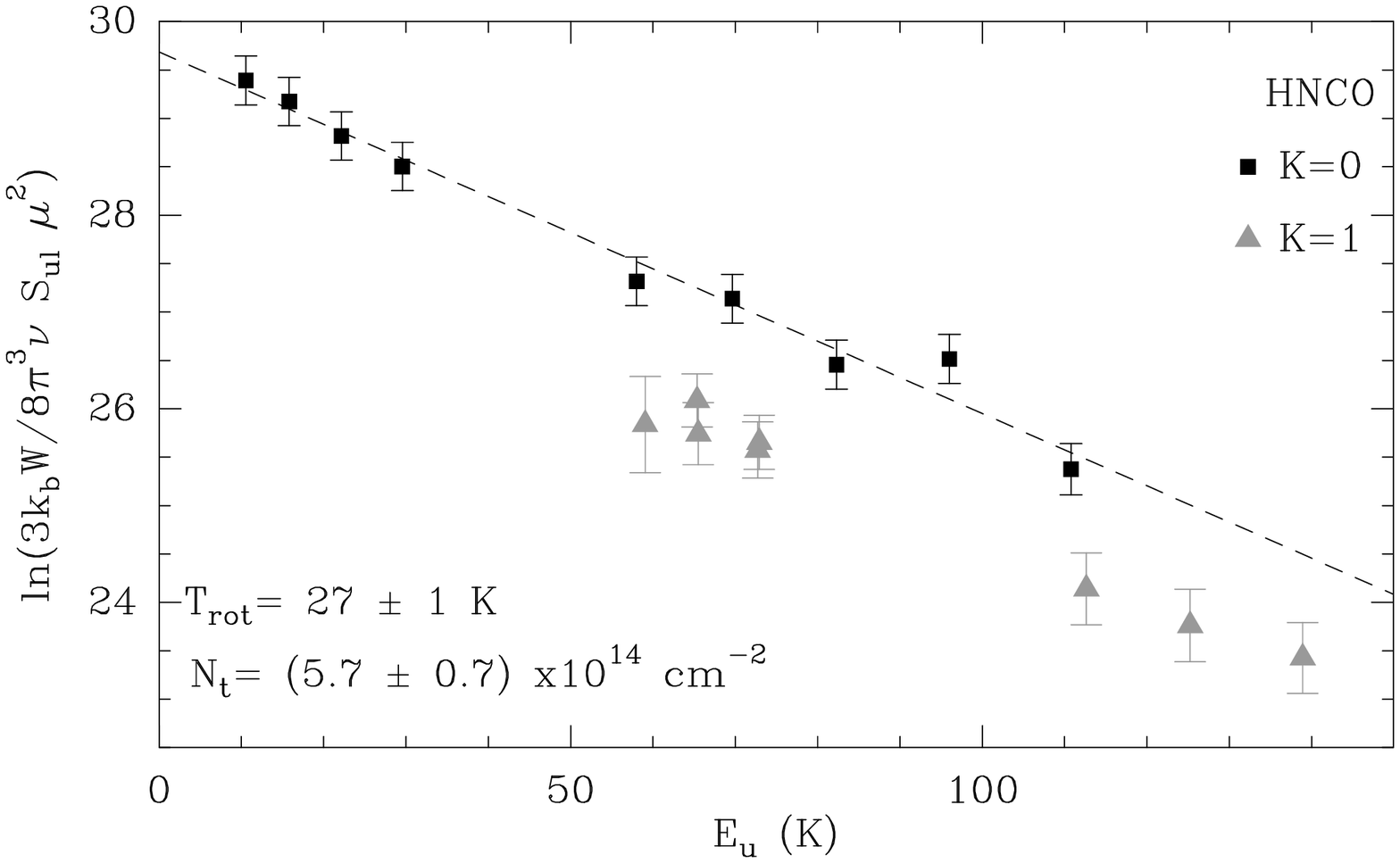}
\includegraphics[width=0.85\hsize]{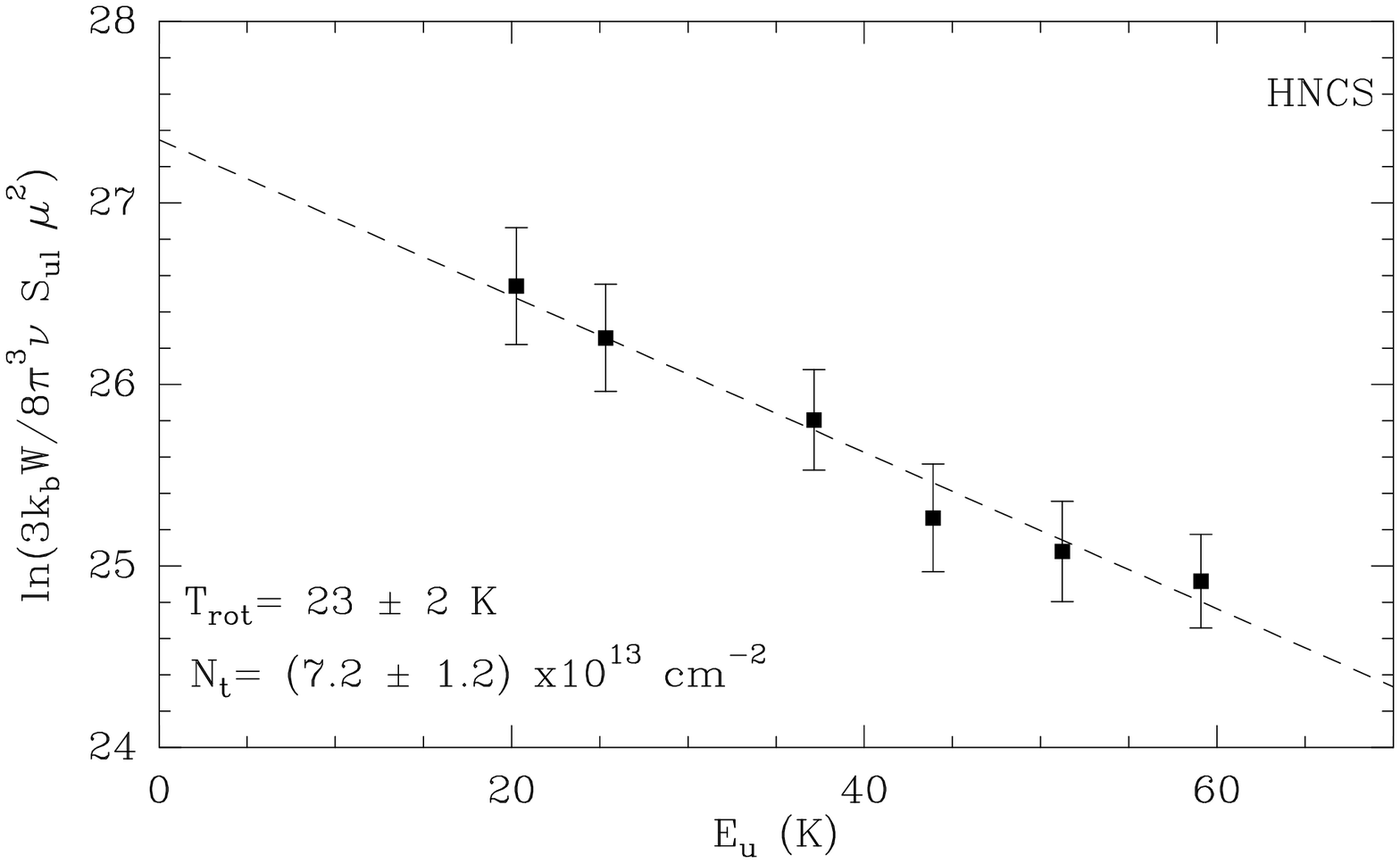}
\includegraphics[width=0.85\hsize]{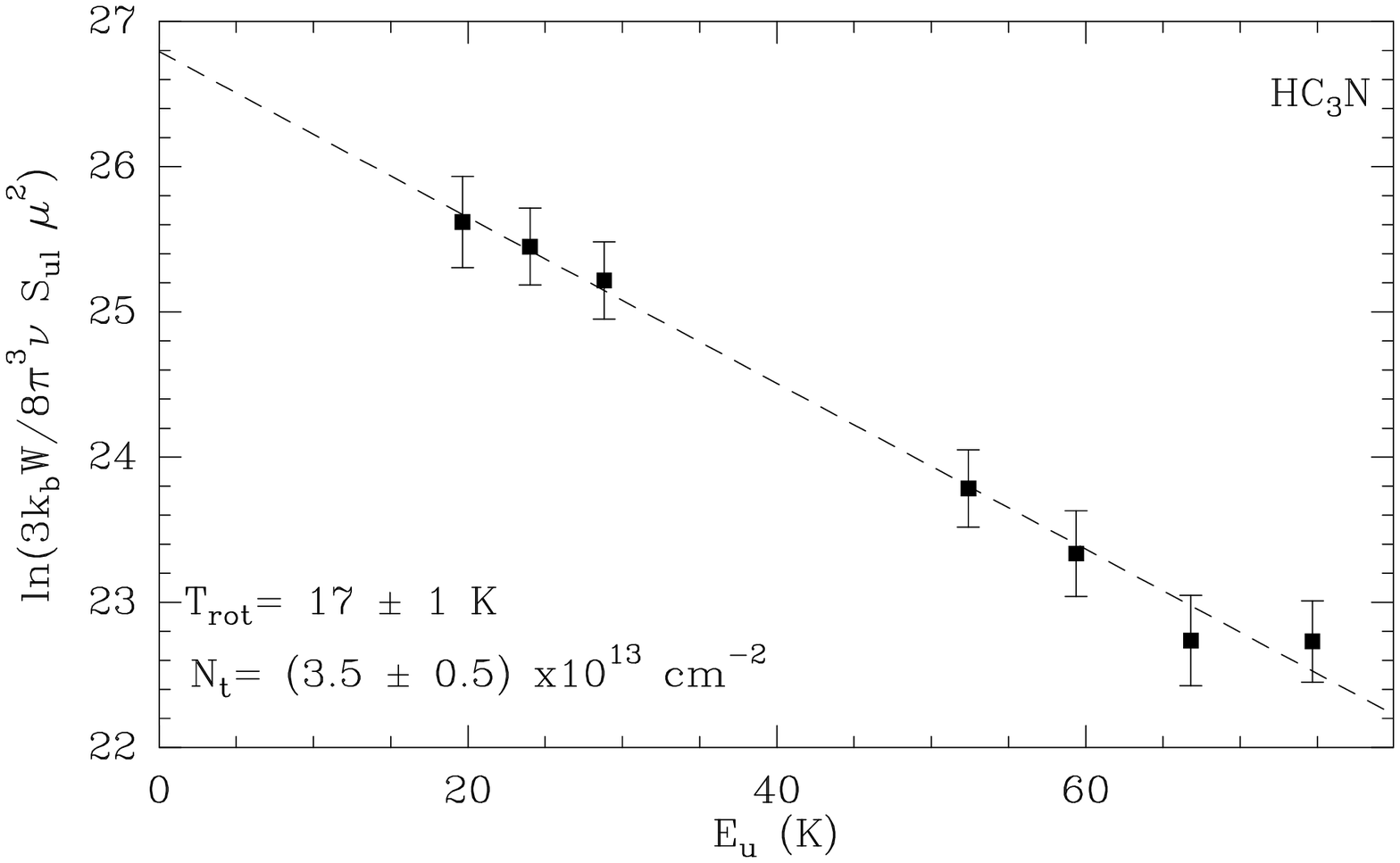}
\includegraphics[width=0.85\hsize]{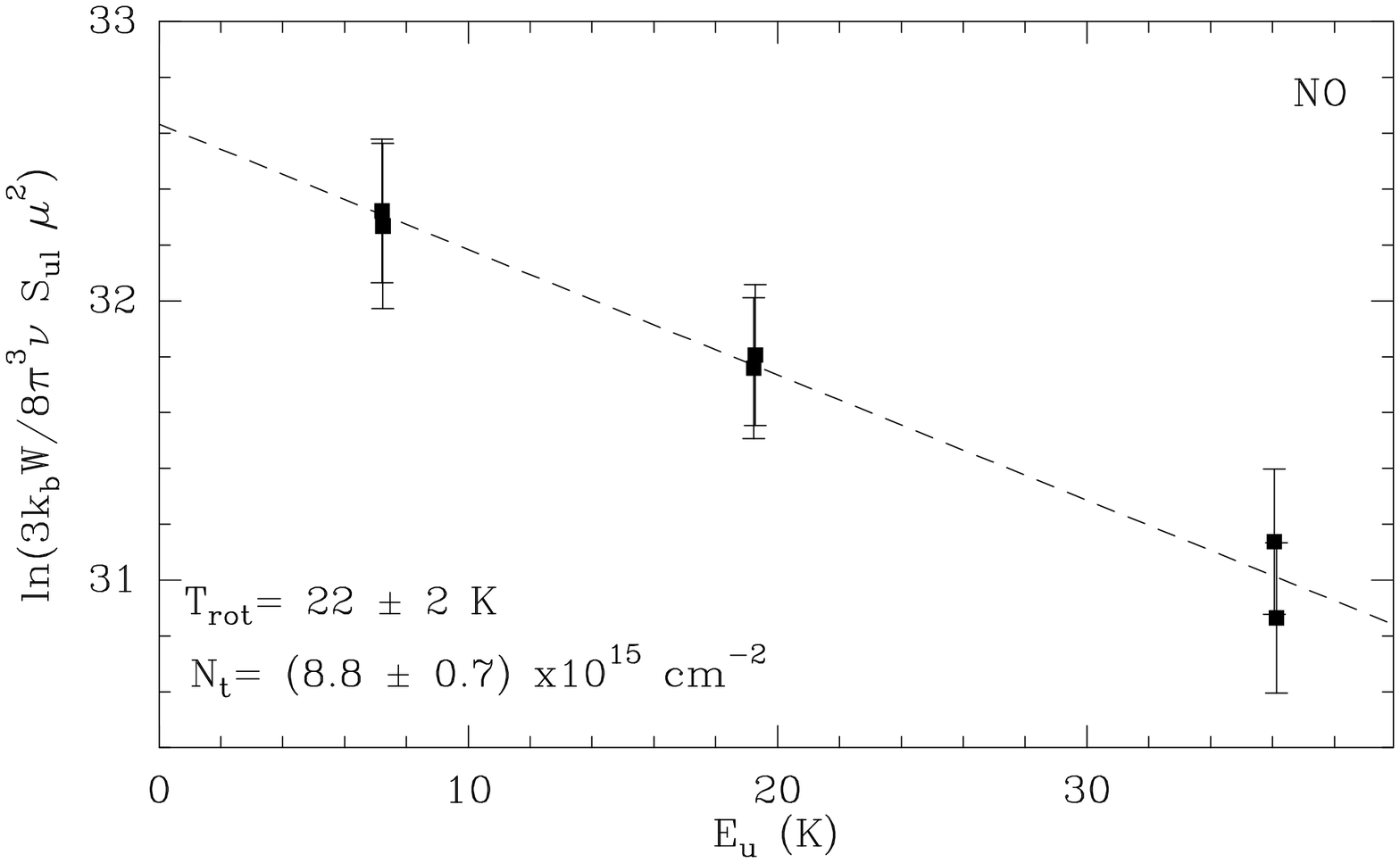}
\caption{Population diagrams of HNCO, HNCS, HC$_3$N, and NO (from top
  to bottom) and best linear fit (dashed line) to the data points
  (filled symbols; from Table~\ref{tab:measures}). In the case of
  HNCO, black squares represent transitions with $K_{\mathrm{a}}$=0 and grey triangles the
  $K_{\mathrm{a}}$=1 transitions.  The values of \trot\ and \ntot\ derived from the
  fit are indicated in the bottom left corner of each box. An opacity
  correction is not applied in these cases because all the transitions
  are optically thin. Error bars as in Fig.~\ref{fig:13co_rot}.}
\label{fig:rotdiags}
\end{figure}

The population diagrams for the molecules reported in this work and
the derived results are shown in Figures
\ref{fig:13co_rot}-\ref{fig:rotdiags} and Table \ref{tab:abun}.  For
\trecem\ we obtain \trot$\sim$13\,K and
\ntot(\trecem)$\sim$2$\times$10$^{17}$\,cm$^{-2}$. For these values of
\trot\ and \ntot, we expect moderate optical depths at the line centre
($\tau_{\mathrm{110\,GHz}}$$\sim$0.5,
$\tau_{\mathrm{220\,GHz}}$$\sim$1.1, and
$\tau_{\mathrm{330\,GHz}}$$\sim$0.9, for a typical linewidth of
FWHM$\sim$35\,\kms), which would lead to underestimating both
\trot\ and \ntot. According to this, an opacity correction factor
$C_{\tau}$=$\ln$($\tau$/(1-e$^{-\tau}$)), as defined by, for instance,
\cite{gol99}, has been introduced in the population diagram, and a new
best fit was obtained. 
This process (of
fitting opacity corrected data points and re-computing $C_{\tau}$) was performed iteratively until convergence was reached. The
opacity-corrected values derived are \trot$\sim$16\,K and
\ntot(\trecem)$\sim$3$\times$10$^{17}$\,cm$^{-2}$.  

In the rotational diagram of HNCO (top panel in
Fig.\,\ref{fig:rotdiags}), we can clearly see that the transitions of
the $K_a$=0 and $K_a$=1 ladders follow two different trends. Both
trends have similar slopes; that is, the $K_{\mathrm{a}}$=0 and
$K_{\mathrm{a}}$=1 data points are arranged in two almost parallel 
straight lines representing similar rotational temperatures.
As we show
in \S\,\ref{sec:nolte} and Appendix\,\ref{sec:app}, the $y$-offset between both lines 
can be explained by non-LTE excitation effects, which are most prominent in
the $K_{\mathrm{a}}$=1 transitions. Using only the HNCO transitions of
the $K_{\mathrm{a}}$=0 ladder, we derive \trot$\sim$27\,K and
\ntot(HNCO)$\sim$6$\times$10$^{14}$\,cm$^{-2}$.  Following the
Eq.\,(\ref{eq:abun}) and using the \trecem\ opacity corrected column
density, we derive a fractional abundance of
$X$(HNCO)\,$\sim$\,1$\times$10$^{-7}$.

The rotational diagrams of HNCS, HC$_3$N, and NO
(Fig.\,\ref{fig:rotdiags}) show a linear trend, which is consistent
with a unique temperature component of \trot$\sim$15-25\,K, in
agreement with the relatively low value obtained from \trecem.  There
is also good agreement with estimates of \trot\ from other molecules
such as SO$_2$ and NH$_3$ from earlier works
\citep[][]{gui86,morr87,san97}. The column densities and fractional
abundances derived range between \ntot=3$\times$10$^{13}$\,cm$^{-2}$ 
and 9$\times$10$^{15}$ and $X$=7$\times$10$^{-9}$ and 
2$\times$10$^{-6}$, respectively (see Table\,\ref{tab:abun}).

Given the relatively low column densities obtained for HNCO, HNCS,
HC$_3$N, and NO, all the transitions detected are optically thin, so no opacity correction is needed. We note that the high
abundance of NO inferred from our data ($\approx$10$^{-6}$) is
comparable to that of SO$_2$ and SO, standing amongst the most
abundant molecules detected to date in this object \citep[see
  e.g.\,][]{morr87,san00,san14}. 
The second-most abundant molecule
reported in this work is HNCO, which is a factor $\sim$20 
less abundant than NO and comparable in abundance to HCN and HNC
\citep{morr87,san97,san00}. The abundance of HNCO is a factor $\sim$10 
larger than that of its sulfur analogue, HNCS.

\begin{table}[hbt!] 
\caption{Column densities (\ntot), rotational temperatures (\trot), and
  fractional abundances relative to H$_2$ ($X$) for the molecules
  detected in \oh\ as derived using the population diagram method
  (\S\,\ref{sec:popdiag}) and from non-LTE excitation analysis
  (\S\,\ref{sec:nolte}). The characteristic angular size for the
  emitting region adopted is $\Omega_{\mathrm{s}}$=4\arcsec$\times$12\arcsec.
}
\label{tab:abun}
\centering    
\begin{tabular}{c c c c}
\hline\hline
Molecule   &  \ntot          & T$_{rot}$            & $X$          \\    
           & (cm$^{-2}$) & (K)                &               \\ 
\hline
\\
\multicolumn{1}{c}{} & ($i$) LTE RESULTS & \multicolumn{2}{c}{}   \
\\
$^{13}$CO & 2.6(0.3)$\times$10$^{17}$ & 16(1) & 5$\times$10$^{-5}$\tablefootmark{\dag} \\ 
HNCO & 5.7(0.7)$\times$10$^{14}$ & 27(1) & 1$\times$10$^{-7}$ \\ 
HNCS & 7.2(1.2)$\times$10$^{13}$ & 23(2) & 1$\times$10$^{-8}$ \\ 
HC$_3$N & 3.5(0.5)$\times$10$^{13}$ & 17(1) & 7$\times$10$^{-9}$ \\ 
NO & 8.8(0.7)$\times$10$^{15}$ & 22(2) & 2$\times$10$^{-6}$ \\ 
\hline 
\\
\multicolumn{1}{c}{} & ($ii$) NON-LTE RESULTS\tablefootmark{\ddag} & \multicolumn{2}{c}{} \\ 
\\
HNCO       &  [4.0-5.2]$\times$10$^{14}$ & [26-28] & [0.8-1]$\times$10$^{-7}$ \\
HNCS       &  [4.9-7.4]$\times$10$^{13}$ & [23-28] & [0.9-1]$\times$10$^{-8}$ \\
HC$_3$N    &  [2.4-3.3]$\times$10$^{13}$ & [45-55] & [5-6]$\times$10$^{-9}$ \\
NO         &  [7.5-8.9]$\times$10$^{15}$ & [28-33] & [1-2]$\times$10$^{-6}$ \\
\hline                                  
\end{tabular}
\tablefoot{ \tablefoottext{\dag}{Adopted based on the estimate of
    X(\trecem) by \cite{morr87}.}\\ \tablefoottext{\ddag}{In
        this case, we provide ranges of column densities, abundances,
        and kinetic temperatures consistent with the
        observations obtained from our non-LTE excitation analysis
        adopting n(H$_2$)=10$^5$\,cm$^{-3}$, for HNCS, HC$_3$N and NO,
        and n(H$_2$)=4$\times$10$^7$\,cm$^{-3}$, for HNCO (see
        \S\,\ref{sec:nolte} and Appendix\,\ref{sec:app}).}\\ }
\end{table}

\subsection{Non-LTE excitation}\label{sec:nolte}

When the local density of molecular hydrogen (\dens) is insufficient
to thermalize the transitions of a given molecule, departures from a
linear relation in the population diagram are expected. For example,
different values of \trot\ may be deduced from different transitions,
leading to a curvature (or multiple slopes) in the distribution of the
data points in the population diagram, which also affects the total
column density inferred. Non-LTE excitation effects on the population
diagrams of the N-bearing molecules detected in \oh\ are investigated
and discussed in Appendix\,\ref{sec:app}. The high nebular
densities in the dominant emitting regions of the outflow
($\sim$10$^5$-10$^6$\,\cm3) indicate that the \trecem\ lines are
thermalized over the bulk of the outflow; 
however, the transitions observed from HNCO, HNCS, HC$_3$N, and NO,
have critical densities of up to $\sim$10$^6$\,cm$^{-3}$ and,
therefore, some LTE deviations may occur. In these cases, the level
propulations of the different species are numerically computed (for
given input values of \tkin, \ntot, and \dens) considering both
collisional and radiative proceses and the well known {\sl LVG}
approximation -- see Appendix\,\ref{sec:app}.

\begin{figure}[hbtp!] 
\centering
\includegraphics[width=0.85\hsize]{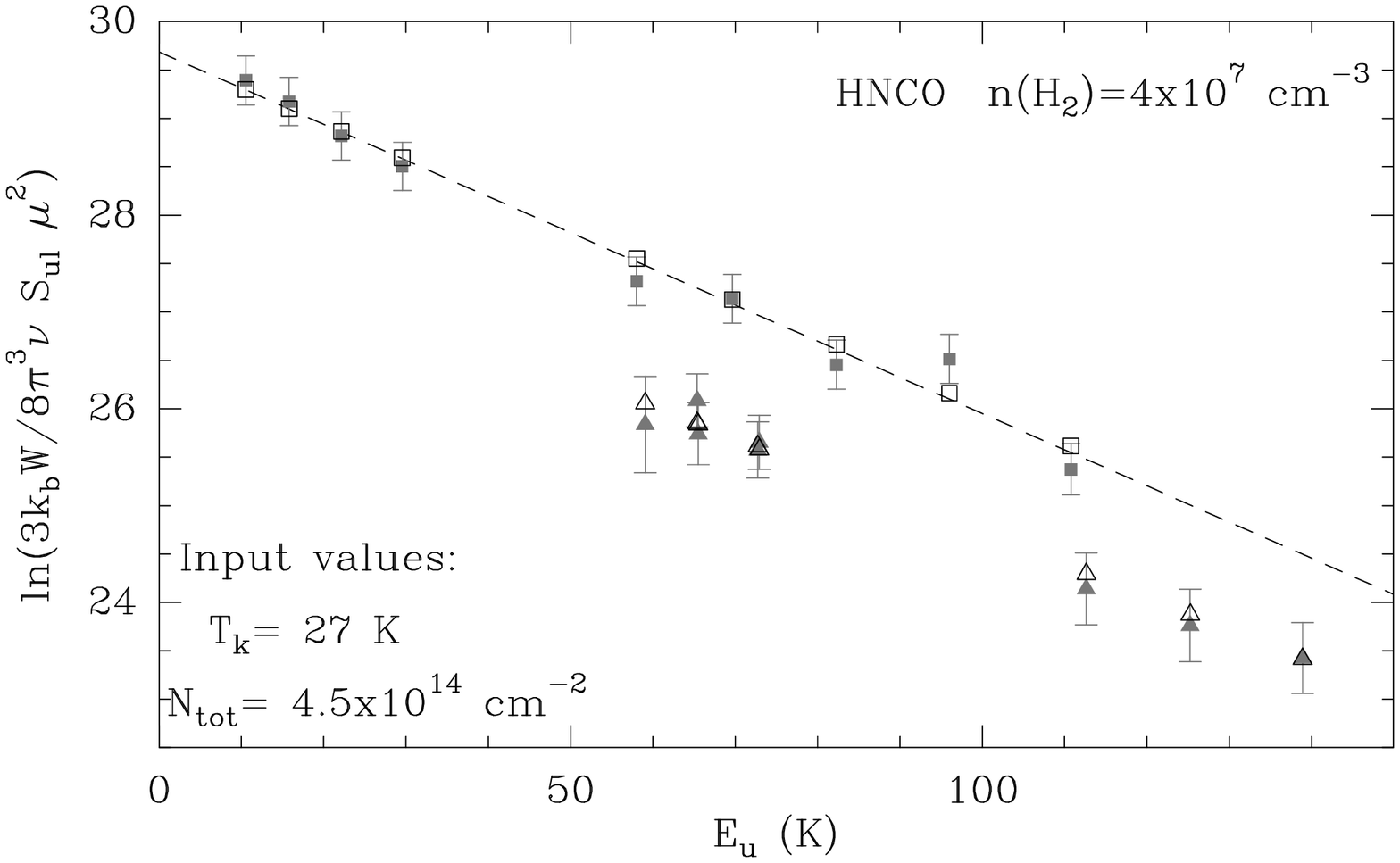} 
\includegraphics[width=0.85\hsize]{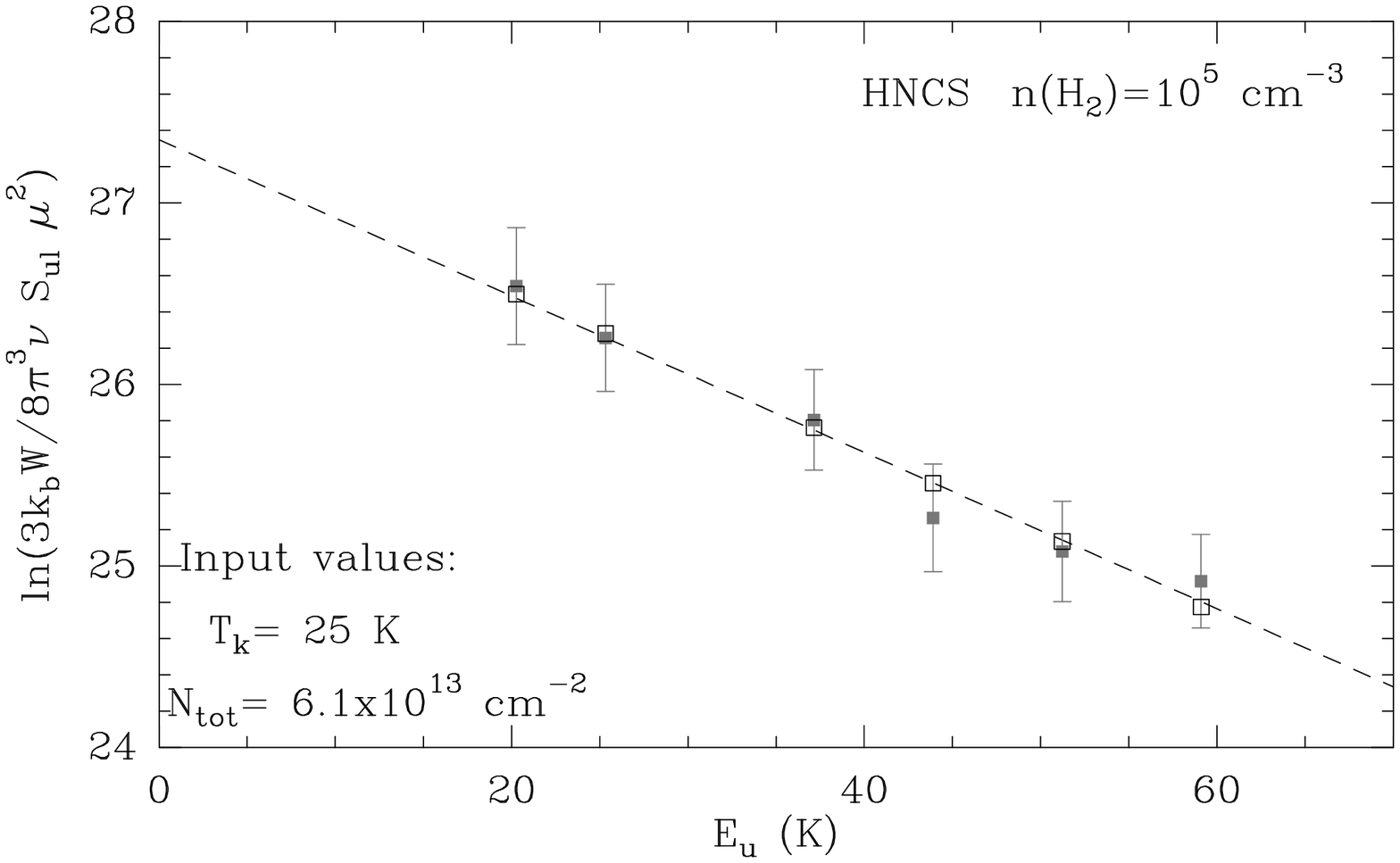}
\includegraphics[width=0.85\hsize]{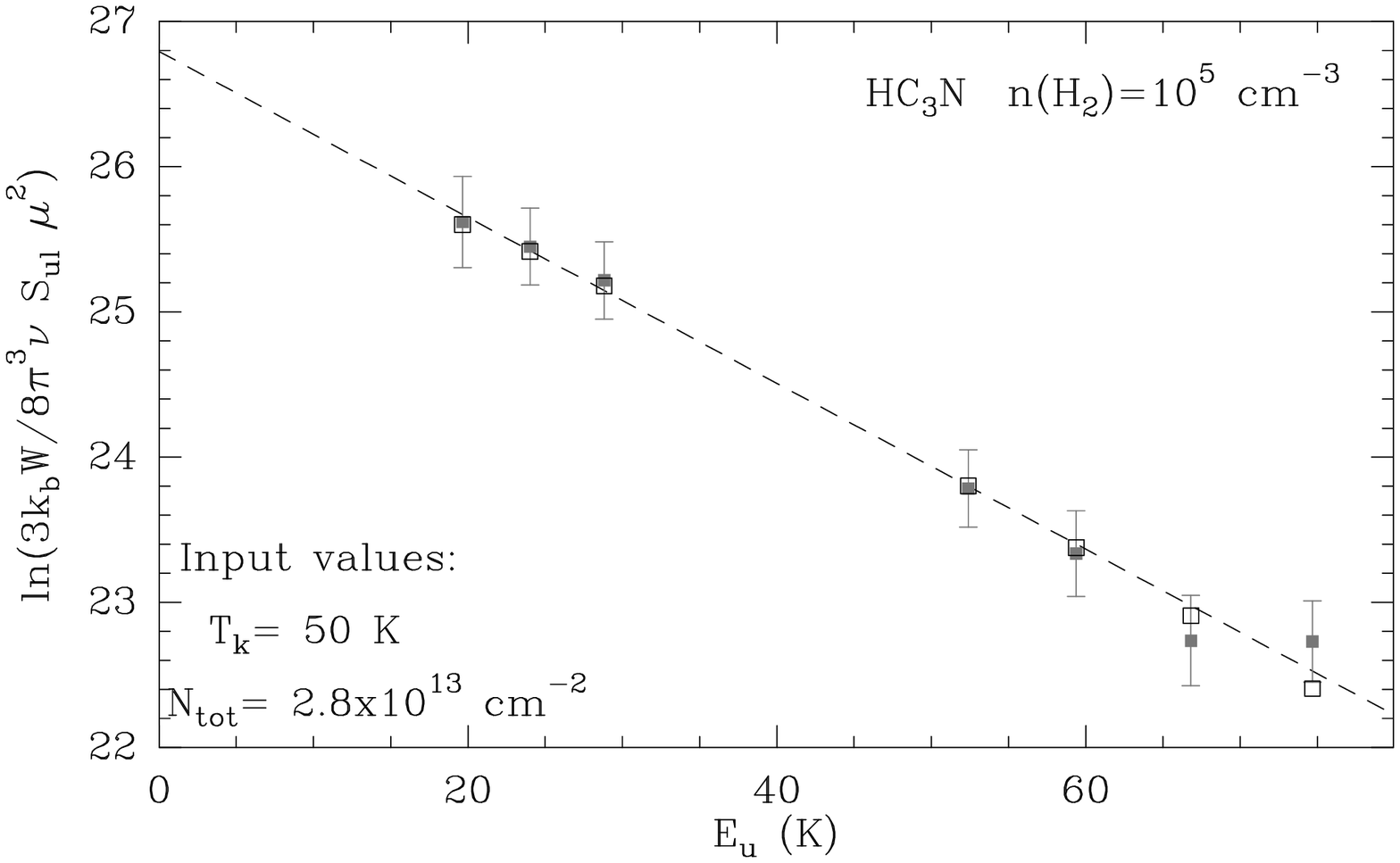}
\includegraphics[width=0.85\hsize]{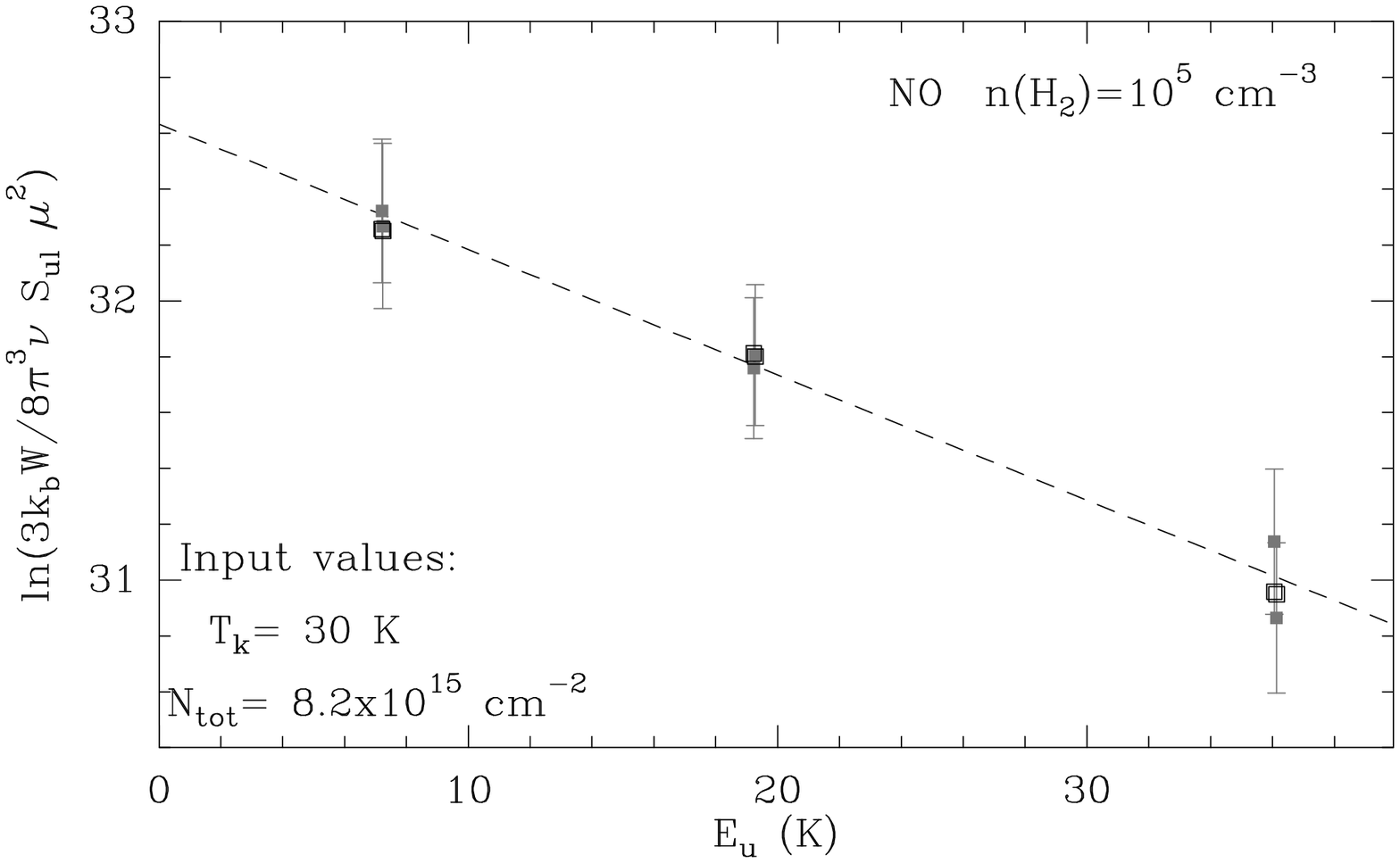}
\caption{Population diagrams of HNCO, HNCS, HC$_3$N, and NO including
  observational data points (filled symbols) and points from
  representative non-LTE models (empty symbols;
  \S\,\ref{sec:nolte}). The main input parameters (\dens, \tkin,
      and \ntot) of the non-LTE models are indicated within the
      boxes, along with the linear fit to the data points (dashed
      line; same as in Fig.\,\ref{fig:rotdiags}). For HNCO, squares
    and triangles represent the $K_{\mathrm{a}}$=0 and
    $K_{\mathrm{a}}$=1 transitions, respectively (as in
    Fig.\,\ref{fig:rotdiags}).}
\label{fig:mxmods}
\end{figure}

In this section, we investigate whether the observations could also be
reproduced by values of \tkin\ and \ntot\ that are different from those deduced
from LTE calculations using moderate \dens\ for which LTE does not
apply. The results from our non-LTE excitation models have been
represented in a population diagram (i.e.\ N$_{\rm u}$/g$_{\rm u}$
vs.\ \eu) together with the observed data-points. The best data-model
fits are shown in Fig.\,\ref{fig:mxmods}, and the range of input values
for \tkin\ and \ntot\ consistent with the observations are given in
Table\,\ref{tab:abun}. 

Except for HNCO (see below), in our non-LTE models we have adopted a
mean characteristic density in the emitting regions of the outflow of
\dens$\sim$10$^5$\,cm$^{-3}$. The lowest densities in \oh,
\dens$\sim$10$^3$-10$^4$\,cm$^{-3}$, are only found at large distances
from the star in the southern lobe (clumps I5 and beyond) that do not
contribute to the emission observed from these N-bearing species
\citep{alc01}.

In the case of HNCO, one notable effect of non-LTE conditions is the
split of the $K_{\mathrm{a}}$=0 and $K_{\mathrm{a}}$=1 ladders into
two almost paralell straight lines in the population diagram
(Appendix\,\ref{sec:app}). The separation between the
$K_{\mathrm{a}}$=0 and $K_{\mathrm{a}}$=1 ladders, which is indeed
observed in \oh\ (e.g.\ Fig.\,\ref{fig:mxmods}, top panel),
progressively reduces as the density increases; when densities
$\ga$10$^8$\cm3 are reached, all transitions reported here are very
close to thermalization, and both the $K_{\mathrm{a}}$=0 and
$K_{\mathrm{a}}$=1 ladders merge into one single straight line. The
non-LTE excitation analysis of HNCO indicates that the observed
separation between the $K_{\mathrm{a}}$=0 and $K_{\mathrm{a}}$=1
ladders in the population diagram of \oh\ requires nebular densities
of \dens$\sim$4$\times$10$^7$\,cm$^{-3}$. This suggests that most
of the observed HNCO emission probably arises at relatively dense
regions in the envelope. Adopting
\dens$\sim$4$\times$10$^7$\,cm$^{-3}$, therefore, we find that the
observations are reproduced well for a range of values of
\ntot(HNCO)$\sim$[4.0-5.2]$\times$10$^{14}$\,cm$^{-2}$ and
\tkin$\sim$26-28\,K, that is, very similar to those obtained
under the LTE approximation. 

Several authors have pointed out the importance of infrared pumping to explain the excitation of HNCO under certain conditions \citep[e.g.][]{chu86,li13}.
We have not taken the effect of IR pumping into account given the complexity of the problem, which is beyond the scope of this paper.
This effect adds additional uncertainties to the HNCO abundance, which could be larger or smaller than the value quoted in Table\,\ref{tab:abun}, but probably by a factor not greater than $\sim$2-5.
Additional discussion about this topic is given in the Appendix\,\ref{sec:app}.

For HNCO and the rest of the molecules, HNCS, HC$_3$N, and NO, one
major effect of non-LTE excitation on the population diagram is a
modification of the slope of the straight line defined by the
data points (N$_{\rm u}$/g$_{\rm u}$ vs.\ \eu) with respect to
the correct value entered as input in the model as \tkin.  In
particular, as can be seen in Fig.\,\ref{fig:mxmods}, for
\dens$\sim$10$^5$\,\cm3, the rotational temperature that one
    would deduce from the population diagram is lower than the input
kinetic temperature (\trot$<$\tkin, sub-thermal excitation). The
largest difference between \tkin\ and \trot\ in our models is found
for HC$_3$N; in this case, values of \tkin\ of up to $\sim$55\,K in
the emitting regions cannot be ruled out.
On the other hand, in general for all species, the column densities
derived from the non-LTE excitation analysis 
are systematically lower than those deduced assuming LTE conditions. 
We note, however, that these differences are typically $\la$30\%.

Finally, as shown in the Appendix\,\ref{sec:app}, for densities of
\dens$\leq$10$^4$\,\cm3, non-LTE level populations of HNCO, HNCS,
HC$_3$N, and NO would result in a double slope in their population
diagrams. (This effect would be most prominent for HC$_3$N.) That this is not observed in our data, described well by a unique
value of \trot, corroborates that the typical densities in the
emitting regions are above 10$^4$\,\cm3.

\section{Chemical modelling} \label{sec:chemmod}
In this section, we present thermodynamical chemical equilibrium (TE) and 
chemical kinetics models to investigate the formation of \nmols\ in O-rich 
CSEs with characteristics similar to those in \oh. The TE calculations should provide a good estimation of
the molecular abundances near the stellar photosphere, up to
$\sim$4-5\,$R_{*}$, given the high densities
($\approx$10$^{14}$-10$^{9}$\,\cm3) and high temperatures
($\sim$2000\,K) expected in these innermost
regions \citep[e.g.][]{tsu73}. As the gas or dust in the stellar wind
expands, the temperature and the density gradually decrease, and the
chemical timescale increases, making chemical kinetics dominate in
determining the molecular abundances. Eventually, the dispersion of
the envelope allows the interstellar UV photons to penetrate through
the outermost layers,
leading to 
the onset of a productive photo-induced chemistry. 

For many important molecules, the abundances established by the
equilibrium chemistry in the dense, hot photosphere are expected to be
greatly modified by various processes (not all understood well) that
operate in the deep envelope layers.
First, the inner wind regions at a few \rs\ are dominated by
non-equilibrium reactions triggered by low-velocity shocks generated
by the stellar pulsation. Second, the formation of dust particles,
which begins farther out at $\sim$5-15\rs, well affects the chemistry a lot
in different ways, for example, depleting refractory species 
from the gas phase (owing to grain adsorption) and powering the
production of other compounds through grain surface reactions. Because
of all this, the molecules produced by these processes in the deep
envelope layers, named `parent', are injected to the
intermediate envelope with initial abundances that might differ
significantly from values predicted by
TE chemical calculations \citep[e.g.][]{che06}.

\subsection{Physical model of the envelope} \label{sec:physmod}

We used two different physical structures as input in our
chemical kinetics models: $a$) a spherical stellar wind with
characteristics similar to those of the slow central nebular
component of \oh; and $b$) a slab of gas (plane-parallel geometry) with
characteristics similar to those of the walls of the hollow lobes
of \oh. For the TE calculations, only the
physical model $a$ has been considered since the conditions for
thermodynamical chemical equilibrium are not met in the lobes
(\dens$\sim$10$^5$\,\cm3 and \tkin$\sim$20\,K).

The physical model $a$ consists of a spherical envelope of gas (and
dust) expanding around the central AGB star of \oh. This has been
taken as a representation of the slow central component of the
outflow, which has been interpreted as the fossil remnant of the old
AGB CSE (\S\,\ref{sec:intro}).  We separately modelled $i$) the
innermost envelope regions (within $\sim$5\rs), where TE conditions
apply, and $ii$) the intermediate/outer envelope (from $\sim$20\,\rs\
to its end), where chemistry is driven by chemical kinetics. The
density, temperature, and velocity are expected to vary across these
two components as a function of the radial distance to the centre
($r$).

The \ioe\ is characterized well observationally (\S\,\ref{sec:intro})
and its main physical parameters are summarized in
Table\,\ref{tab:oh231}.
For modelling purposes, the \ioe\ has been chosen to begin at
$\sim$20\rs, that is to say, well beyond the dust condensation radius (\rd) where
the full expansion velocity of the gas (by radiation pressure onto
dust) has been reached. Throughout this slow central component, we
adopt a characteristic constant expansion velocity of \vexp=20\,\kms.
The gas kinetic temperature has been approximated by a power law that
varies with the radius as $r^{-0.7}$ (typical of AGB CSEs,
e.g.\,\cite{che92}).
The density in the \ioe\ is given by the law of
conservation of mass,
which results in a density profile varying as $\propto r^{-2}$\mloss/\vexp.  
The outer radius of the slow central component of the molecular
outflow of \oh\ is $\sim$7$\times$10$^{16}$\,cm (3\arcsec\ at
$d$=1500pc; see Sect.\,1), which has been adopted in our model. This
value is a factor $\sim$10 lower than the CO photodissociation
radius given the AGB mass-loss rate and expansion velocities measured
in this object and adopting $X$(CO)=3$\times$10$^{-4}$,
following \cite{mam88} and \cite{pla90}. This indicates that the outer
radius of the envelope in the equatorial direction represents a real
density cut-off in the AGB wind marked by the beginning of the heavy
AGB mass loss.

The inner envelope, where the TE calculations were done, begins at
the stellar photosphere and ends at the dust condensation radius
($R_{\rm c}$=5\rs), i.e.\ before dust acceleration takes place. For
these regions we use \vexp=5\,\kms, in agreement with the line widths
of vibrationally excited \water\ emission lines in the far-IR in this
object \citep{san14} and typical values in other AGB
stars \citep[][and references therein]{marce12}.  For the gas kinetic
temperature, we use the same power law as in the \ioe.  The densities
in these innermost regions near the stellar photosphere were
calculated based on theoretical arguments considering hydrostatic
equilibrium (in the static stellar atmosphere, up to $\sim$1.2\rs) and
pulsation induced, low-velocity shocks (in the dynamic atmosphere,
from $\sim$1.2\rs\ and up to \rd$\sim$5\rs -- see
e.g. \cite{marce12}, and references therein for a complete
formulation and discussion).  According to this, in our model, the
density steeply varies from $\sim$10$^{14}$\cm3 at \rs\ to
$\sim$10$^{12}$\,\cm3 at 1.2\rs; beyond this point, it decreases to
$\sim$10$^{9}$\,\cm3 at $\sim$5\rs.  Although the density and
temperature in the innermost nebular regions of \oh\ are not as well
constrained observationally as in the \ioe, the laws adopted reproduce the physical conditions in the SiO maser emitting
regions
well \citep[$\approx$10$^{9-10}$\,\cm3 and $\sim$1000-1500\,K at
$\sim$2-3\,\rs;][]{san02}.
\begin{table} 
\caption{Parameters of the central Mira-type star and 
slow, core envelope component of \oh\,
used for the chemical models (\S\,\ref{sec:chemmod}.)}
\label{tab:oh231}
\centering    
\begin{tabular}{l c c}
\hline\hline
Parameter & Value & Reference      \\ 
\hline   
Distance ($d$)                             &  1500\,pc                                   & b \\
Stellar radius (R$_{*}$)                   &  4.4x10$^{13}$\,cm                           & g  \\ 
Stellar effective temperature (T$_{*}$)    &  2300\,K                                    & c,d,h  \\
Stellar luminosity (L$_{*}$)               &  10$^{4}$\,\ls                               & g  \\
Stellar mass (M$_{*}$)                     &  1\,\msun                                   & g  \\ 
AGB CSE expansion velocity (\vexp)       &  20\,\kms                            &  a,e,f,g,i \\
AGB mass loss rate (\mloss)                    &  10$^{-4}$\,\my                              &  e,a,f \\
Gas kinetic temperature (T$_{\mathrm{k}}$) &  T$_*(r/R_{*})^{-0.70}$\,K &  i \\
\hline                        
\hline                                  
\end{tabular}
\tablefoot{a: \cite{alc01}, b: \cite{choi12}, c: \cite{coh81}, d: \cite{kas92}, e: \cite{morr87},
f: \cite{san97}, g: \citep{san02}, h: \cite{san04}, i: this work.  }
\end{table}

The walls of the lobes, where the molecular abundances have also been
predicted using our chemical kinetics model, have been approximated by
a slab of gas (with a plane-parallel geometry; model $b$ above). We used a wall thickness of $\sim$1\arcsec, and characteristic H$_2$
number density and kinetic temperature of \dens$\sim$10$^5$\,\cm2
and \tkin=20\,K constant in the slab, which is representative of the
clumps (I2 \& I4) at the base of the lobes \citep{alc01}.  Within the
slab, we consider the gas to be static, since the expansion velocity
gradient across the lobe walls must be small. 
The sources of ionization and dissociation adopted in our model are cosmic rays and the
interstellar ultraviolet radiation field (see Appendix\,\ref{sec:app_uv} for more
details on the ionization/dissociation sources adopted in our model).

\subsection{Thermodynamical chemical equilibrium (TE) model}\label{sec:chemeq}

Although the innermost envelope regions where TE
conditions hold are not directly probed by the mm-wavelength data
presented in this paper, it is useful to investigate if 
the new N-bearing molecules detected in \oh\ could form in substantial
amounts in these regions or if, alternatively, they need to be 
produced farther out in the stellar wind. 
Our TE code is described in \cite{tej91} and used, for example, in \cite{marce07}.
The computations are performed for more than 600
different species (electron, atoms, and molecules) following the method
described in \cite{tsu73}, and see also \cite{marceTesis}.
The code requires thermochemical information of each molecule
(\cite{chs98}\footnote{NIST-JANAF thermochemical
tables \tt \tiny{http://kinetics.nist.gov/janaf/}},
\cite{mcb02} and \cite{bur05}), as well as the initial elemental
abundances (given in Table\,\ref{tab:abuneq}) and the physical
conditions in the inner layers of the envelope, as described in the
beginning of Sect.\,\ref{sec:chemmod}. 

For HNCS, thermochemical information is not available, so
this molecule has not been modelled. A rough guess of the HNCS
abundance can be obtained scaling from the abundance of its O-bearing analogue,
HNCO, by a factor similar to the oxygen-to-sulphur ratio,
O/S$\sim$37 \citep[][]{asp09}. Excluding HNCS from our chemical
network may, in principle, result in an overestimate of the HNCO,
HC$_3$N, and NO abundances in our model since some of the elements
that would have gone into HNCS are now in other N-, C-, and S-bearing
compounds. Given that HNCS is neither a major carrier of these atoms
nor a key molecule of the gas-phase chemistry, we expect this effect
to be weak.

\begin{table} 
\caption{Elemental abundances used in the TE model (\S\,\ref{sec:chemeq}) taken from \cite{asp09}.}
\label{tab:abuneq}
\centering    
\begin{tabular}{c c}
\hline\hline
Element & Abundance \\
\hline   
H  &  12.00    \\ 
He &  10.93    \\ 
C  &  8.43     \\ 
O  &  8.69     \\ 
N  &  7.83     \\ 
S  &  7.12     \\
\hline
\hline
\end{tabular}
\tablefoot{The abundances are given in the usual logarithmic astronomical scale
where H is defined to be $\log{\rm H}$=12.00 and $\log{\rm
X}$=$\log{\rm (N_X/N_H)}$+12), where N$_X$ and N$_H$ are the
number densities of elements X and H, respectively.
}
\end{table}

\begin{figure}[hbtp!] 
\centering
\includegraphics[width=0.9\hsize]{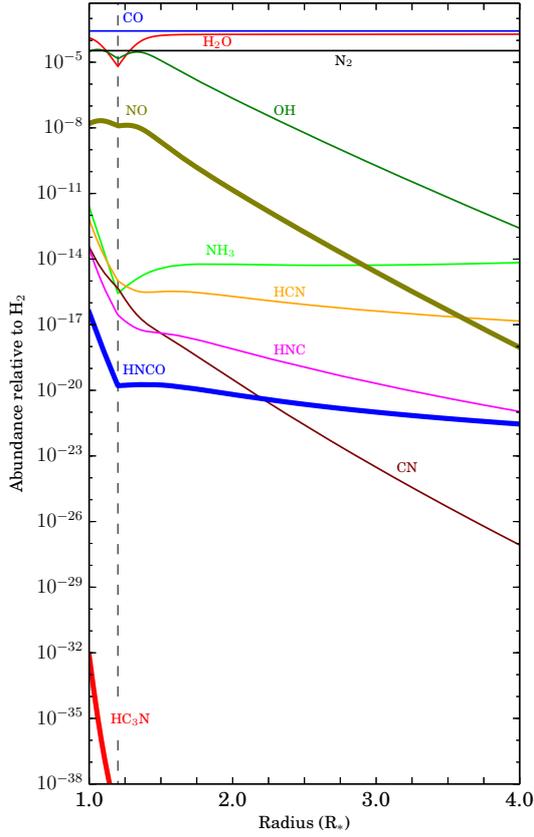}
\caption{Thermodynamical chemical equilibrium model predictions for the 
innermost layers of an O-rich AGB CSE with characteristics presumably
similar to those in \oh\ (\S\,\ref{sec:chemmod}). The spatial
distribution of the N-bearing molecules detected in this work and
other relevant species are plotted. The dashed vertical line denotes the
boundary between the static and the dynamic atmosphere, at
$\sim$1.2\rs.}
\label{fig:chemeq}
\end{figure}

The spatial distribution of the molecular abundances near the star
predicted by our TE chemistry model is shown in
Fig.\,\ref{fig:chemeq}. In all cases, the model abundances are several
orders of magnitude lower than those derived from the observations
(Table\,\ref{tab:abun}).  The largest model-data discrepancy is found
for HC$_3$N, which is formed with an extremely marginal peak
fractional abundance of $X$(HC$_3$N)$<$10$^{-32}$. NO is (after N$_2$)
the most abundant N-compound in the hot photosphere, although the TE
peak abundance is two orders of magnitude lower than observed in \oh.
We may safely conclude that the high abundances derived from the
observations for these molecules cannot be explained as a result of
equilibrium processes at the innermost parts of the envelope where, in
contrast, abundant parent molecules, such as CO, N$_2$, and H$_2$O,
are efficiently formed.

\subsection{Chemical kinetics model}\label{sec:chemkin}
Our chemical kinetics model is based on that of \cite{marce06}, which
has been widely used to model the chemistry across the different
envelope layers of the prototype C-rich star IRC\,10216 \citep[see
also][]{marce07,marce08,marce10} and, most recently, the O-rich YHG
IRC+10420 \citep{qui13}.  The chemical network in our code includes
gas-phase reactions, cosmic rays, and photoreactions with interstellar
UV photons, but it does not incorporate reactions involving dust grains,
X-rays, or shocks.  Chemical reactions considered in this network are
mainly obtained from the UMIST database \citep{woo07}. The network has
been updated with the latest kinetic rates and coefficients for HNCO
\citep{qua10}. As mentioned in Sect.\,\ref{sec:chemeq}, HNCS has not been 
modelled because thermochemical parameters or reaction kinetic rates are
not available for this molecule. Considering that oxygen is
more abundant than sulphur (by a factor O/S$\sim$37;
Table\,\ref{tab:abuneq}), we expect the HNCS abundance to be lower
than that of its O-analogue, HNCO.

There are two major inputs to our chemical kinetics model, namely, the
physical model of the envelope (models $a$ and $b$, respectively
in \S\,\ref{sec:physmod}) and the initial abundances of the `parent'
species, formed in deeper layers, which are injected into the
envelope. Once incorporated into the outflowing wind, these parent
species become the basic ingredients for the formation of new
(`daughter') molecules.  The initial abundances of the parent species
used in our model are given in Table\,\ref{tab:abchemkin}. These
abundances come from thermodynamical chemical equilibrium calculations
and observations in the inner regions of O-rich envelopes (references
are provided in this table).  In the case of NH$_3$, which is a basic
parent molecule for the formation of N-bearing species, we have
adopted the value at the high end of the range of abundances
observationally determined for a few O-rich CSEs,
X(NH$_3$)$\sim$[0.2-3]$\times$10$^{-6}$ \citep{men10}.  We note that,
as also pointed out by these authors, the formation of circumstellar NH$_3$
is particularly enigmatic since the observed abundances exceed the
predictions from conventional chemical models by many orders of
magnitude.

\begin{figure}[hbtp!] 
\centering
\includegraphics[width=1\hsize]{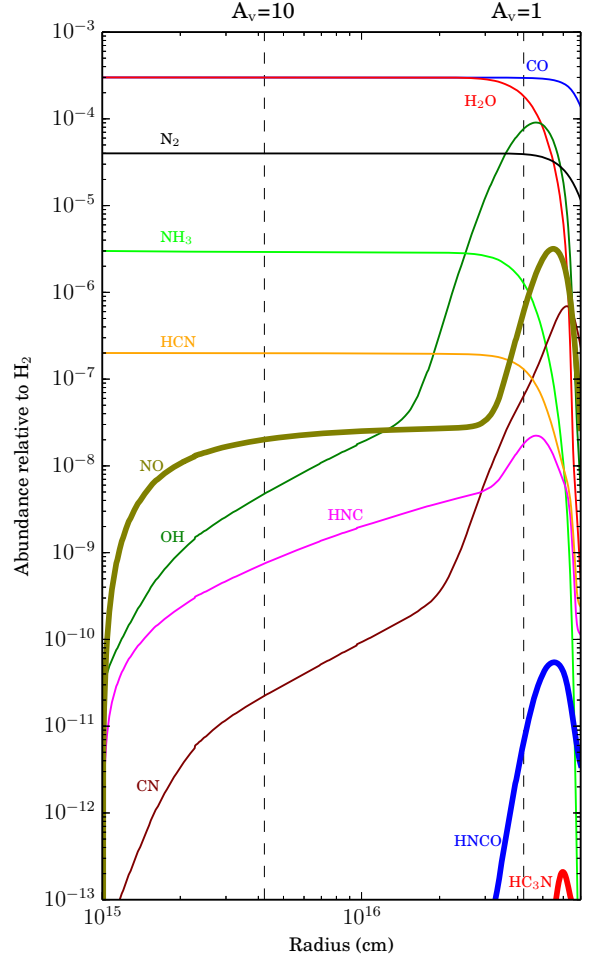}
\caption{Chemical kinetics model predictions for the intermediate/outer layers of an O-rich AGB CSE with physical properties similar to those of the 
slow central parts of \oh\ (model input $a$, \S\,\ref{sec:chemmod}
and Table\,\ref{tab:abchemkin}). The species represented are as in
Fig.\,\ref{fig:chemeq}. }
\label{fig:chemkin}
\end{figure}

Our chemical kinetics model has been used first to investigate
the formation of \nmols\ in the \ioe\ of an O-rich AGB star similar to
the slow central component of \oh\ (model $a$ described
in \S\,\ref{sec:physmod}, see also Table\,\ref{tab:oh231}). The spatial
distribution of the model molecular abundances as a function of the
distance to the centre are shown in Fig.\,\ref{fig:chemkin}. As the
gas in the envelope expands, parent molecules start to be exposed to
the interstellar UV radiation and photochemistry drives the formation
of new species. Penetration of photons through deeper layers is
gradually blocked by the dust extinction\footnote{The visual optical
extinction in magnitudes ($A_V$) is related to the 
H column density by
$N_{H}$=1.87$\times$10$^{21}$$\times$$A_V$\,\cm2, when adopting the
standard conversion from \cite{boh78}.}.
At the very inner layers of
the \ioe, parent species are preserved with their initial abundances, and
at the very outermost layers, all molecules are finally fully
dissociated (destroyed).

As seen in Fig.\,\ref{fig:chemkin}, the peak abundances
of all the N-molecules detected in this work are significantly lower
than observed, except for NO (by 3-4 orders of magnitude).  The peak fractional
abundance for NO predicted by the model,
$X$(NO)$\sim$3$\times$10$^{-6}$, is in principle consistent with the
average value measured in \oh. According to our calculations, this
molecule is expected to form rather efficiently in the winds of O-rich
CSEs mainly via the gas-phase reaction

\begin{equation} 
 \label{eq:no}
\mathrm{N + OH \longrightarrow NO + H}
\end{equation}
and, therefore, NO should be common amongst O-rich evolved
stars. Detection of NO emission lines is, however, hampered by the low
dipole moment of this molecule ($\mu$=0.16\,Debyes).

As derived from our model, the main chemical routes that would form HNCO and  
HC$_3$N in an O-rich CSE are 

\begin{eqnarray} 
\label{eq:hnco}
\mathrm{CN + OH \longrightarrow NCO + H}\\
\mathrm{H_3^+ + NCO \longrightarrow HNCO^+ + H_2}\\
\mathrm{HNCO^+ + H_2 \longrightarrow HNCOH^+ + H}\\
\mathrm{HNCOH^+ + e^- \longrightarrow HNCO + H}
\end{eqnarray}

\noindent 
and\begin{equation} 
  \label{eq:hc3n}
\mathrm{CN + C_2H_2 \longrightarrow HC_3N + H} 
.\end{equation}
However, the standard processes considered here are not sufficient to
reproduce the abundances observed in the particular case of \oh.

\begin{table}[ht!] 
\caption{Initial abundances relative to H$_2$ for representative elements 
and parent molecules used as input for the chemical kinetics models.}
\label{tab:abchemkin}
\centering    
\begin{tabular}{c c c}
\hline\hline
Species & Abundance & Reference \\
\hline                        
He      &  0.17                & a   \\
H$_2$O  &  3.0$\times$10$^{-4}$ & b,TE  \\
CO      &  3.0$\times$10$^{-4}$ & c,TE  \\
CO$_2$  &  3.0$\times$10$^{-7}$ & d  \\
NH$_3$  &  3.0$\times$10$^{-6}$ & e  \\
N$_2$   &  4.0$\times$10$^{-5}$ & TE \\
HCN     &  2.0$\times$10$^{-7}$ & f,g\\ 
H$_2$S  &  7.0$\times$10$^{-8}$ & h  \\
SO      &  9.3$\times$10$^{-7}$ & f  \\
SiO     &  1.0$\times$10$^{-6}$ & i  \\
SiS     &  2.7$\times$10$^{-7}$ & j \\                                                  
\hline                                  
\end{tabular}
\tablefoot{a:\,\cite{asp09}, b:\,\cite{mae08}, c:\,\cite{tey06}, d:\,\cite{tsu97}, e:\,\cite{men95}, f:\,\cite{buj94}, g:\,\cite{sch13}, h:\,\cite{ziu07}, i:\,\cite{gon03}, j:\,\cite{sch07} 
TE: From thermodynamical chemical equilibrium calculations (\S\,\ref{sec:chemeq}). 
}
\end{table}
 
\begin{figure}[hbtp!] 
\centering
\includegraphics[width=1\hsize]{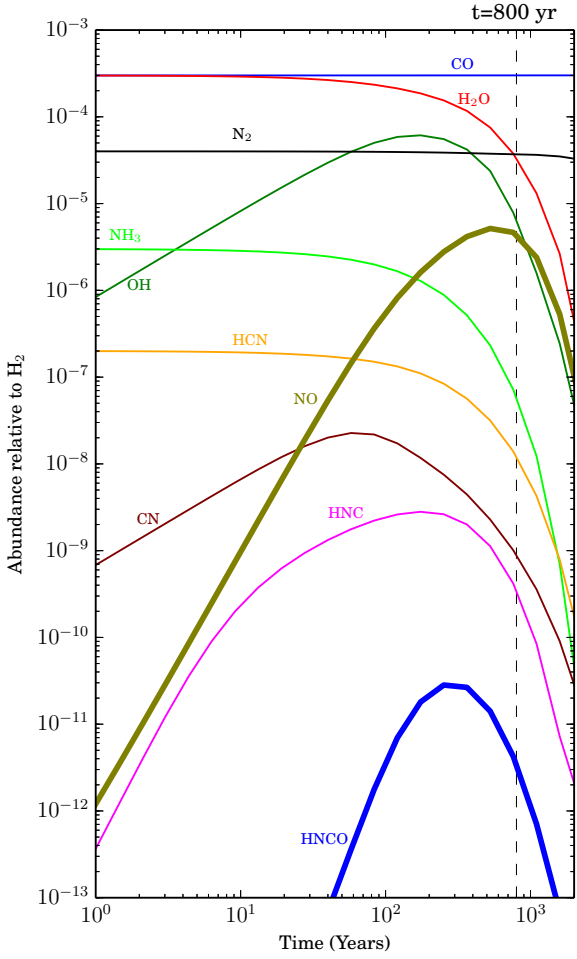}
\caption{Evolution with time of the molecular fractional abundances 
predicted by our chemical kinetics model for a rectangular slab of gas
(plane-parallel geometry) with physical properties similar to those of
the walls of the hollow lobes of the molecular outflow of \oh\ (model
input $b$, \S\,\ref{sec:chemmod}). The model presented is for a
representative gas cell in the middle of the lobe walls
($A_V$=1\,mag). The species represented are as in
Fig.\,\ref{fig:chemeq} and \ref{fig:chemkin}, except for HC$_3$N,
which has a predicted maximum abundance ($\approx$10$^{-19}$) well
below the lower limit of the $y$-axis. }
\label{fig:outkin}
\end{figure}

We have investigated whether deeper penetration of interstellar UV
radiation through the lobe walls of \oh, which are on average more
tenous than the central regions, can result in a significant
production of HNCO and HC$_3$N,  which could explain the observations.
The input physical model for the lobe walls (a rectangular gas slab,
model $b$) is described in Sect.\,\ref{sec:physmod}. The total
extinction through the lobe walls is $A_V$$\sim$2.3\,mag, taking their thickness, mean H$_2$ number density, and the
standard \dens/$A_v$ conversion into
account \citep{boh78}.  The variation with
time of the fractional abundances predicted by the model for a
representative cell in the middle of the lobe walls ($A_V$$\sim$1\,mag)
are shown in Fig.\,\ref{fig:outkin}.  

As in the central nebular regions (model $a$), the abundances of HNCO
(and probably HNCS) and HC$_3$N in the lobes never reach values
comparable to those observationally determined. Except for NO,
the abundances predicted in the lobes after $\sim$800\,yr, which is
the dynamical age of the molecular flow of \oh, are lower than those
expected in the slow central parts.  We find that NO reaches a
fractional abundance of $\sim$4$\times$10$^{-6}$ in
$\sim$800\,yr.  This value is comparable to the average NO abundance
deduced from the observations and to the value found in the slow
central component (model $a$, Fig.\,\ref{fig:chemkin}).

\subsection{Enhanced N elemental abundance}\label{sec:nenhance}

We have considered whether an overabundance of the elemental nitrogen
could result in fractional abundances of N-bearing compounds in better
agreement with the observations. Such an elemental N-enrichment could 
result from hot bottom burning (HBB) process for stars with masses
$\ga$3\,\msun\, and it has been proposed to explain the high abundance
of NO deduced in the molecular envelope of the yellow hypergiant (YHG)
star IRC$+$10420 \citep[][and references therein]{qui13}.

As a first step, we ran our TE model again, increasing
the elemental nitrogen abundance by a factor 40 -- a larger enrichment
factor is not expected \citep{boo93}.
In the inner layers of the envelope, our TE model shows that the N$_2$
fractional abundance increases proportionally, i.e.\, also by a factor
40. Other N-bearing molecules, such as NH$_3$, HCN, NO, HNCO, and
HC$_3$N, are less sensitive to the initial N-abundance, and they increase
their abundances by a smaller factor, $\sim$5-10. As expected, given
the very large discrepancy between TE model and data abundances, this
factor is insufficient to explain the observations in \oh.

As a second step, we ran our chemical kinetics
model (case $a$) again but modified the initial abundances of relevant
parent species (N$_2$, NH$_3$, and HCN) according to the TE
predictions: N$_2$ in increased by a factor 40 and NH$_3$ and HCN by a
factor 7 relative to the values in Table \ref{tab:abchemkin}. Our TE
calculations also show that these are the N-containing molecules that are most
sensitive to the initial abundance of N. From our chemical kinetics model, we
find that neither HNCO (and presumably HNCS) nor HC$_3$N experiment a
significant variation, maximum abundance of HNCO is
$\sim$8$\times$10$^{-10}$ and $\sim$8$\times$10$^{-13}$ for HC$_3$N,
which are still low compared to the values derived from the
observations. As NO concerns, we find a model peak abundance of
$\sim$3$\times$10$^{-5}$ in the outer layers of the slow central
component, which would be in excess of the value derived from the
observations in \oh. We therefore believe that a resonable enhancement of
the elemental abundance of nitrogen, if it exists, would not reproduce the abundances of the N-molecules discussed by us
satisfactorily. On
the one hand, HNCO, HNCS, and HC$_3$N are underestimated, and on the
other, NO (and maybe others, such as NH$_3$) would be
significantly overestimated.

\section{Discussion} \label{sec:concl}
Except maybe for NO, the relatively large abundances of
the N-molecules detected in this work cannot be explained as thermodynamical 
chemical equilibrium or photodissociation products in the outflow of \oh,
even in the case of HNCO after considering IR pumping effects (\S\,\ref{sec:nolte}, Appendix\,\ref{sec:app}). 
In principle, the inability of the model to reproduce the observed
abundances of HNCO, HNCS, and HC$_3$N could be attributed to the
simplicity of the physico-chemical scenario adopted, for example, to
the lack of certain molecule formation routes (e.g.\ involving dust
grains).
However, if these were the main reasons for the model-data
discrepancies in \oh\ (and provided that these unknown chemical routes
prove to be an efficient way of producing HNCO, HNCS, and HC$_3$N in O-rich
environments, which remains to be seen), then these molecules should
be present with comparable abundances in other O-rich envelopes of
similar characteristics.
 
These molecules have been searched for but not detected by our team
(and others) in two of the strongest molecular emitters and best
studied O-rich envelopes: the AGB star IK\,Tau (Velilla et al., in
prep) and the red supergiant VY\,CMa (Quintana-Lacaci et al., in
prep).
In these objects, NO is detected with a fractional abundance of
$X$(NO)$\approx$10$^{-7}$, in agreement with the model predictions,
but only upper limits are obtained for the rest of the N molecules
discussed here, $\la$10$^{-9}$-$\la$10$^{-10}$. The upper limits
estimated for IK\,Tau and VY\,CMa are consistent with the low
abundances predicted by the models, which suggests that the
physico-chemical scenario used by us is an acceptable representation
of a standard O-rich AGB CSE.
We recall that the notable chemical differences between \oh\ and other
O-rich AGB CSE are widely recognized and are not restricted to the
N-molecules detected here but affect most of the species identified in
this object, including C-, N-, and S-containing ones such as
HCN, H$_2$CO, H$_2$S, SO, SO$_2$,  which are undetected or
detected with much lower abundances in `normal' O-rich AGB CSE, as
pointed out by previous works (\S\,\ref{sec:intro}).

In principle, there is no reason to suspect a particularly intense
interstellar UV radiation field or peculiar dust properties or content
that could (or not) account for the unique, rich chemistry of
\oh\ compared to its O-rich relatives. The main difference between
\oh\ and `normal' O-rich AGB CSEs
is the presence of a fast ($\sim$400\,\kms) accelerated outflow in the
former. Given the formation history of such an outflow, possibly as
the result of a sudden jet+`AGB CSE' interaction $\sim$800\,yr ago
(\S\,\ref{sec:intro}), fast shocks have probably played a major role
not only in the physics but also in the chemistry of \oh. Molecules
are expected to be initially destroyed by the high-speed axial shocks
produced in the jet+'AGB CSE' interaction \citep[e.g.][]{neu90}.
At present, the shocked material has cooled down sufficiently to allow
molecule reformation, which probably happened very quickly, in
less than $\sim$150\,yr, 
under non-equilibrium conditions. Moreover, additional atoms
(Si, S, etc.)  may have been extracted by the shocks from the dust
grains and released into the gas phase \citep{morr87,lin92}, altering
the proportions of the different elements available for molecule
regeneration in the post-shocked gas. Both non-equlibrium conditions
and non-standard elemental proportions in the post-shocked gas are
crucial factors determining the abundances of the second-generation
molecules in \oh.

Shocks could also have been decisive in defining the chemistry of the
slow central parts of the envelope around \oh. As already pointed by,
for example, \cite{lin92}, the maximum expansion velocities measured towards
the central nebular regions, \vexp$\sim$35\,\kms, are higher than for
normal OH/IR stars, which indicates that some acceleration is likely
to have occurred. In fact, it may be possible for shocks developed in
the jet+‘AGB CSE’ interaction to move sideways and backward (with
moderate velocities lower than those reached along the jet axis)
compressing the gas in the equatorial plane and shaping the central,
torus-like structure of \oh.  Although these moderate-velocity
($\la$40\,\kms) shocks are not expected to destroy molecules (at least
not fully),
the compression and heating of these equatorial regions would result
in a profound chemical mutation with respect to normal unperturbed AGB
CSEs (for example, activating certain endothermic reactions, or
exothermic reactions with barriers, otherwise forbidden).

\section{Summary and conclusions} \label{sec:summary}
We have reported the first detection of the N-bearing molecules \nmols\ in
the circumstellar envelope of the O-rich evolved star \oh\ based on
single-dish observations with the \iram\ telescope.  HNCO and HNCS are
first detections in circumstellar envelopes; HC$_3$N is a first
detection in an O-rich environment; NO is a first detection in a CSE
around a low-to-intermediate-mass, evolved star.
From the observed profiles, we deduce the presence of these species in
the slow central parts of the nebula, as well as at the base of the
fast bipolar lobes.

The intense, low-velocity components of the HNCO $K_{\rm a}$=0, HNCS,
and HC$_3$N profiles have similar widths (FWHM$\sim$20-30\,\kms) and
velocity peaks (\vlsr$\sim$28-29\,\kms). Previous SO emission mapping
\citep{san00} shows the presence of an equatorial expanding disk or torus
around the central star that produces double-peaked (at \vlsr=28 and
40\,\kms) spectral profiles in many SO transitions (and also in other
molecules). The coincident peak velocity of HNCO, HNCS, and HC$_3$N
transitions with the blue peak of the disk/torus feature suggests that
part of the low-velocity emission from these molecules may arise at
this equatorial structure.  The HNCO $K_{\rm a}$=1 transitions are
narrower, with FWHM$\sim$13\,\kms, and may arise in regions closer in
to the central source.

The profiles of the NO lines are broader (FWHM$\sim$40-50\,\kms) and
are centred on somewhat redder velocities \vlsr$\sim$40\,\kms. As
explained in Sect.\,\ref{sec:obsres}, this is not only due to the
hyperfine structure of the NO transitions, but it also indicates that a
significant part of the NO emission is produced in regions with high
expansion velocities; in particular, the contribution to the emission
from clump I4 at the base of the southern lobe is notable. 
Broad profiles 
(FWHM$\sim$40-90\,\kms) are also found for mm-wave transitions of
HCO$^+$ \citep{san00} and other molecular ions recently discovered by
us in \oh\ (S\'anchez Contreras et al.\ 2014, in prep.). This suggests
a similar spatial distribution of these species with enhanced
abundances in the high-velocity gas relative to the low-velocity
nebular component at the centre.

We derived typical rotational
temperatures of $\sim$15-30\,K, in agreement with previous estimates
of the kinetic temperature in the CO flow \citep{alc01}. Non-LTE
effects are expected to be moderate, given the relatively high
densities of the dominant emitting regions ($\ga$10$^5$\,\cm3).
Nevertheless, in the case of HC$_3$N, moderate sub-thermal excitation
is possible in the most tenuous parts of the outflow, and somewhat
higher temperatures of $\sim$45 to 55\,K cannot be ruled out.
Adopting a characteristic size of the emitting nebula of
$\sim$4\arcsec$\times$12\arcsec, we obtained column densities of
\ntot(\trecem)$\sim$3$\times$10$^{17}$\,\cm2,
\ntot(HNCO)$\sim$6$\times$10$^{14}$\,\cm2,
\ntot(HNCS)$\sim$7$\times$10$^{13}$\,\cm2,
\ntot(HC$_3$N)$\sim$3$\times$10$^{13}$\,\cm2, and
\ntot(NO)$\sim$9$\times$10$^{15}$\,\cm2.

The beam-averaged fractional abundances in \oh\ obtained are (in
decreasing order) $X$(NO)$\sim$[1-2]$\times$10$^{-6}$,
$X$(HNCO)$\sim$[0.8-1]$\times$10$^{-7}$,
$X$(HNCS)=[0.9-1]$\times$10$^{-8}$, and
$X$(HC$_3$N)=[5-7]$\times$10$^{-9}$. We note the large abundance of
NO, which is comparable to that of, e.g., SO and SO$_2$ (already known
to be dominant in \oh). Our measurement implies that NO is one of the
most abundant N-containing molecule in this object. Also
remarkable is the relatively large abundance of HNCO, closely following
that of major carriers of carbon in \oh, apart from CO and \trecem,
such as HCN, H$_2$CO, and CS, and comparable to and even larger than
that of HNC and HCO$^+$ \citep{morr87,lin92,san97,san00,vel13,san14}.

We modelled thermodynamical equilibrium and non-equilibrium
kinetically driven chemistry to investigate the production of
HNCO, HC$_3$N, and NO in \oh. HNCS cannot be modelled because of the lack
of thermochemical parameters and reactions rates.  We modelled the slow central component and the lobe walls
separately.

We found that none of the molecules HNCO, HC$_3$N, or NO are formed in
significant amounts in the vicinity of the AGB star (up to
$\sim$4\rs), where thermodynamical equilibrium conditions prevail
(Fig.\,\ref{fig:chemeq}). In these regions, the vast majority of N
atoms are locked in N$_2$, followed by NO.

In the intermediate/outer layers of the slow central component of the
envelope (from $\sim$10$^{15}$ to $\sim$10$^{17}$ \,\cm), the model
fails to reproduce the large abundances observed in \oh, except for NO
(Fig.\,\ref{fig:chemkin}). The model-data discrepancies cannot be explained 
by a reasonable enhancement of the elemental nitrogen abundance (as
a result of HBB processes).

In the lobes, our chemistry model indicates that the only molecule
that reaches fractional abundances comparable to the values
observationally determined is NO.
For HNCO (and probably HNCS) and
HC$_3$N, the model abundances in the lobes are more than five orders of
magnitude lower than the observed average values. 

Based on this and previous works, the rich chemistry of \oh, which is
unparalleled amongst AGB and post-AGB envelopes, is corroborated. New
detection of \nmols\ add to the list of N-bearing molecules present in
its molecular outflow with high abundances. This could be the best
example of a shocked environment around an evolved star, and \oh\ therefore stands out as a 
reference target for studying non-equilibrium, shock-induced chemical processes in oxygen-rich environments.

\begin{acknowledgements}
We acknowledge the \iram\, staff for the support and help kindly
given during the observations presented in this article, in particular
to M.\ Gonz\'alez.  We also acknowledge the help provided by
J.\ R.\ Pardo during the different observational runs in which he took part.  
This work was done at the Astrophysics
Department of the Centro de Astrobiolog\'{i}a (CAB-INTA/CSIC) and the 
Molecular Astrophysics Department of the Instituto de Ciencias de
Materiales de Madrid (ICMM-CSIC).
We acknowledge the
Spanish MICINN/MINECO for funding support through grants
AYA2009-07304, AYA2012-32032, and the ASTROMOL Consolider project
CSD2009-00038.  L.V. acknowledges the Spanish MINECO for funding
support through FPI2012 short stay programme (ref. EEBB-I-13-06211) and
the Laboratoire D'Astrophysique de Bordeaux (LAB-CNRS) for hosting
this stay under the supervision of Dr.\ Marcelino Ag\'undez.
L.V.\ also acknowledges the support of the Universidad Complutense de
Madrid through the PhD programme.  M.A.\ acknowledges the support from
the European Research Council (ERC Grant 209622: E$_3$ ARTHS). This
research made use of the IRAM GILDAS software, the
JPL Molecular Spectroscopy catalogue,
the Cologne Database for Molecular Spectroscopy, the SIMBAD database,
operated at the CDS, Strasbourg, France, NASA's Astrophysics Data
System, and Aladin.
\end{acknowledgements}


\clearpage
\newpage
\appendixpage
\appendix
\section{Non-LTE effects on the population diagrams of HNCO, HNCS, HC$_3$N, and NO.}
\label{sec:app}

The population diagram method 
is a common tool for deriving physical
gas conditions from molecular line observations \citep[see][for a classic
  reference]{gol99}. It relies on two major assumptions: $i$)
optically thin emission and $ii$) LTE conditions. The latter
assumption (LTE) implies that the populations of all levels are
described by the Boltzmann distribution with a unique rotational
temperature, \trot\footnote{Since we are dealing with pure rotational
  transitions in the ground vibrational state, we use the term
  rotational temperature, \trot, instead of the more general
  designation as excitation temperature, \tex.}, which is equal to the
kinetic temperature of the gas (\tkin=\trot). In this case, for a given
molecule and a series of transitions $u \raw l$, a plot of the natural
logarithm of the upper state column density per statistical weight
($N_u/g_u$) versus the energy above the ground (\eu), the so-called 
population diagram, will yield a straight line with a slope 1/\trot.

\cite{gol99} have numerically investigated how the optical depth and
deviations from LTE affect the temperature and column density derived
using the population diagram technique for two molecules: HC$_3$N and
CH$_3$OH. In this appendix, we perform a similar analysis for 
\nmols,\ but in this case, we focus on non-LTE excitation effects since the lines detected in \oh\ are
optically thin. (For HC$_3$N we compared our results, analysing
both optical depth and non-LTE excitation effects, with those by
\cite{gol99} and found an excellent agreement.)

In low-density regions\footnote{Both collisional and radiative
  processes can excite molecules, and for each transition a critical
  density can be defined where the two processes are equally important
  ($n_{\rm crit}$=$A_{ul}$/$\gamma_{ul}$). At lower densities
  radiation dominates, while at higher densities collisions drive the
  level populations to thermodynamic equilibrium.}, LTE may not be a
valid approximation, and therefore, the level populations, which may
not longer be described well by the Boltzmann distribution, have to be
numerically computed by solving the statistical equilibrium
and radiation transport equations. 
The excitation analysis presented in this Appendix
has been done using MADEX \citep[Molecular and Atomic Database
  and EXcitation code,][]{madex}. This is a code that solves the
molecular excitation 
(including collisional and radiative excitation
mechanisms)
and radiation transfer
problem under the large velocity gradient (LVG) formalism.
It contains up-to-date spectroscopy (rest frequencies, level energies,
line strength/Einstein coefficients, etc.) and collisional rates
available from the literature for more than 5000 different molecular
and atomic entries including isotopologues and vibrationally excited states.
MADEX computes molecule-H$_2$ collisional rates from those
  available in the literature by adopting other collision partners, such
  as He or para-$H_2$.
MADEX also evaluates the partition function of the molecule (using 
a large enough number of levels to obtain accurate values of the partition function
even at high temperatures) and predicts the emergent spectra. 

\begin{figure*}[hbt!] 
\centering
\includegraphics[width=0.45\hsize]{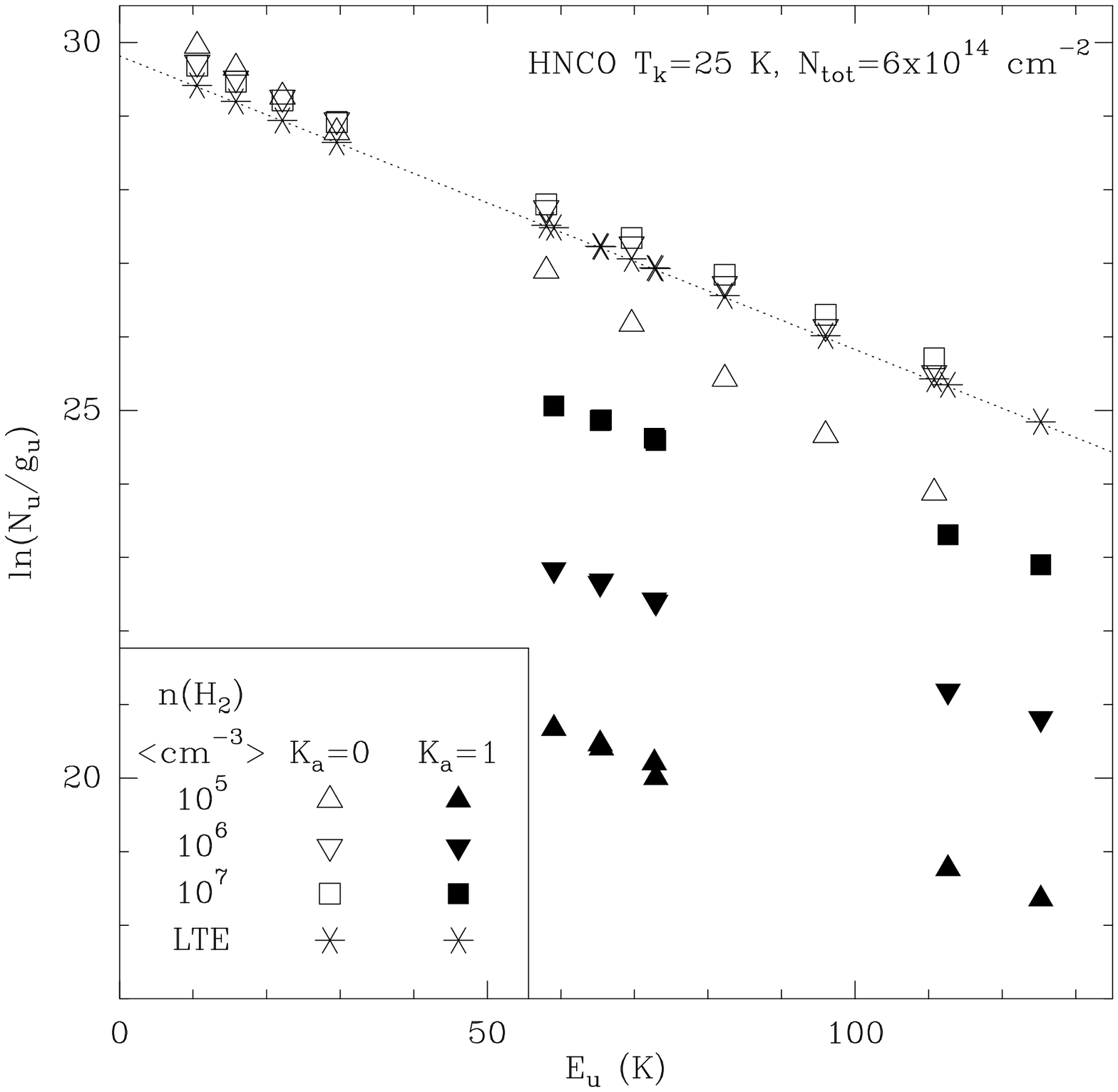}\hspace{0.75cm}
\includegraphics[width=0.45\hsize]{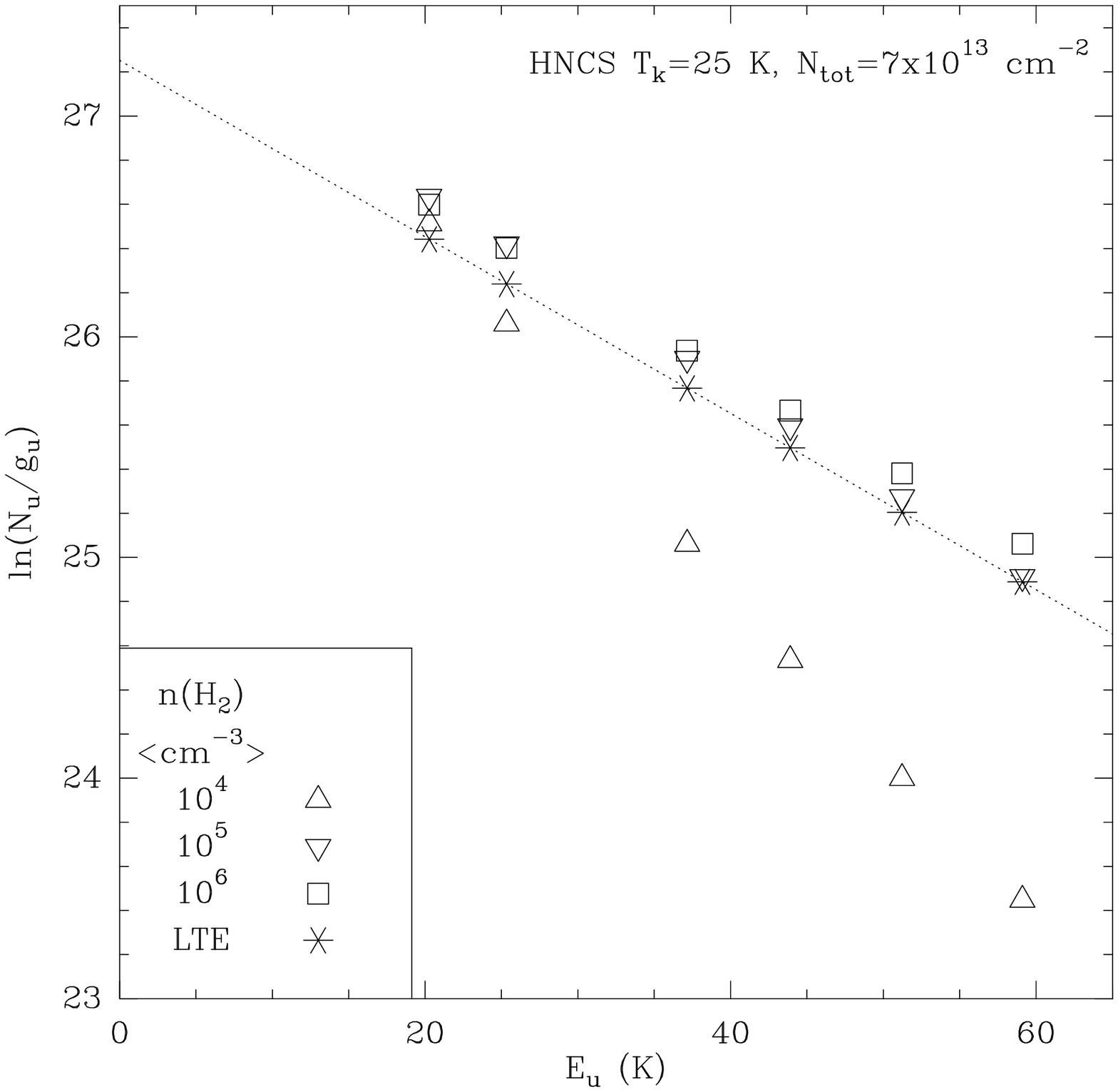} \\
\vspace{0.75cm}
\includegraphics[width=0.45\hsize]{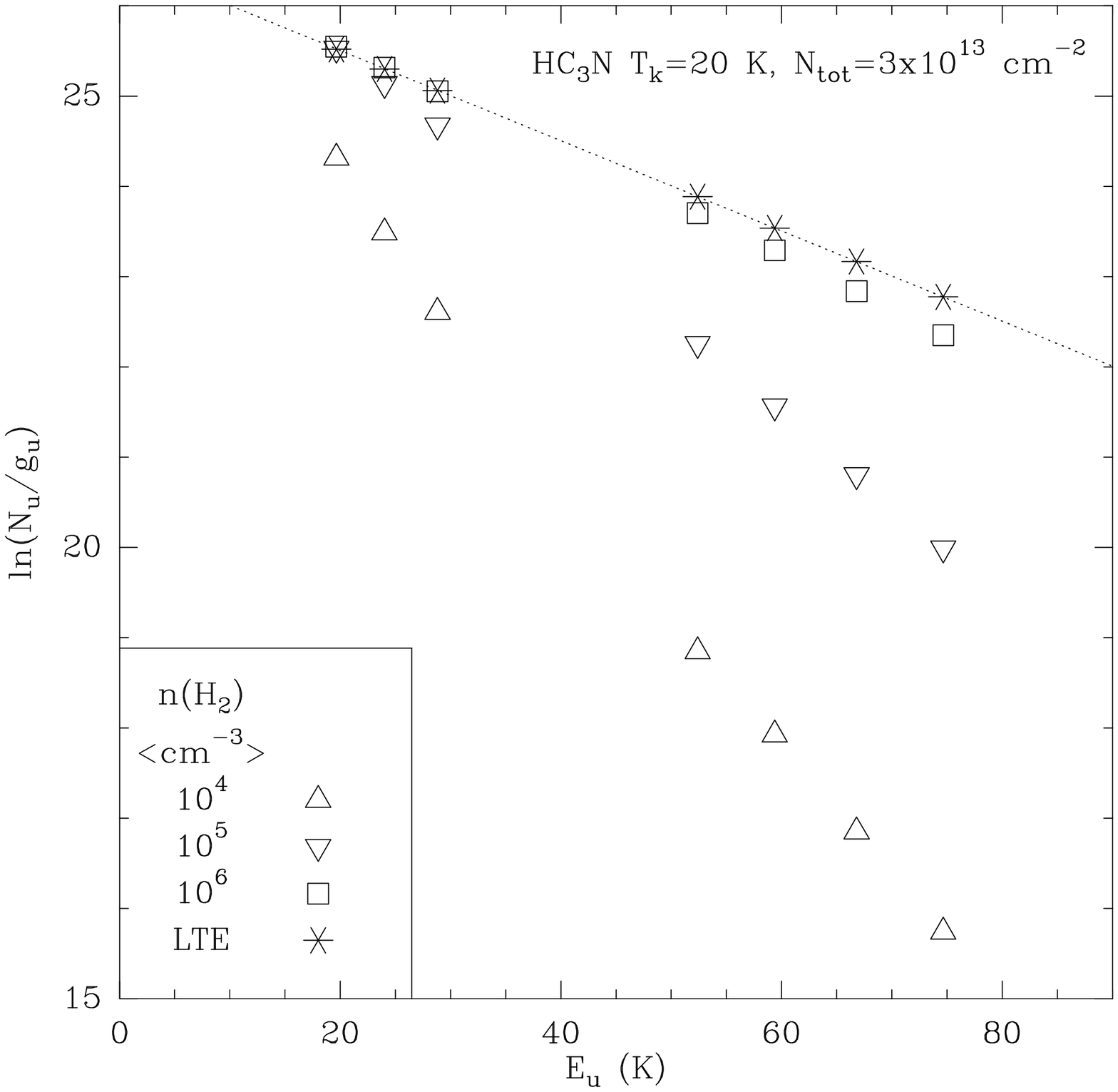}\hspace{0.75cm}
\includegraphics[width=0.45\hsize]{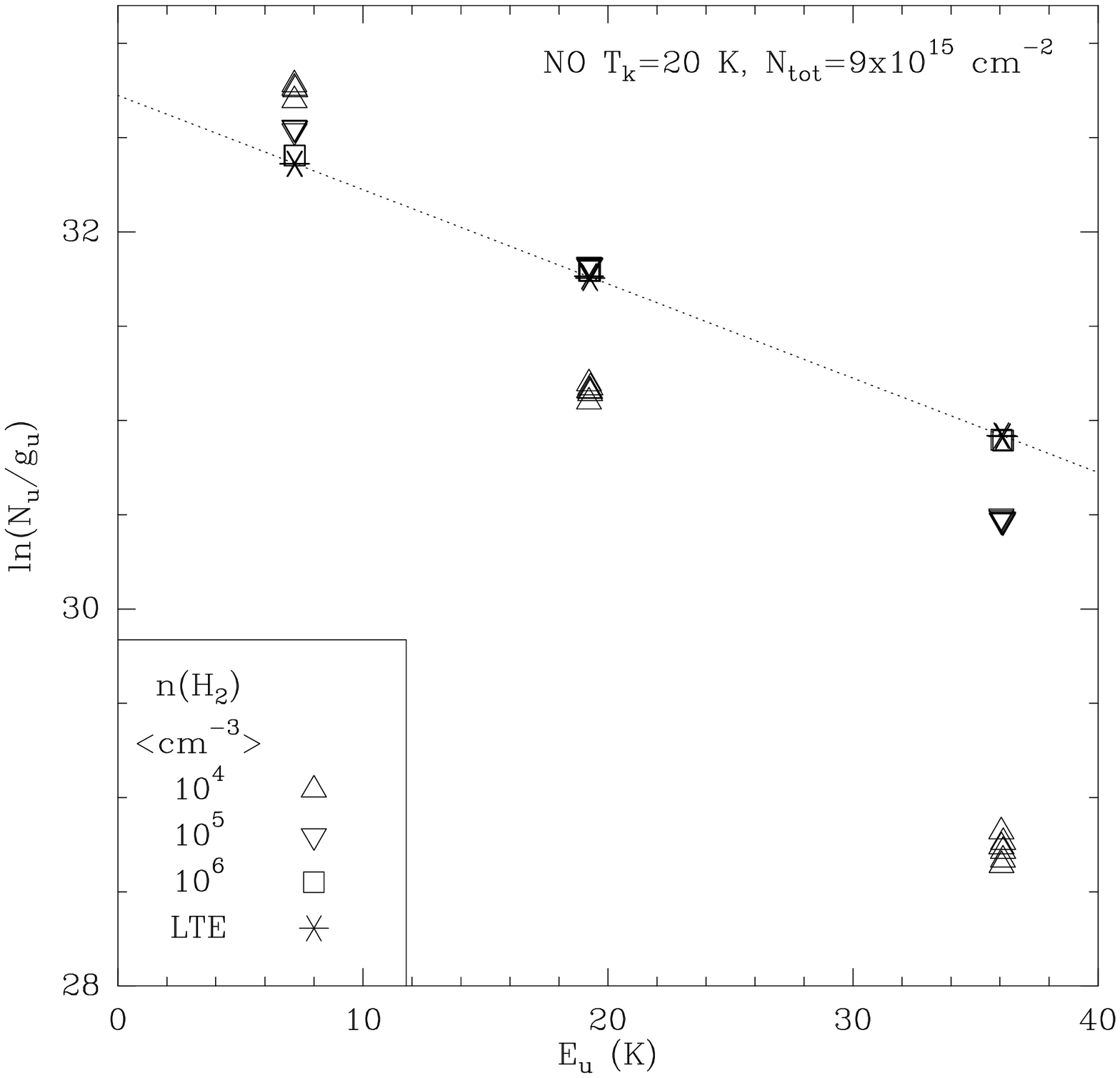} 
\caption{Population diagrams of \nmols\ for a range of molecular
  hydrogen densities (\dens, bottom left corner) and a given input
  value for the kinetic temperature and total column density
  (\tkin\ and \ntot; top right corner); the dotted line connects the
  LTE points. Only the transitions detected in this work are
  represented in this diagram (Table\,\ref{tab:measures}). For HNCO,
  the transitions within the $K_{\mathrm{a}}$=0 and $K_{\mathrm{a}}$=1
  ladders are indicated by empty and filled symbols, respectively. For
  NO, we plot the three hyperfine components with the highest values
  of the Einstein coefficient (A$_{\rm ul}$) for each of the $\Pi^-$
  and $\Pi^+$ doublets at 250 and 350\, GHz (i.e.\ at \eu=19.2 and
  36.1\,K), the three hyperfine components with the largest A$_{\rm ul}$
  of the $\Pi^+$ band at 150\, GHz (i.e.\ at \eu=7.2\,K) and the $\Pi^-$(3/2,3/2)-(1/2,1/2) line, which is
  spectrally isolated in our data, at 150\,GHz (\eu=7.2\,K). These
  calculations have been done with a LVG radiative
  transfer code (MADEX) -- see text in this appendix.}
\label{fig:ap1}
\end{figure*}

To examine what happens 
if some or all of the transitions of \nmols\ are not thermalized, we computed the level populations for a range of
densities, \dens, and a given input value of \tkin\ and \ntot\ for
each molecule. The adopted values of \tkin\ and \ntot\ are similar to those
obtained from the LTE analysis (\S\,\ref{sec:popdiag}). We assumed
a linear velocity gradient d$(\ln V)$/d$(\ln r)$=1 and typical line
widths of FWHM=20-40\,\kms, as observed towards
\oh\ (\S\,\ref{sec:intro} and \ref{sec:obsres}). The resulting
population diagrams based on our non-LTE excitation calculations, including the LTE theoretical points, are shown in
Fig.\,\ref{fig:ap1}.

For HNCO we used collisional rates from Green (1986)\footnote{\tt
  http://data.giss.nasa.gov/mcrates/}, computed for the lowest 164
levels at temperatures from 30 to 350\,K. For the non-LTE analysis, we considered rotational levels in the ground vibrational state
($v$=0) up to $J_{\rm max}$=18 for both the $K_{\mathrm{a}}$=0
and $K_{\mathrm{a}}$=1 ladder transitions (and also up to
$K_{\mathrm{a}}$=4, which has been included in the calculations). This implies a maximum upper state energy of 830\,K.  As shown in
Fig.\,\ref{fig:ap1} (top left box), for the lowest density models
(\dens=10$^5$\cm3), there is a notable separation between the
$K_{\mathrm{a}}$=0 and $K_{\mathrm{a}}$=1 ladders, which follow two
straight lines with different slopes.  The slope of the $K_{\mathrm{a}}$=0 ladder
implies a rotational temperature of \trot$\sim$16\,K, which is lower than the real input
kinetic temperature (sub-thermal excitation), while the slope of the $K_{\mathrm{a}}$=1 ladder
implies a rotational temperature of \trot$\sim$30\,K.  The $y$-offset between
the two $K_{\mathrm{a}}$ ladders decreases as the density increases,
until they merge in a single line at densities \dens$\ga$10$^8$\cm3, with \trot=\tkin.

Collisional rates are not available for HNCS, therefore we
used those of HNCO after applying the standard reduced mass
correction.
As for HNCO, we included rotational levels of up to $J_{\rm
  max}$=18 (within the $K_{\mathrm{a}}$=0, 1, and 2 ladders), which implies maximum upper state energies of
\eu$\sim$355\,K. As expected, deviations from LTE affect the population
diagram of HNCS similarly to HNCO. In the case of HNCS, only the
$K_{\mathrm{a}}$=0 ladder has been plotted since these are the only ones detected.
For the lowest density model, \dens=10$^4$\,\cm3, the inferred values
of \trot$\sim$10-12\,K deviate significantly from the real input
values (\tkin=25\,K).

Collisional rates for HC$_3$N are from \cite{wer07}, which include 51
levels and are computed for temperatures between 5\,K and 100\,K.
Collisional rates for NO are adopted from \cite{klo08}, computed for
98 rotational levels and temperatures between 10 and 500\,K. The
highest rotational levels included in our non-LTE excitation
calculations for HC$_3$N an NO are $J_{\rm max}$=30 (\eu$^{\rm
  max}$$\sim$200\,K) and $J_{\rm max}$=13/2 (\eu$^{\rm max}$$\sim$300\,K),
respectively. The resulting non-LTE models for HC$_3$N and NO
(Fig.\,\ref{fig:ap1}), as for the other species, clearly show steeper
slopes (\trot$<$\tkin) for the lowest density models. Another effect
of non-LTE excitation is that the different level populations cannot
be described by a unique rotational temperature, which translates into a
different slope for the low-$J$ and high-$J$ levels in the population
diagram. This effect is most notable for HC$_3$N: the temperatures
implied are \trot$\sim$5\,K for the three lowest-$J$ levels and
\trot$\sim$7\,K for the four highest-$J$ levels. The HC$_3$N column
densities that one would derive using low-$J$ or high-$J$ levels are
also different, in particular,
\ntot(low-$J$)=3.3$\times$10$^{13}$\,\cm3
$>$\ntot=3.0$\times$10$^{13}$\,\cm3 $>$
\ntot(high-$J$)=7.2$\times$10$^{12}$\,\cm3. In the lowest density
model of NO, the implied \trot\ ($<$\tkin=26\,K) does not vary
appreciably across the different levels (less than 6\%).

For all molecules, at the highest densities considered, the
populations of the levels considered are thermalized or very close to
thermalization, and the non-LTE and LTE predictions converge. In both
cases, the data points in the population diagram can be satisfactorily
fit by a straight line with \trot=\tkin. 
We consider of interest to provide a summary with the range of critical
densities for the transitions analysed in this work. These are for a
representative temperature of 25\,K, which is common in the dominant
emitting components of AGB CSEs and PPNs \citep{buj01}. For HNCO the
critical densities are in the range of
\nc=[4$\times$10$^5$-1$\times$10$^7$]\,cm$^{-3}$ for the $K_{\mathrm{a}}$=0
transitions, and \nc=[1$\times$10$^6$-1$\times$10$^7$]\,cm$^{-3}$ for
the $K_{\mathrm{a}}$=1 transitions. For HNCS, HC$_3$N, and NO, the critical densities
are \nc=[5$\times$10$^5$-2$\times$10$^6$]\,cm$^{-3}$,
\nc=[1$\times$10$^6$-3$\times$10$^6$]\,cm$^{-3}$, and
\nc=[2$\times$10$^4$-2$\times$10$^6$], respectively. In all these
cases, the lowest value of \nc\ in the ranges given above corresponds
to the lowest-$J$ and/or lowest-E$_{\mathrm{u}}$ transition.

Finally, we briefly comment on the possible effect of
the pumping of the rotational levels of HNCO by the infrared (IR)
photons emitted by the star and the central dust region \citep[see e.g.][]{kua96}.
This IR emission would cause the radiative pumping of levels inside the ground vibrational state and the vibrationally excited states of HNCO.
The latter would eventually produce a fluorescence effect between the ground state and the six vibrational modes of HNCO.
The radiative pumping inside the rotational levels of the ground state alone was found to be dominant in the HNCO excitation in Sgr B2 \citep{chu86}.
In this region, the typical densities are relatively low, n(H$_2$)$\sim$10$^4$\,cm$^{-3}$, but the central source is optically thick at 100\,\microns\ and extended.
The IR pumping through vibrationally excited states, has also noticeable effects in the emerging intensities of some rotational lines for certain molecules like H$_2$O IRC+10216 \citep{marce06}. 
Recently, it has been pointed out that the time variability of the IR pumping can produce intensity variations in the high rotational lines of abundant molecules in the envelopes of Mira-type stars, such as CCH towards IRC+10216 \citep{cer14}.
Thus, to summarize, transitions with high upper level energies or those that are severely underexcited by collisions could be more sensitive to IR pumping. 
When HNCO levels were populated mainly by IR radiation, the column density implied by the
rotational diagram method could be different than the true value; 
the rotational temperature would also be different than the gas kinetic temperature (but see next paragraph).

Our previous discussion of the excitation state of the observed lines of HNCO shows that they are practically thermalized to the temperatures expected in the studied nebula.
The population diagram of HNCO (Fig.\,\ref{fig:rotdiags}) covers a wide range of upper energies, and it clearly shows a single slope for both the K$_a$=0 and K$_a$=1 transitions.
This should imply that the line intensities are described well by one rotational temperaure, so that the column density and the derived abundance should be close to the true value.
Also, the range of derived kinetic temperatures is consistent with previous estimations of the rotational temperatures and the kinetic temperature in the CO flow \citep[see e.g.][]{gui86,morr87,alc01}.
Finally, the bulk of the emission of HNCO (\S\,\ref{sec:obsres}) arises from the central dense region of the CSE (I3) and the base of the southern lobe (I4).
These regions are characterized by densities of n(H$_2$)$\geq$10$^6$-10$^7$\,cm$^{-3}$ in the central region and n(H$_2$)$\geq$10$^5$\,cm$^{-3}$\,in the lobes \citep{alc01,buj02}.
This leads to densities close or higher than the critical densities for HNCO K$_a$=0 and K$_a$=1 transitions. 
Therefore, the effects of vibrational cascades should be minor in our case, but we cannot rule out uncertainties in the abundance of HNCO less than a factor 2-5.
On the other hand, we have seen that the excitation via the various vibrational states is extremely complex for these relatively heavy molecules. 
A detailed study of these intricate phenomena is obviously beyond the scope of this paper.

\section{Comparison with other astrophysical environments}
\label{sec:chem}

\subsection{HNCO}\label{sec:hnco}
Isocyanic acid has been detected in different environments with a
variety of physical conditions, including SgrB2 \citep{sny72}, the
Taurus Molecular Cloud TMC-1 \citep{bro81}, external galaxies
\citep{ngu91}, the shocked-outflow of the young stellar object L1157
\citep{rod2010}, hot cores \citep{chu86,mar08}, 
translucent clouds \citep{tur99}, etc. 
Nevertheless, prior to this work, HNCO was not 
detected in any CSE around evolved stars, either oxygen or
carbon rich.

Formation of HNCO through gas-phase and grain surface chemistry from
different chemical pathways has been studied by several authors
\citep{igl77,tur99,zic00,gar08,mar2010,qua10}. This molecule was first 
proposed as a high density tracer \citep{jac84} and, more recently, a
shock tracer \citep[][and references therein]{rod2010}.

The fractional abundance of HNCO varies somewhat in different sources,
typically within the range $\approx$10$^{-10}$ -10$^{-9}$. The highest
fractional abundance (relative to H$_2$) of HNCO has been measured
towards the shocked region of the L1157 outflow, L1157-B2, where
$X$(HNCO)$\sim$9.6$\times\,10^{-8}$
\citep[][]{rod2010}. Interestingly, this value of the abundance is
comparable to what is estimated towards \oh.

\subsection{HNCS}\label{sec:hncs}
Isothiocyanic acid was first detected in SgrB2 \citep{fre79} and has
been recently observed in TMC-1 \citep{ada10}. Detection of HNCS in a
circumstellar envelope has not been reported previously to this work.

The formation of HNCS in the cold core TMC-1 and the hot core in SgrB2
has been studied theoretically through gas-phase, ion-molecule
chemistry and grain surface reactions \citep[][]{ada10}.
Prior to this work, the highest abundance
of HNCS had been found towards SgrB2 and TMC-1, with a value of
$\approx\,10^{-11}$.  In \oh\, we derive a fractional abundance of
HNCS that is about 1000 times higher.

\subsection{HC$_3$N}\label{sec:hc3n}
Cyanoacetylene is detected in assorted environments, including SgrB2
\citep{tur71}, H\,II regions, dark clouds \citep{morr76}, the Orion
molecular cloud \citep{gol82}, the Perseus globules \citep{bac86}, in
the Taurus molecular clouds \citep{cer84}, protoplanetary disks
\citep{cha12}, etc. HC$_3$N is also detected in several C-rich CSEs,
including the well-known AGB star IRC$+$10216 and the protoplanetary
nebula CRL\,618 \citep[e.g.][]{morr75,buj94,aud94,cer00,par04}; however,
this molecule has not been identified before in an O-rich CSE.

HC$_3$N is considered to be a high density tracer \citep{morr76} and
in CSEs, particularly in IRC$+$10216, it is distributed in a hollow
spherical shell around the central star, which is a major product of
photodissociation in the outer parts of the envelope
\citep[][]{aud94}. The observed abundance in IRC$+$10216 is
$X$(HC$_3$N)$\sim$1$\times$10$^{-6}$ agrees with theoretical
predictions in C-rich CSEs \citep{marce10}. We infer
$X$(HC$_3$N)$\sim$7$\times\,10^{-9}$ towards \oh. This value is much
lower than in IRC$+$10216 by virtue of the O-rich vs.\ C-rich
nature of both sources.

\subsection{NO}\label{sec:no}
Nitric oxide has been previously detected in several astrophysical
environments, including molecular clouds \citep{ger92}, SgrB2
\citep{hal01}, and pre-protostellar cores \citep{aky07}. Recently, NO
has been detected for the first time in the CSE around the yellow
hypergiant (YHG) IRC$+$10420, which is a massive ($\sim$50\,\msun) evolved star
with a N-rich chemistry \citep{qui13}, with a fractional abundance of
$X$(NO)$\approx$10$^{-5}$. 

Chemical models presented by \cite{qui13} in IRC$+$10420 predict
that indeed NO forms very efficiently by photochemistry (mainly
through the reaction N + OH $\longrightarrow$ NO + H) in the outer
circumstellar layers, where it reaches a maximum abundance of
$\approx$10$^{-6}$.
In the case of IRC$+$10240, nitrogen enrichment
due to hot bottom burning has been proposed to explain the NO
abundance observed, which is larger by a factor 10 than predicted 
by these models adopting the solar nitrogen abundance.

\section{Ionization and dissociation sources in our chemistry models}\label{sec:app_uv}

The sources of ionization and dissociation adopted in our model are cosmic rays and the
interstellar ultraviolet radiation field. The cosmic-ray ionization
rate adopted is 1.2$\times$10$^{-17}$\,\invsec\ \citep{dal06}. 
The intensity of the UV
field assumed is the Draine field (X) or Go=1.7 in units of the Habing
field (Go=1.6$\times$10$^{−3}$\,\intensity; Habing, 1968,
Draine \& Salpeter, 1978). The ISM UV field illuminates the
nebula externally.

We also evaluated two possible additional sources of internal
ionization and dissociation in \oh: 1) the UV radiation by the A0 main sequence
companion of the primary AGB star (QX Pup) at the nucleus of the
nebula, and 2) the high-energy radiation generated by the cooling of
hot gas behind the fast shock fronts ahead of the lobe tips
(Sect.\,\ref{sec:intro}). In both cases, the effect on the chemistry is
not expected to be predominant and has not been considered in our model.

First, the UV radiation field emitted by the $\sim$10~000\,K
main-sequence companion cannot penetrate
very deep through the dense dusty wind of the AGB mass-losing star
except, maybe, along the direction of the lobes owing to the
lower extinction by dust along the outflow cavities \citep[of the
  order of $A_V$$\sim$1\,mag;][]{san04}.
However, there is observational evidence against the presence of a
substantial amount of ionized or atomic gas in the stellar vicinity, hence against the existence of an intense stellar UV field
that could have a noticeable effect on the chemistry in the inner
regions of the lobes: ($i$) the lack of H$_\alpha$ emission (or any
other recombination or forbidden lines in the optical) from the
nucleus of \oh\ rules out an emergent ionized region around the star
\citep{coh85,rei87,san00opt,san04}; and ($ii$) the lack of low-excitation
atomic emission lines in the far-infrared, e.g.\,[\ion{O}{I}] emission
at 63.2 and 145.5$\mu$m (unpublished \hso\ archive data) indicates the
absence of a photodissociation region (PDR) in the nebula centre.
Moreover, considering the temperature and luminosity of the warm
companion \citep[$\sim$10~000K and $\sim$5-30\ls;][]{san04}, it can be
demonstrated that the stellar UV flux that reaches to a point
located at the inner edges of the lobes ($\sim$4\arcsec\ from the star)
with a visual extinction of A$_V$=1.0\,mag,
would be smaller than, or at most comparable to, the ISM
UV field. Therefore, including the stellar UV radiation as a source of
internal illumination of the lobe walls in the model will yield very
similar results (and even smaller molecular abundances for `daughter'
species) to those obtained assuming external illumination by the ISM
UV field. 

 Second, the shocks that are currently active in \oh\ are those traced
 by H$_\alpha$ emission, which arises in two bubble-like structures of
 shock-excited gas surrounding the molecular outflow
 \citep{rei87,san00opt,buj02}.  These fast shocks may have been generated
 by interaction between the dense, fast molecular outflow and the
 tenuous ambient material. The characteristics
 of the exciting shocks have been studied in detail by
 \cite{san00opt}. In particular, these authors compare the relative
 intensities of the different optical lines observed with the
 diagnostic diagrams by \cite{dop95}, which can
 distinguish between shocks with or without a photoionized preshock
 region. These diagrams not only confirm the shock nature of the
 emission but also indicate that the emission from a photoionized
 precursor region is either weak or absent. We note, moreover, that
 even in the improbable case that sufficiently intense UV radiation
 from the current shocks is produced, its effect on the chemistry will
 be almost exclusively limited to the outflow regions immediately
 behind the shock fronts, that is, the molecular clumps at the very
 end of the lobes of \oh. However, given the beam size of our
 observations (Fig.\ref{fig:oh231}), the contribution of these molecular clumps to
 the total emission by the N-bearing molecules reported in this work
 is insignificant.


\begin{thebibliography}{}

\bibitem[Adande et al.(2010)]{ada10} Adande, G.~R., Halfen, 
D.~T., Ziurys, L.~M., Quan, D., \& Herbst, E.\ 2010, \apj, 725, 561 

\bibitem[Ag{\'u}ndez 
\& Cernicharo(2006)]{marce06} Ag{\'u}ndez, M., \& Cernicharo, J.\ 2006, \apj, 650, 374 

\bibitem[Ag{\'u}ndez et al.(2007)]{marce07} Ag{\'u}ndez, M., 
Cernicharo, J., \& Gu{\'e}lin, M.\ 2007, \apjl, 662, L91 

\bibitem[Ag{\'u}ndez et 
al.(2008)]{marce08} Ag{\'u}ndez, M., Fonfr{\'{\i}}a, J.~P., Cernicharo, J., Pardo, J.~R., \& Gu{\'e}lin, M.\ 2008, \aap, 479, 493 

\bibitem[Ag{\'u}ndez(2009)]{marceTesis} Agundez, M.\ 2009, Ph.D.~Thesis, Universidad Aut\'onoma de Madrid

\bibitem[Ag{\'u}ndez et al.(2010)]{marce10} Ag{\'u}ndez, M., 
Cernicharo, J., \& Gu{\'e}lin, M.\ 2010, \apjl, 724, L133 

\bibitem[Ag{\'u}ndez et al.(2012)]{marce12} Ag{\'u}ndez, M., Fonfr{\'{\i}}a, J.~P., Cernicharo, J., et al.\ 2012, \aap, 543, A48 

\bibitem[Akyilmaz et 
al.(2007)]{aky07} Akyilmaz, M., Flower, D.~R., Hily-Blant, P., Pineau Des For{\^e}ts, G., \& Walmsley, C.~M.\ 2007, \aap, 462, 221 

\bibitem[Alcolea et al.(2001)]{alc01} Alcolea, J., Bujarrabal, V.,
  S{\'a}nchez Contreras, C., Neri, R., \& Zweigle, J.\ 2001, \aap,
  373, 932

\bibitem[Asplund et al.(2009)]{asp09} Asplund, M., Grevesse, N., Sauval, A.~J., \& Scott, P.\ 2009, \araa, 47, 481 

\bibitem[Audinos et al.(1994)]{aud94} Audinos, P., Kahane, C., \& Lucas, R.\ 1994, \aap, 287, L5 

\bibitem[Bachiller 
\& Cernicharo(1986)]{bac86} Bachiller, R., \& Cernicharo, J.\ 1986, \aap, 168, 262 

\bibitem[Balick 
\& Frank(2002)]{bal02} Balick, B., \& Frank, A.\ 2002, \araa, 40, 439 



\bibitem[Bohlin et al.(1978)]{boh78} Bohlin, R.~C., Savage, 
B.~D., \& Drake, J.~F.\ 1978, \apj, 224, 132 


\bibitem[Boothroyd et al.(1993)]{boo93} Boothroyd, A.~I., 
Sackmann, I.-J., \& Ahern, S.~C.\ 1993, \apj, 416, 762 


\bibitem[Bowers \& Morris(1984)]{bow84} Bowers, P.~F.,
  \& Morris, M.\ 1984, \apj, 276, 646   

\bibitem[Brown(1981)]{bro81} Brown, R.~L.\ 1981, \apjl, 248, 
L119 

\bibitem[Bujarrabal et 
al.(1994)]{buj94} Bujarrabal, V., Fuente, A., \& Omont, A.\ 1994, \aap, 285, 247 

\bibitem[Bujarrabal et 
al.(2001)]{buj01} Bujarrabal, V., Castro-Carrizo, A., Alcolea, J., \& S{\'a}nchez Contreras, C.\ 2001, \aap, 377, 868 

\bibitem[Bujarrabal et 
al.(2002)]{buj02} Bujarrabal, V., Alcolea, J., S{\'a}nchez Contreras, C., \& Sahai, R.\ 2002, \aap, 389, 271 

\bibitem[Bujarrabal et al.(2012)]{buj12} Bujarrabal, V., Alcolea, J.,
  Soria-Ruiz, R., et al.\ 2012, \aap, 537, A8

\bibitem[Burcat \& Rucsic(2005)]{bur05} Burcat, A. \& Ruscic, B. 2005,\ 'Third millenium ideal gas and condensed phase
thermochemical database for combustion with updates from active thermochemical tables',\ ANL-05/20 and TAE 960
Technion-IIT,\ Aerospace Engineering, and Argonne National Laboratory, Chemistry Division, September 2005.

\bibitem[Cabezas et al.(2013)]{cab13} Cabezas, C., 
Cernicharo, J., Alonso, J.~L., et al.\ 2013, \apj, 775, 133 

\bibitem[Carter et al.(2012)]{car12} Carter, M., Lazareff, B., Maier,
  D., et al.\ 2012, \aap, 538, A89

\bibitem[Castro-Carrizo et al.(2010)]{cc10} Castro-Carrizo, A., et al.\ 2010, \aap, 523, A59  

\bibitem[Cernicharo et 
al.(1984)]{cer84} Cernicharo, J., Guelin, M., \& Askne, J.\ 1984, \aap, 138, 371 

\bibitem[Cernicharo et al.(1985)]{cernicharo1} Cernicharo, J. 1985, Internal IRAM report (Granada:IRAM)

\bibitem[Cernicharo 
\& Guelin(1987)]{cer87} Cernicharo, J., \& Guelin, M.\ 1987, \aap, 183, L10 

\bibitem[Cernicharo et 
al.(2000)]{cer00} Cernicharo, J., Gu{\'e}lin, M., \& Kahane, C.\ 2000, \aaps, 142, 181 

\bibitem[Cernicharo(2012)]{madex} Cernicharo, J.\ 2012, EAS 
Publications Series, 58, 251 

\bibitem[Cernicharo et al.(2014)]{cer14} Cernicharo, J., 
Teyssier, D., Quintana-Lacaci, G., et al.\ 2014, arXiv:1410.5852 

\bibitem[Cohen(1981)]{coh81} Cohen, M.\ 1981, \pasp, 93, 288  

\bibitem[Cohen et al.(1985)]{coh85} Cohen, M., Dopita, M.~A., 
Schwartz, R.~D., \& Tielens, A.~G.~G.~M.\ 1985, \apj, 297, 702  

\bibitem[Chapillon et al.(2012)]{cha12} Chapillon, E.,
Dutrey, A., Guilloteau, S., et al.\ 2012, \apj, 756, 58 

\bibitem[Chase(1998)]{chs98} Chase, M.~W.\ 1998, 'NIST-JANAF Thermochemical Tables',\ J.\ Phys.\ Chem.\ Ref.\ Data,\ Monograph.\ 9,\ 4th ed.,\ Melville:\ AIP

\bibitem[Cherchneff et al.(1992)]{che92} Cherchneff, I., 
Barker, J.~R., \& Tielens, A.~G.~G.~M.\ 1992, \apj, 401, 269 

\bibitem[Cherchneff(2006)]{che06} Cherchneff, I.\ 2006, \aap, 456, 1001 


\bibitem[Choi et al.(2012)]{choi12} Choi, Y.~K., Brunthaler, 
A., Menten, K.~M., \& Reid, M.~J.\ 2012, IAU Symposium, 287, 407   

\bibitem[Churchwell et al.(1986)]{chu86} Churchwell, E., 
Wood, D., Myers, P.~C., \& Myers, R.~V.\ 1986, \apj, 305, 405 

\bibitem[Dalgarno(2006)]{dal06} Dalgarno, A.\ 2006, 
Proceedings of the National Academy of Science, 103, 12269 

\bibitem[Desmurs et al.(2007)]{des07} Desmurs, J.-F., Alcolea, J., Bujarrabal, V., S{\'a}nchez Contreras, C., \& Colomer, F.\ 2007, \aap, 468, 189 

\bibitem[Dopita 
\& Sutherland(1995)]{dop95} Dopita, M.~A., \& Sutherland, R.~S.\ 1995, \apj, 455, 468 

\bibitem[Draine 
\& Salpeter(1978)]{dra78} Draine, B.~T., \& Salpeter, E.~E.\ 1978, \nat, 271, 730 

\bibitem[Feast et al.(1983)]{fea83} Feast, M.~W., Catchpole, 
R.~M., Whitelock, P.~A., et al.\ 1983, \mnras, 203, 1207 

\bibitem[Frerking et al.(1979)]{fre79} Frerking, M.~A., 
Linke, R.~A., \& Thaddeus, P.\ 1979, \apjl, 234, L143 

\bibitem[Garrod et al.(2008)]{gar08} Garrod, R.~T., Weaver, 
S.~L.~W., \& Herbst, E.\ 2008, \apj, 682, 283 

\bibitem[Gerin et 
al.(1992)]{ger92} Gerin, M., Viala, Y., Pauzat, F., \& Ellinger, Y.\ 1992, \aap, 266, 463 

\bibitem[Goldsmith et al.(1982)]{gol82} Goldsmith, P.~F., 
Snell, R.~L., Deguchi, S., Krotkov, R., 
\& Linke, R.~A.\ 1982, \apj, 260, 147 

\bibitem[Goldsmith 
\& Langer(1999)]{gol99} Goldsmith, P.~F., \& Langer, W.~D.\ 1999, \apj, 517, 209

\bibitem[Gonz{\'a}lez Delgado et 
al.(2003)]{gon03} Gonz{\'a}lez Delgado, D., Olofsson, H., Kerschbaum, F., et al.\ 2003, \aap, 411, 123 

\bibitem[Guilloteau et al.(1986)]{gui86} Guilloteau, S., Lucas, R.,
  Omont, A., \& Nguyen-Q-Rieu 1986, \aap, 165, L1
  
\bibitem[Habing(1968)]{hab68} Habing, H.~J.\ 1968, \bain, 19, 
421 

\bibitem[Halfen et al.(2001)]{hal01} Halfen, D.~T., Apponi, 
A.~J., \& Ziurys, L.~M.\ 2001, \apj, 561, 244 

\bibitem[Iglesias(1977)]{igl77} Iglesias, E.\ 1977, \apj, 
218, 697 

\bibitem[Jackson et al.(1984)]{jac84} Jackson, J.~M., 
Armstrong, J.~T., \& Barrett, A.~H.\ 1984, \apj, 280, 608 

\bibitem[Jones et al.(1950)]{jon50a} Jones, L.~H., Shoolery, 
J.~N., Shulman, R.~G., \& Yost, D.~M.\ 1950, \jcp, 18, 990 

\bibitem[Jones \& Badger(1950)]{jon50b} Jones, L.~H., \& Badger, R.~M.\ 1950, \jcp, 18, 1511 


\bibitem[Jura \& Morris(1985)]{jur85} Jura, M., \& Morris, M.\ 1985, \apj, 292, 487 

\bibitem[Kastner et al.(1992)]{kas92} Kastner, J.~H., 
Weintraub, D.~A., Zuckerman, B., et al.\ 1992, \apj, 398, 552 

\bibitem[K{\l}os et al.(2008)]{klo08} K{\l}os, J., Lique, F., 
\& Alexander, M.~H.\ 2008, Chemical Physics Letters, 455, 1 

\bibitem[Kramer(1997)]{kra97} Kramer, C.\ 1997, Internal IRAM report (Granada:IRAM)

\bibitem[Kuan 
\& Snyder(1996)]{kua96} Kuan, Y.-J., \& Snyder, L.~E.\ 1996, \apj, 470, 981 

\bibitem[Li et 
al.(2013)]{li13} Li, J., Wang, J.~Z., Gu, Q.~S., \& Zheng, X.~W.\ 2013, \aap, 555, AA18 

\bibitem[Lindqvist et 
al.(1992)]{lin92} Lindqvist, M., Olofsson, H., Winnberg, A., \& Nyman, L.~A.\ 1992, \aap, 263, 183 

\bibitem[Lique et 
al.(2009)]{liq09} Lique, F., van der Tak, F.~F.~S., K{\l}os, J., Bulthuis, J., \& Alexander, M.~H.\ 2009, \aap, 493, 557 

\bibitem[Maercker et 
al.(2008)]{mae08} Maercker, M., Sch{\"o}ier, F.~L., Olofsson, H., Bergman, P., \& Ramstedt, S.\ 2008, \aap, 479, 779 

\bibitem[Mamon et al.(1988)]{mam88} Mamon, G.~A., Glassgold, 
A.~E., \& Huggins, P.~J.\ 1988, \apj, 328, 797 

\bibitem[Marcelino et 
al.(2010)]{mar2010} Marcelino, N., Br{\"u}nken, S., Cernicharo, J., et al.\ 2010, \aap, 516, A105 

\bibitem[Mart{\'{\i}}n et al.(2008)]{mar08} Mart{\'{\i}}n, 
S., Requena-Torres, M.~A., Mart{\'{\i}}n-Pintado, J., 
\& Mauersberger, R.\ 2008, \apj, 678, 245 


\bibitem[McBride et al.(2002)]{mcb02} McBride, B.~J., Zehe, M.~J., \&\ Gordon, S.\ 2002, 'NASA Glenn coefficients for calculating thermodynamic
properties of individual species',\ NASA report TP-2002-211556

\bibitem[Menten \& Alcolea(1995)]{men95} Menten, K.~M., \& Alcolea, J.\ 1995, \apj, 448, 416 

\bibitem[Menten et al.(2010)]{men10} Menten, K.~M., Wyrowski, F.,
  Alcolea, J., et al.\ 2010, \aap, 521, L7

\bibitem[Milam et al.(2009)]{mil09} Milam, S.~N., Woolf, 
N.~J., \& Ziurys, L.~M.\ 2009, \apj, 690, 837  


\bibitem[Morris et al.(1975)]{morr75} Morris, M., Gilmore, W., 
Palmer, P., Turner, B.~E., \& Zuckerman, B.\ 1975, \apjl, 199, L47 

\bibitem[Morris et al.(1976)]{morr76} Morris, M., Turner, 
B.~E., Palmer, P., \& Zuckerman, B.\ 1976, \apj, 205, 82 

\bibitem[Morris et al.(1987)]{morr87} Morris, M., Guilloteau, S.,
  Lucas, R., \& Omont, A.\ 1987, \apj, 321, 888

\bibitem[M{\"u}ller et al.(2005)]{mul05} M{\"u}ller, 
H.~S.~P., Schl{\"o}der, F., Stutzki, J., 
\& Winnewisser, G.\ 2005, Journal of Molecular Structure, 742, 215 

\bibitem[Neri et al.(1998)]{neri98} Neri, R., Kahane, C., Lucas, R.,
  Bujarrabal, V., \& Loup, C.\ 1998, \aaps, 130, 1

\bibitem[Neufeld(1990)]{neu90} Neufeld, D.~A.\ 1990, 
Molecular Astrophysics, 374 

\bibitem[Nguyen-Q-Rieu et 
al.(1991)]{ngu91} Nguyen-Q-Rieu, Henkel, C., Jackson, J.~M., \& Mauersberger, R.\ 1991, \aap, 241, L33 

 \bibitem[Omont et al.(1993)]{omo93} Omont, A., Lucas, R., Morris, M.,
   \& Guilloteau, S.\ 1993, \aap, 267, 490

\bibitem[Pardo et al.(2001)]{pardo1} Pardo, J.R., Cernicharo, J. \&
  Serabyn, E. 2001, IEEE Tras. Antennas and Propagation, (49, 12)

\bibitem[Pardo et al.(2004)]{par04} Pardo, J.~R., Cernicharo, 
J., Goicoechea, J.~R., \& Phillips, T.~G.\ 2004, \apj, 615, 495 

\bibitem[Pickett et al.(1998)]{pick98} H. M. Pickett, R. L. Poynter, E. A. Cohen, M. L. Delitsky,
 J. C. Pearson, and H. S. P. Muller,\ 1998, Journal of Quantitative Spectroscopy and Radiative Transfer  
60, 883-890

\bibitem[Planesas et al.(1990)]{pla90} Planesas, P., Bachiller, R., Martin-Pintado, J., \& Bujarrabal, V.\ 1990, \apj, 351,
263 

\bibitem[Quan et al.(2010)]{qua10} Quan, D., Herbst, E., 
Osamura, Y., \& Roueff, E.\ 2010, \apj, 725, 2101 

\bibitem[Quintana-Lacaci et 
al.(2013)]{qui13} Quintana-Lacaci, G., Ag{\'u}ndez, M., Cernicharo, J., et al.\ 2013, \aap, 560, L2 

\bibitem[Ramstedt \& Olofsson(2014)]{ram14} Ramstedt,
S., \& Olofsson, H.\ 2014, \aap, 566, A145

\bibitem[Reipurth(1987)]{rei87} Reipurth, B.\ 1987, \nat, 
325, 787 

\bibitem[Rodr{\'{\i}}guez-Fern{\'a}ndez et 
al.(2010)]{rod2010} Rodr{\'{\i}}guez-Fern{\'a}ndez, N.~J., Tafalla, M., Gueth, F., \& Bachiller, R.\ 2010, \aap, 516, A98 

\bibitem[Sabin et al.(2014)]{sab14} Sabin, L., Zhang, Q., 
Zijlstra, A.~A., et al.\ 2014, \mnras, 438, 1794 

\bibitem[Sahai \& Trauger(1998)]{sah98} Sahai, R., \& Trauger, J.~T.\ 1998, \aj, 116, 1357 

\bibitem[S{\'a}nchez Contreras et al.(1997)]{san97} S{\'a}nchez
  Contreras, C., Bujarrabal, V., \& Alcolea, J.\ 1997, \aap, 327, 689

\bibitem[S{\'a}nchez Contreras et al.(2000a)]{san00opt} S{\'a}nchez
  Contreras, C., Bujarrabal, V., Miranda, L.~F., \&
  Fern{\'a}ndez-Figueroa, M.~J.\ 2000, \aap, 355, 1103 (a)
  
\bibitem[S{\'a}nchez Contreras et al.(2000b)]{san00} S{\'a}nchez
  Contreras, C., Bujarrabal, V., Neri, R., \& Alcolea, J.\ 2000, \aap,
  357, 651 (b) 

\bibitem[S{\'a}nchez Contreras et 
al.(2002)]{san02} S{\'a}nchez Contreras, C., Desmurs, J.~F., Bujarrabal, V., Alcolea, J., \& Colomer, F.\ 2002, \aap, 385, L1 

\bibitem[S{\'a}nchez Contreras et al.(2004)]{san04} 
S{\'a}nchez Contreras, C., Gil de Paz, A., 
\& Sahai, R.\ 2004, \apj, 616, 519 


\bibitem[S{\'a}nchez Contreras et al.(2011)]{san11} S{\'a}nchez
  Contreras, C., Velilla Prieto, L., Cernicharo, J., et al.\ 2011, IAU
  Symposium, 280, 327P

\bibitem[S{\'a}nchez Contreras 
\& Sahai(2012)]{san12} S{\'a}nchez Contreras, C., \& Sahai, R.\ 2012, \apjs, 203, 16 

\bibitem[S{\'a}nchez Contreras et al.(2014)]{san14} S{\'a}nchez
  Contreras, C., Velilla, L., Alcolea, J., et al.\ 2014, Asymmetrical
  Planetary Nebulae VI conference, Proceedings of the conference held
  4-8 November, 2013.~Edited by C.~Morisset, G.~Delgado-Inglada and
  S.~Torres-Peimbert.~Online at
  {\tt http://www.astroscu.unam.mx/apn6/PROCEEDINGS/},  id.88

\bibitem[Sch{\"o}ier et 
al.(2007)]{sch07} Sch{\"o}ier, F.~L., Bast, J., Olofsson, H., \& Lindqvist, M.\ 2007, \aap, 473, 871 

\bibitem[Sch{\"o}ier et 
al.(2013)]{sch13} Sch{\"o}ier, F.~L., Ramstedt, S., Olofsson, H., et al.\ 2013, \aap, 550, A78 

\bibitem[Snyder 
\& Buhl(1972)]{sny72} Snyder, L.~E., \& Buhl, D.\ 1972, \apj, 177, 619 

\bibitem[Solomon et al.(1971)]{sol71} Solomon, P., Jefferts, 
K.~B., Penzias, A.~A., \& Wilson, R.~W.\ 1971, \apjl, 163, L53 

\bibitem[Tejero \& Cernicharo(1991)]{tej91} Tejero, M., Cernicharo,
 J., \ 1991, Modelos de equilibrio termodin{\'a}mico aplicados a
 envolturas circunestelares de estrellas evolucionadas (Madrid:IGN)%

\bibitem[Teyssier et 
al.(2006)]{tey06} Teyssier, D., Hernandez, R., Bujarrabal, V., Yoshida, H., \& Phillips, T.~G.\ 2006, \aap, 450, 167 

\bibitem[Tsuji(1973)]{tsu73} Tsuji, T.\ 1973, \aap, 23, 411 

\bibitem[Tsuji et 
al.(1997)]{tsu97} Tsuji, T., Ohnaka, K., Aoki, W., \& Yamamura, I.\ 1997, \aap, 320, L1 

\bibitem[Turner(1971)]{tur71} Turner, B.~E.\ 1971, \apjl, 
163, L35 

\bibitem[Turner et al.(1999)]{tur99} Turner, B.~E., Terzieva, 
R., \& Herbst, E.\ 1999, \apj, 518, 699 

\bibitem[Ukita 
\& Morris(1983)]{uki83} Ukita, N., \& Morris, M.\ 1983, \aap, 121, 15 

\bibitem[Velilla Prieto et al.(2013)]{vel13} Velilla Prieto, L.,
  S{\'a}nchez Contreras, C., Cernicharo, J., et al.\ 2013, Highlights
  of Spanish Astrophysics VII, 676

\bibitem[Wernli et 
al.(2007)]{wer07} Wernli, M., Wiesenfeld, L., Faure, A., \& Valiron, P.\ 2007, \aap, 464, 1147 
  
\bibitem[Westenberg \& Wilson (1950)]{wes50} Westenberg, A.~A. \& Wilson, E. Bright \ 1950, Journal of the American
Chemical Society, 72, 1, 199-200

\bibitem[Woodall et 
al.(2007)]{woo07} Woodall, J., Ag{\'u}ndez, M., Markwick-Kemper, A.~J., \& Millar, T.~J.\ 2007, \aap, 466, 1197 


\bibitem[Zijlstra et al.(2001)]{zij01} Zijlstra, A.~A., 
Chapman, J.~M., te Lintel Hekkert, P., et al.\ 2001, \mnras, 322, 280 

\bibitem[Zinchenko et 
al.(2000)]{zic00} Zinchenko, I., Henkel, C., \& Mao, R.~Q.\ 2000, \aap, 361, 1079 

\bibitem[Ziurys(2006)]{ziu06} Ziurys, L.~M.\ 2006, 
Proceedings of the National Academy of Science, 103, 12274 

\bibitem[Ziurys et al.(2007)]{ziu07} Ziurys, L.~M., Milam, 
S.~N., Apponi, A.~J., \& Woolf, N.~J.\ 2007, \nat, 447, 1094 

\bibitem[Ziurys et al.(2009)]{ziu09} Ziurys, L.~M., 
Tenenbaum, E.~D., Pulliam, R.~L., Woolf, N.~J., 
\& Milam, S.~N.\ 2009, \apj, 695, 1604 

\end{thebibliography}
\end{document}